\documentclass[preprint,12pt, 3p, sort&compress]{elsarticle}
\usepackage{fullpage}
\usepackage[T1]{fontenc}
\usepackage{amssymb,amsmath,amsthm,amsfonts,amsbsy,mathrsfs}
\usepackage{calligra}
\usepackage[caption=false]{subfig}

\usepackage{graphicx,color}
\usepackage{siunitx}
\usepackage{setspace}

\usepackage[version=3]{mhchem}

\usepackage{caption}
\usepackage{multirow}
\usepackage{booktabs}
\usepackage[shortlabels]{enumitem}

\usepackage[colorlinks   = true, urlcolor     = cyan, linkcolor    = blue, citecolor   = red]{hyperref}
\usepackage{bm}
\usepackage{pict2e}
\usepackage{epsfig}
\usepackage{epstopdf}
\usepackage{comment}
\usepackage{accents}
\usepackage{ulem}
\usepackage{url}
\usepackage[ruled,vlined]{algorithm2e}

\usepackage{tikz}
\usepackage{pgfplots}
\usepackage{pgfplotstable}
\usetikzlibrary{shapes.geometric, arrows.meta, arrows}
\usetikzlibrary{babel}
\usetikzlibrary{positioning}
\usetikzlibrary{calc}

\pgfplotsset{compat = newest}
\usepgfplotslibrary{external}
\usetikzlibrary{external}
\pgfplotsset{
    compat=newest,
    table/header=false,
    title style={font=\small},
    tick label style={font=\scriptsize},
    label style={font=\scriptsize},
    legend style={font=\scriptsize},
    legend cell align=left
}

\usepackage{tabularx,multirow,booktabs}
\newcolumntype{L}{>{\raggedright\arraybackslash}X}
\newcolumntype{C}{>{\centering\arraybackslash}X}
\newcolumntype{R}{>{\raggedleft\arraybackslash}X}

\newcommand\blfootnote[1]{%
  \begingroup
  \renewcommand\thefootnote{}\footnote{#1}%
  \addtocounter{footnote}{-1}%
  \endgroup
}

\journal{arXiv}

\newcommand{\rz}{\mathbb{R}}

\newcommand{\bfa}{{\bf a}}
\newcommand{\bfb}{{\bf b}}

\newcommand{\bfk}{{\bf k}}

\newcommand{\bfr}{{\bf r}}

\newcommand{\bfx}{{\bf x}}

\newcommand{\bfz}{{\bf z}}

\newcommand{\bfC}{{\bf C}}

\newcommand{\bfH}{{\bf H}}

\newcommand{\bfR}{{\bf R}}

\newcommand{\bfT}{{\bf T}}

\newcommand{\beq}{\begin{equation}}
\newcommand{\eeq}{\end{equation}}
\newcommand{\beqs}{\begin{eqnarray}}
\newcommand{\eeqs}{\end{eqnarray}}
\newcommand{\beql}{\begin{equation} \label}

\newcommand{\half}{\frac{1}{2}}

\newcommand{\calD}{{\cal D}}

\newcommand{\calF}{{\cal F}}
\newcommand{\calG}{{\cal G}}

\newcommand{\calN}{{\cal N}}

\newcommand{\calP}{{\cal P}}

\newcommand{\R}{\mathbb{R}}
\newcommand{\Z}{\mathbb{Z}}

\newcommand{\hpd}[3]{\frac{\partial^{#3}#1}{\partial#2^{#3}}}

\let\oldFootnote\footnote
\newcommand\nextToken\relax

\renewcommand\footnote[1]{%
    \oldFootnote{#1}\futurelet\nextToken\isFootnote}

\newcommand\isFootnote{%
    \ifx\footnote\nextToken\textsuperscript{,}\fi}

\definecolor{mycolor1}{rgb}{1.00000,0.00000,1.00000}

\begin{document}
\onehalfspacing

\begin{frontmatter}
\title{Carbon Kagome Nanotubes --- quasi-one-dimensional nanostructures with flat bands}

\author[uclamse]{Hsuan Ming Yu\corref{contrib}}
\ead{kingforever10@ucla.edu}
\author[uofm]{Shivam Sharma\corref{contrib}}
\ead{sharm608@umn.edu}
\author[uclaece]{Shivang Agarwal}
\ead{shivangag97@ucla.edu}
\author[uclamse]{Olivia Liebman}
\ead{oliebman@ucla.edu}
\author[uclamse]{Amartya S.\ Banerjee\corref{cor1}}
\ead{asbanerjee@ucla.edu}
\cortext[cor1]{Corresponding author. Tel:+1-763-656-7830. E-mail: asbanerjee@ucla.edu}
\cortext[contrib]{Authors contributed equally}
\address[uclamse]{Department of Materials Science and Engineering, University of California, Los Angeles, CA 90095, U.S.A}
\address[uofm]{Department of Aerospace Engineering and Mechanics, University of Minnesota, Minneapolis, MN 55455, U.S.A}
\address[uclaece]{Department of Electrical and Computer Engineering, University of California, Los Angeles, CA 90095, U.S.A}
\begin{abstract}
In recent years, a number of bulk materials and heterostructures have been explored due their connections with exotic materials phenomena emanating from flat band physics and strong electronic correlation. The possibility of realizing such fascinating material properties in simple realistic nanostructures is particularly exciting, especially as the investigation of exotic states of electronic matter in wire-like geometries is relatively unexplored in the literature. Motivated by these considerations, we introduce in this work carbon Kagome nanotubes (CKNTs) --- a new allotrope of carbon formed by rolling up Kagome graphene, and investigate this material using specialized first principles calculations. We identify two principal varieties of CKNTs --- armchair and zigzag, and find both varieties to be stable at room temperature, based on ab initio molecular dynamics simulations. CKNTs are metallic and feature dispersionless states (i.e., \textit{flat bands}) near the Fermi level throughout their Brillouin zone, along with an associated singular peak in the electronic density of states. We calculate the mechanical and electronic response of CKNTs to torsional and axial strains, and show that CKNTs appear to be more mechanically compliant than conventional carbon nanotubes (CNTs). Additionally, we find that the electronic properties of CKNTs undergo significant electronic transitions --- with emergent partial flat bands and tilted Dirac points --- when twisted. We develop a relatively simple tight-binding model that can explain many of these electronic features. We also discuss possible routes for the synthesis of CKNTs. Overall, CKNTs appear to be unique and striking examples of realistic elemental quasi-one-dimensional materials that may display fascinating material properties due to strong electronic correlation. Distorted CKNTs may provide an interesting nanomaterial platform where flat band physics and chirality induced anomalous transport effects may be studied together.
\blfootnote{Abbreviations: CKNT - Carbon Kagome Nanotube, CNT - Carbon Nanotube, QBC - quadratic band crossing, TB - tight-binding, 1D - quasi-one-dimensional, 2D - quasi-two-dimensional.}
\end{abstract}
 
\end{frontmatter}
\section{Introduction}
\label{sec:introduction}

Over the past two decades, the design, discovery and characterization of nanomaterials and nanostructures with special features in their electronic band structure has gained prominence. Exemplified famously by the case of linear dispersive relations in graphene (associated with massless Dirac fermions \citep{novoselov2005two, TwoDimGasMasslessFermions_Geim_2005}), such  features often point to the existence of exotic electronic states, and the possibility of realizing unconventional electromagnetic, transport and optical properties in real systems. In recent years, materials and structures featuring dispersionless electronic states or \textit{flat bands} have been heavily investigated due to the fact that electrons associated with such states have quenched kinetic energies\footnote{While linearly dispersive states are associated with extremely high charge carrier mobilities, flat band electrons are massive 
and effectively have zero group velocity.} (are spatially localized), and interact in the strong correlation regime\citep{balents2020superconductivity, yin2019negative, derzhko2015strongly, elias2021flat}. This 
manner of interaction leads to a variety of fascinating materials phenomena, including superconductivity \citep{balents2020superconductivity,iglovikov2014superconducting}, ferromagnetism \citep{pons2020flat}, Wigner crystallization \citep{wu2007flat, shayegan2022wigner}, and the fractional quantum Hall effect \cite{wang2012fractional,tang2011high, sun2011nearly, neupert2011fractional}.

Along these lines,  moir\'e superlattices in twisted bilayers \citep{andrei2021marvels, wang2020correlated, cao2018unconventional, pons2020flat, choi2021correlation, cao2018correlated} and materials with tailored atomic lattices \citep{ArtificialFlatBandSystems_Leykam, OneDimLatticeStrips_Huda, li2022lieb, Barreteau_2017_kagome} have received much attention since they feature flat bands and rich electron physics. In order to observe and maintain desirable electronic properties, such  systems usually involve some degree of engineering, and more often, a fine control over important system parameters (e.g. bilayer twists at specific \textit{magic angles} \citep{tarnopolsky2019origin}, or a critical magnetic field strength in quasi-two-dimensional (2D) network structures \citep{naud2001aharonov, vidal1998aharonov}). Therefore, a strand of recent investigations has focused on the synthesis and characterization of materials which feature electronic flat bands due to their natural atomic arrangements \citep{ortiz2019new, li2021dirac, kang2020dirac, liu2018giant, liu2019magnetic, neupert2022charge, jiang2021unconventional, han2016correlated, liu2021screening}. Exotic electronic states hosted by such materials can be stable with respect to perturbations such as changes in temperature or applied strains, and they can often display such states without external fields --- features which make them suitable for device applications. Alongside these experimental studies, computational investigations have also  probed elemental versions of such materials, i.e., stable bulk or 2D nanomaterials made of a single species of atom that can feature unusual electronic states by virtue of their atomic arrangements alone \citep{ZHONG201665_kagome, zhong2016coexistence, partoens_kagome, Sarikavak-Lisesivdin2020, zhang_marvin_prl, FerromagnetismAndWignerCrystallization, Sarikavak-Lisesivdin2020}. The present study extends this particular line of work to the important case of quasi-one-dimensional (1D) nanomaterials featuring flat bands, which have generally received far less attention in the literature.\footnote{There has been recent work (see e.g. \citep{kennes2020one, elias2021flat, li2023one, schleder2023one}) on quasi-two-dimensional systems and heterostructures, featuring one-dimensional flat bands, or flat bands along specific directions of a two-dimensional Brillouin zone. In contrast, the systems studied here are all elemental quasi-one-dimensional (1D) nanostructures that feature flat bands throughout the entirety of their (one-dimensional) Brillouin zone.}  

The possibility of realizing flat band physics and exotic states of electronic matter in wire-like geometries is particularly exciting.  While well known theoretical considerations \citep{Hohenberg_1967, mermin1966absence} appear to preclude the existence of long range order in low dimensional systems (necessary, e.g. in realizing superconducting states), such restrictions do not necessarily apply to the quasi-one-dimensional structures considered here  \citep{SuperconductivityIn1D}. In fact, there are reasons to expect that the screw transformation symmetries and quantum confinement effects often associated with such systems can actually result in enhancement of collective or correlated electronic properties \citep{Ceperly_Martin_Interacting_Electrons_Book, James_OS, kim2022strain}, and that such properties are likely to be manifested in manners that are quite different from bulk phase materials. In particular, materials such as the ones considered in this work can be \textit{chiral} --- due to intrinsic or applied twists --- and therefore, feature anomalous transport (the Chiral Induced Spin Selectivity, or CISS effect \citep{aiello2020chirality, naaman2015spintronics}). The exploration of simultaneous manifestations of such effects along with correlated electron physics has begun fairly recently \citep{nakajima2023giant, linder2015superconducting, sidorenko2018functional, qin2017superconductivity}. We posit that the carbon nanostructures explored here are likely to emerge as a possible material platform for such studies in the future.

Due to its unique allotrope forming features and versatility \citep{hirsch2010era, zhang2019art, hoffmann2016homo}, carbon is particularly attractive as a building block of novel materials. A large number of computational studies have recently been devoted to 2D and bulk allotropes of carbon displaying Dirac cones, flat bands and non-trivial topological states \citep{li2020three, wang2015phagraphene, zhong2016coexistence, chen2020topological, yan2021newly, You_Gu_2019, shi2021high, zhou2014s, maruyama2017interplay}. On the other hand, although several 1D allotropes of carbon are well known \citep{endo2013carbon, iijima1993single, wakabayashi1999electronic, son2006energy, yu2008first, Zhang2017_y_carbon, FamiliesOfCNTs_Legoas, coluci2004new, AlphaGraphyne_Kang_Lee,  coluci2004theoretical, GraphyneNanotubes_Kang, yang2017novel}, none of these systems are usually associated with flat band physics. As far as we can tell, there have been only a few earlier attempts at producing and investigating flat bands in realistic 1D nanomaterials of carbon: partially flat bands in zigzag graphene nanoribbons \citep{wakabayashi2009electronic}, spin polarized flat bands in hydrogenated carbon nanotubes  \citep{yang2009itinerant}, and moir\'e type flat bands in chiral carbon nanotubes with collapsed structures \citep{MoireSuperlatticeFlatBands, arroyo2023universality} or incommensurate double wall geometries \citep{koshino2015incommensurate}. Our contribution aims to address this particular gap in the literature by studying a family of realistic 1D carbon nanostructures that naturally feature flat bands throughout their  Brillouin zone. The flat bands in the structures presented here arise out of geometric and orbital frustration and without the aid of dopant or major structural instability effects, as employed in the aforementioned studies. Moreover, as we demonstrate, these dispersionless electronic states show fascinating transitions as the structures are subjected to strains, while also proving to be robust and retaining many desirable characteristics in some respects.

To obtain a 1D carbon nanostructure with flat bands, our starting point is a planar sheet of Kagome graphene. We then ``roll up'' this 2D material along different directions to obtain Carbon Kagome Nanotubes (CKNTs). Kagome graphene and related bulk structures have recently received much attention in the materials literature \citep{zhang_marvin_prl, Sarikavak-Lisesivdin2020, FerromagnetismAndWignerCrystallization, ZHONG201665_kagome, partoens_kagome,Liu_2014_kagome, wavrunek2022mechanical}, and also in the physics literature, where the material is often identified as a ``decorated honeycomb'' or ``star lattice'' structure \citep{ ruegg2010topological, lopez2020magnetism, lopez2022topological, merino2021unconventional, chen2018quantum, rhim2020quantum, Barreteau_2017_kagome}. Although Kagome graphene remains to be experimentally synthesized, successful synthesis of a variant of this material, i.e., nitrogen-doped graphene on a silver substrate --- a 2D material with a Kagome pattern, has been carried out recently \cite{pawlak2021surface, li2022selective}. Moreover, synthesis of novel complex nanotube structures in general (see e.g. \citep{korde2022single}) and through the roll-up of 2D sheets in particular, is fairly common (see e.g. \citep{wang2021bamboo}), thus suggesting that CKNTs can be synthesized in the near future.

In this work, we introduce CKNTs, and carry out a thorough and systematic first principles characterization of this material  in terms of its structural, mechanical and electromechanical properties. Wherever relevant, we provide comparisons of the properties of CKNTs against those of conventional carbon nanotubes (CNTs). All CKNTs studied here are metallic and feature flat bands (throughout their Brillouin zone) near the Fermi level, along with an associated singular peak in the electronic density of states. We show in particular that CKNTs appear to be more mechanically compliant when compared against CNTs, and that their electronic properties undergo significant electronic transitions --- with emergent partial flat bands and Dirac points --- when subjected to torsional strains. Our studies are made possible largely due to a suite of recently developed symmetry adapted electronic structure calculation techniques \citep{My_PhD_Thesis, banerjee2021ab, yu2022density, ghosh2019symmetry, banerjee2016cyclic, ML_HelicalDFT, agarwal2022solution}, that allow ab initio calculations of 1D materials and their deformed states to be carried out accurately and efficiently. We also develop a $\pi$-electron based tight binding model that includes up to next-nearest-neighbor interactions,  which is able to capture many of the electronic properties of CKNTs, as revealed via first principles data. 

The rest of the paper is organized as follows: Section \ref{sec:materials&method} describes the geometry of the materials under study (subsection \ref{sec:kagome_graphene}), as well as various aspects of the specialized first principles computational techniques used in this work (subsection \ref{sec:computational_methods}). Section \ref{sec:results&discussion}
presents results, touching on structural (subsection \ref{subsec:structural_properties}), mechanical (subsection \ref{subsec:mechanical_properties}), electronic (subsection \ref{subsec:electronic_prop_undistorted}) and electromechanical 
 (subsection \ref{subsec:electromechanical_prop}) aspects. We discuss possible routes to the synthesis of CKNTs in subection \ref{subsec:Synthesis} and conclude in Section \ref{sec:conclusions}. Details of the tight binding model for CKNTs are presented in Appendix \ref{sec:app_TBM}.

\section{Material and Methods}
\label{sec:materials&method}
In this section, we introduce the geometry of Carbon Kagome nanotubes (CKNTs) and their construction from Kagome graphene through the  ``roll up'' procedure commonly employed in other similar types of nanomaterials \citep{Dresselhaus_CNT_textbook, Dumitrica_James_OMD}. We also provide an outline of the various computational and theoretical methods used in our study.
\subsection{From Kagome Graphene to Carbon Kagome Nanotubes}
\label{sec:kagome_graphene}
\begin{figure}[htbp]
    \centering
    \includegraphics[trim={1cm 1cm 3cm 0.5cm}, clip, width=0.5\linewidth]{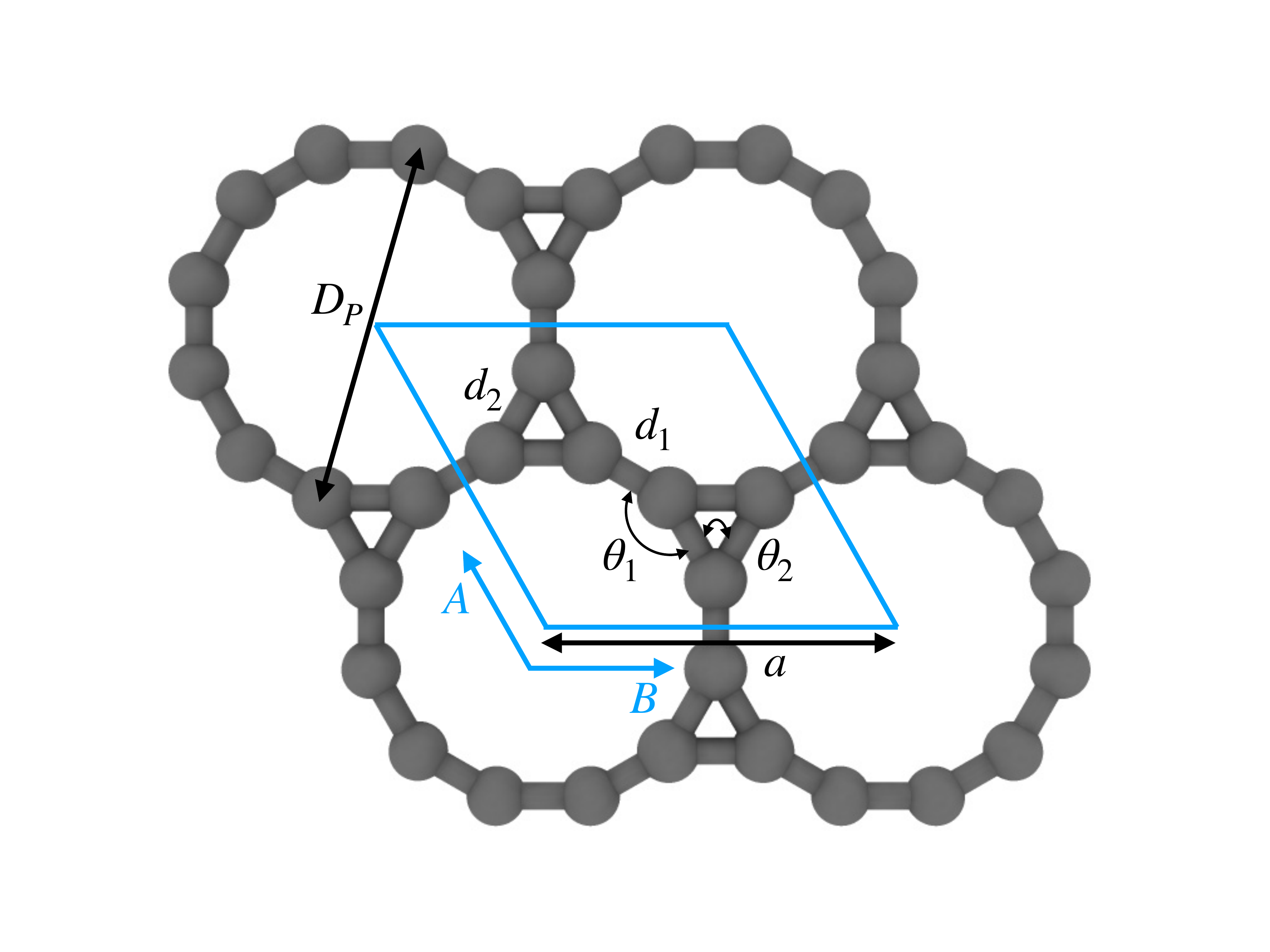}
    \caption{Unit cell of Kagome Graphene with various structural parameters indicated. The angle $\theta_1$ is $150^{\circ}$, while $\theta_2$ is $60^{\circ}$. The other parameters can be found in Table \ref{tab:ckl_parameters}.}
    \label{fig:hexagonal_ckl}
\end{figure}

Several recent studies have explored the structure of Kagome graphene sheets and related bulk structures  \citep{partoens_kagome,Sarikavak-Lisesivdin2020,Liu_2014_kagome,FerromagnetismAndWignerCrystallization,ZHONG201665_kagome,Barreteau_2017_kagome}. As a starting point, we first consider the geometry of Kagome graphene. The hexagonal unit cell of this 2D material consists of 6 carbon atoms that form a pair of equilateral triangles, as shown in Fig.~\ref{fig:hexagonal_ckl}. 
We used the planewave code ABINIT \citep{Gonze_ABINIT_1, romero2020abinit, gonze2016recent} to optimize the geometry of this structure. Our calculations employed norm conserving Troullier Martins pseudopotentials \citep{Troullier_Martins_pseudo}, an energy cutoff of $50$ Ha, $21 \times 21 \times 1$ k-point sampling and Fermi-Dirac smearing of $0.001$ Ha. These parameters were  sufficient to produce accurate energies, forces and cell stresses for the pseudopotentials chosen \citep{yu2022density}.  We employed both Perdew-Wang \citep{Perdew_Wang} local density approximation (LDA) and Perdew-Burke-Ernzerhof (PBE)\citep{perdew1996generalized} generalized gradient approximation (GGA) exchange correlation functionals. At the end of the relaxation procedure, the atomic forces were typically of the order of $10^{-5}$ Ha/Bohr, while the cell stresses were of the order of $10^{-8}\,\text{Ha/Bohr}^3$. 

\begin{figure}[htbp]
    \centering
    \includegraphics[trim={1cm 1cm 3cm 0.5cm}, clip, width=0.6\linewidth]{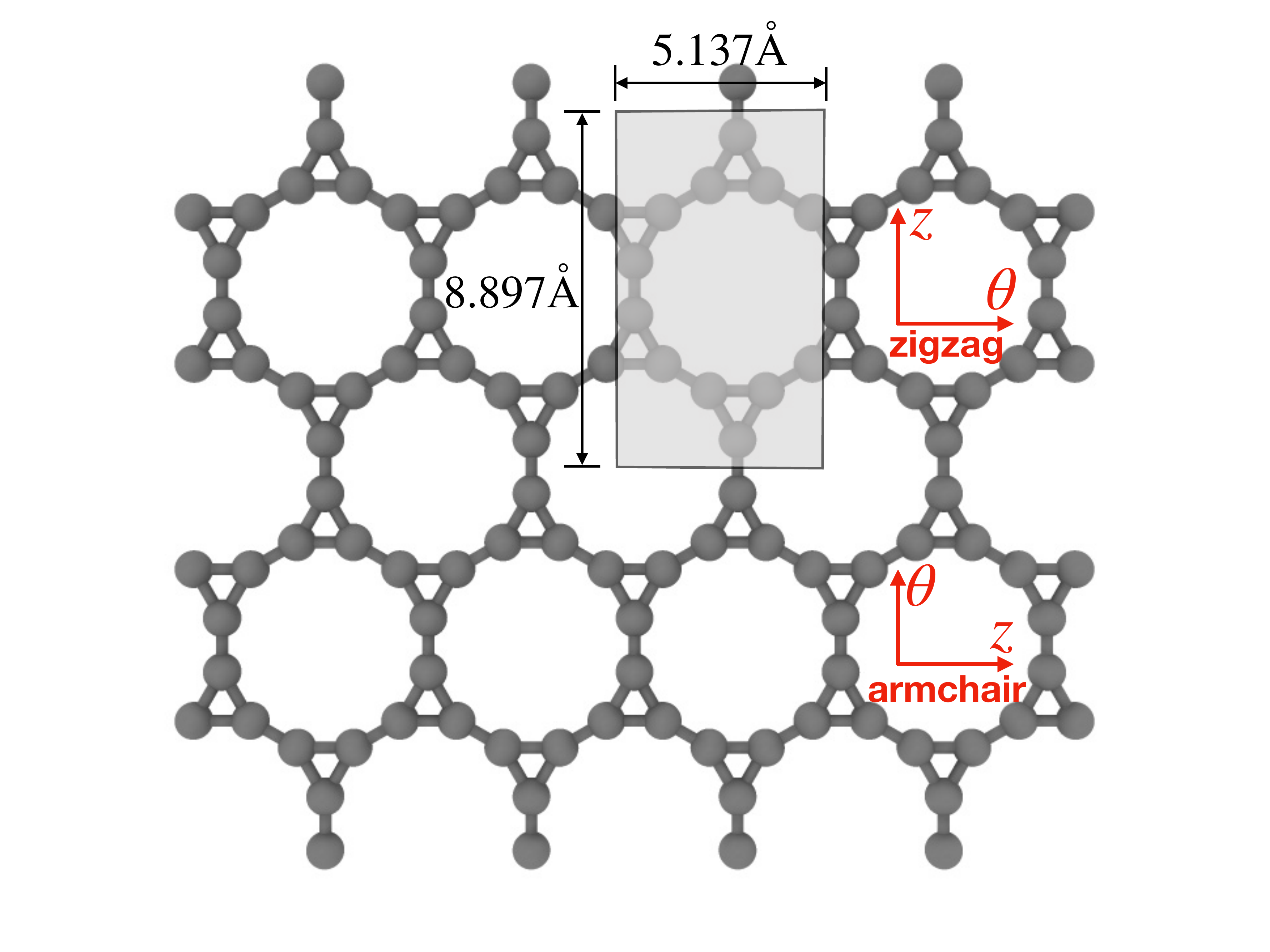}
    \caption{Roll-up construction of CKNTs, starting from a sheet of Kagome graphene. $\theta$ denotes the direction of roll up, while $z$ denotes the tube axis direction. The $12$ atoms shown in the shaded region are the representative atoms in the fundamental domain used for Helical DFT \citep{banerjee2021ab, yu2022density} calculations. The domain size parameters illustrated above correspond to calculations based on LDA exchange-correlation.}
    \label{fig:orth_ckl_rolling}
\end{figure}

\begin{figure}[htbp]
  \centering
    \subfloat[][Armchair (6,6)  CKNT]{\includegraphics[width=.3\linewidth]{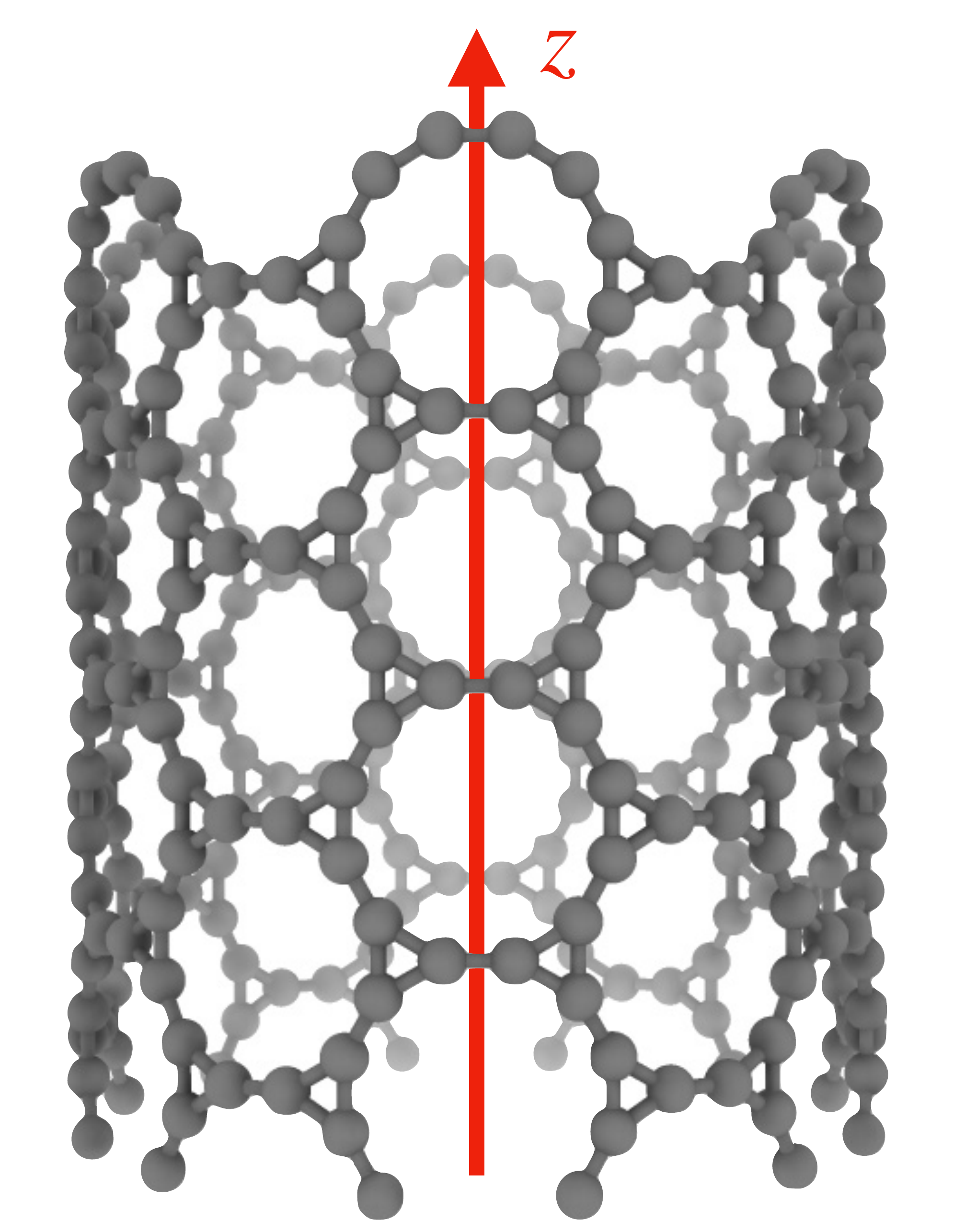}\label{fig:CKL_armchair}} \hspace{4cm}
    \subfloat[][Zigzag (9,0)  CKNT]{\includegraphics[width=.3\linewidth]{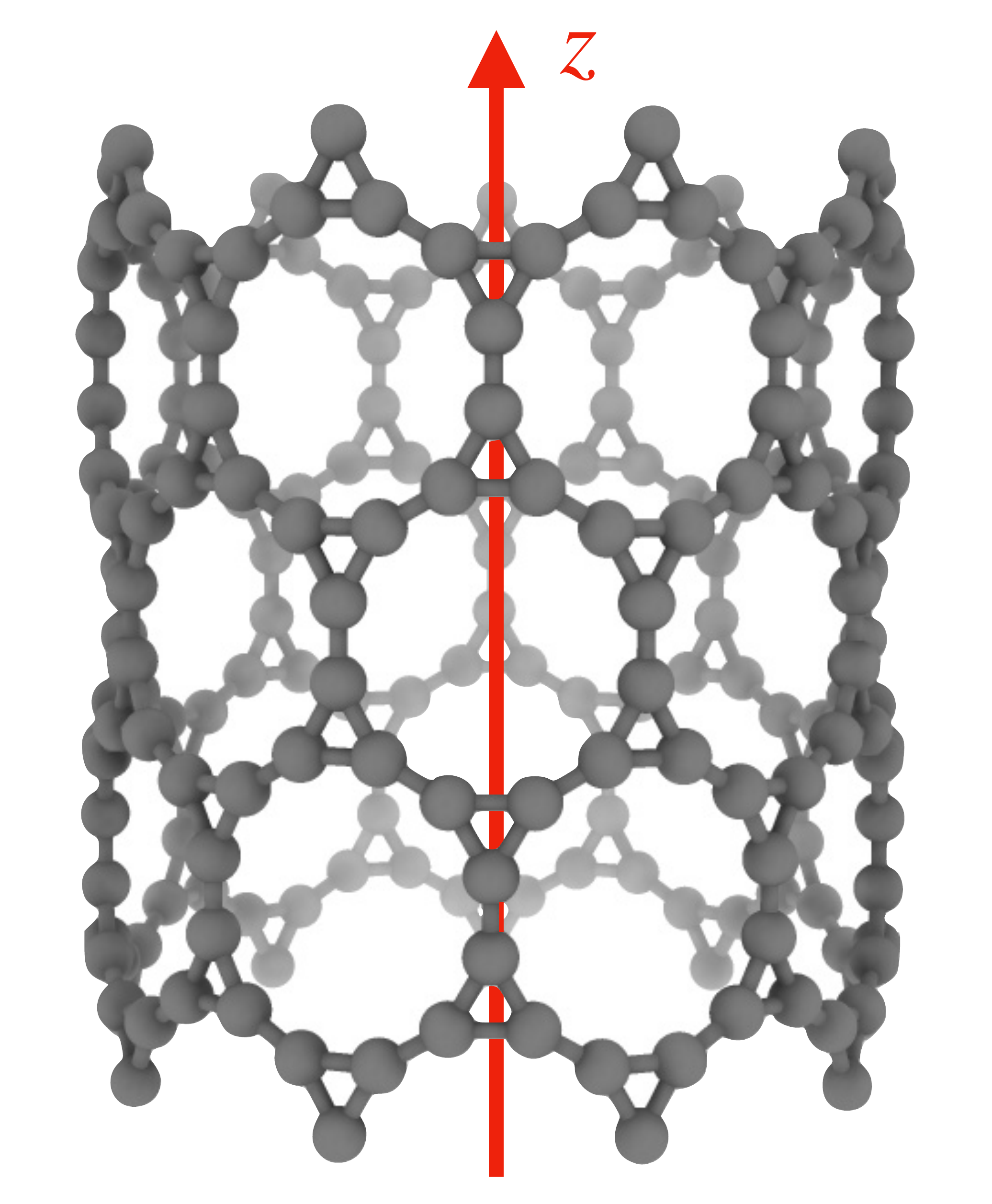}\label{fig:CKL_zigzag}}
  \caption{Two varieties of CKNTs investigated in this work: (a) Armchair $(n,n)$ and (b) Zigzag $(n,0)$ tubes. The tube radii are $0.85$ nm and $0.74$ nm, respectively for the above examples. $n$ is the cyclic symmetry group order about the tube axis.}
\label{fig:CKNTs}
\end{figure}

Table \ref{tab:ckl_parameters} shows that the optimized structural parameters obtained by us are in very good agreement with the literature. As expected \citep{van1999correcting}, the LDA bond lengths are somewhat shorter than those obtained through functionals involving gradient corrections, although the variations observed are quite minor overall. Notably, the intertriangle C-C bond length was found to be slightly smaller than the bond length corresponding to the triangle sides. Additionally, the calculated bond angles were found to be $60^{\circ}$ (in-triangle) and $150^{\circ}$ (inter-triangle) almost perfectly, consistent with the literature.  
\begin{table*}[htbp]
\centering
\begin{tabular}[b]{ |c|c|c| }
  \hline
  Parameters & Current work & Literature\\[3pt]
  \hline
  $a$ (\r{A}) & $5.1370^a(5.1662)^b$ & $5.2085^c, 5.2087^d, 4.46^e$\\
  $d_{1}$ (\r{A}) & $1.3402^a(1.3400)^b$ & $1.3559^c, 1.3567^d, 1.50^e$\\
  $d_{2}$ (\r{A}) & $1.4078^a(1.4206)^b$ & $1.4305^c, 1.4299^d, 1.53^e$\\
  $D_{p}$ (\r{A}) & $5.3090^a(5.3331)^b$ & $5.3817^c, 5.3829^d$\\

  \hline
\end{tabular}
\caption{Optimized structural parameters of Kagome graphene. Superscripts denote parameters obtained using: (a) LDA functional (this work), (b) GGA functional (this work), (c) SGGA-PBE functional (reference \citep{Sarikavak-Lisesivdin2020}), (d) SGGA-PBE functional  with Grimme D3 correction (reference \citep{Sarikavak-Lisesivdin2020}), and (e) GGA functional using the bulk structure (reference \citep{zhang_marvin_prl}).}
\label{tab:ckl_parameters}
\end{table*}
Next, following the construction of carbon nanotubes from graphene \citep{CNT_hamada,CNT_sinnott, Dresselhaus_CNT_textbook}, we rolled up the optmized flat Kagome graphene structures into seamless cylinders and arrived at carbon Kagome nanotubes (see Fig.~\ref{fig:orth_ckl_rolling}). Depending on the direction of rolling, the tubes maybe armchair, zigzag or chiral, with non-negative integers $(n,m)$ denoting the chirality indices. In this work, we focus exclusively on armchair (i.e., $(n,n)$) and zigzag (i.e., $(n,0)$) nanotubes (illustrated in Fig.~\ref{fig:CKNTs}). The index $n$ for such achiral tubes indicates the degree of cyclic symmetry about the tube axis.
An investigation of chiral CKNTs is the scope of future work. Notably, the replacement of hexagons in conventional CNTs, by dodecagonal rings in CKNTs, results in a structure with more porous sidewalls, and suggests the use of this material in filtration \citep{roy2012formation}, desalination \citep{corry2008designing} and electrochemical storage aplications \citep{frackowiak2002electrochemical, frackowiak1999electrochemical, nutzenadel1998electrochemical}.

\subsection{Computational Methods}
\label{sec:computational_methods}
We outline the computational methods employed in this work to study CKNTs. These include specialized, real-space symmetry adapted first principles simulations (using the Helical DFT code \citep{banerjee2021ab, yu2022density}) and periodic plane-wave calculations (based on the ABINIT \citep{Gonze_ABINIT_1, gonze2016recent, romero2020abinit}, Quantum Espresso \citep{Quantum_Espresso_1, giannozzi2017advanced, giannozzi2020quantum} and PWDFT \citep{hu2015dgdft, lin2012adaptive, jia2019parallel,hu2017adaptively} codes), to augment the Helical DFT results. A tight-binding model which is able to replicate many of the electronic properties revealed by the above first principles calculations is detailed separately in \ref{sec:app_TBM}. In what follows, the atomic unit system with $m_e = 1, \hbar = 1, \frac{1}{4 \pi \epsilon_0} = 1$ is used throughout, unless mentioned otherwise.  We will denote the standard orthonormal basis of $\R^3$  with $\textbf{e\textsubscript{X}}$, $\textbf{e\textsubscript{Y}}$, $\textbf{e\textsubscript{Z}}$. Lowercase boldface letters will denote vectors in three dimensions, while uppercase boldface will denote  matrices.
\subsubsection{Helical DFT calculations}
\label{sec:helical dft}

The majority of the first principles calculations in this work have been carried out using Helical DFT \citep{banerjee2021ab, yu2022density} --- a symmetry-adapted real-space formulation of Kohn-Sham Density Functional Theory \citep{KohnSham_DFT}. We now highlight a few important features and technical details of this specialized computational technique. A key advantage of the methodology is its ability to efficiently simulate quasi-one-dimensional materials by exploiting global structural symmetries, which  enables it to reduce calculations to just a few representative atoms within the simulation cell. The nanotubes of interest to this work, both in pristine or deformed configurations, can be conveniently described using helical and cyclic symmetries about a common axis \citep{James_OS, hakobyan2012objective, Dumitrica_James_OMD}. Specifically, for a nanotube with axis $\textbf{e\textsubscript{Z}}$, if $\calP = \{\bfr_1, \bfr_2,\ldots, \bfr_M: \bfr_i \in \rz^3\}$ are the coordinates of the representative atoms in the simulation cell (or \textit{fundamental domain}), then the collection of coordinates of the entire structure can be expressed as:
\begin{align}
\mathcal{S} = \underset{\mu=0,1,\cdots,\mathfrak{N}-1}{\bigcup_{\zeta \in \mathbb{Z}}}\!\!\bigcup^{M}_{i=1} {\textbf{R}}_{(2\pi \zeta\alpha+\mu\Theta)}\mathbf{r}_{i}+\zeta\tau\,\textbf{e\textsubscript{Z}}\,.
\label{Eqn:sim_atoms}
\end{align}
Thus, the symmetry group of the nanotube consists of the collection of isometries (i.e., rotations and translations):
\beqs
\mathcal{G} = \Big\{ \Upsilon_{\zeta,\mu}=\big(\bfR_{(2\pi \zeta\alpha+\mu\Theta)}|
\,\zeta\tau \textbf{e\textsubscript{Z}}):&& \nonumber\\ 
 && \hspace{-1.7cm}\zeta \in \mathbb{Z},\mu=0,1,\dots,\mathfrak{N} - 1\Big\}\,.
\eeqs
Here, each symmetry operation $\Upsilon_{\zeta,\mu}$ consists of a rotation operation  about $\textbf{e\textsubscript{Z}}$ through the angle $2\pi \zeta\alpha+\mu\Theta$ (denoted via the action of the rotation matrix ${\textbf{R}}_{(2\pi \zeta\alpha+\mu\Theta)}$ above), along with simultaneous translation by $\zeta\tau$ about the same axis. The quantity $\mathfrak{N}$ is a natural number that captures cyclic symmetries in the nanotube, with the angle $\Theta=2\pi/\mathfrak{N}$ (i.e., $\mathfrak{N}$ is the same as $n$ in armchair $(n,n)$ and zigzag $(n,0)$ nanotubes). The scalar $\alpha$ is related to the applied or intrinsic twist in the structure, with the amount of twist per unit length measured as $\beta = 2\pi\alpha/\tau$. The parameter $\tau$ is related to the pitch of screw transformation symmetries in the nanotube, and variations in this quantity enable extensions and compressions about the tube axis to be captured.

For the CKNTs considered in this work, we have employed an orthogonal unit cell with $12$ representative atoms, as shown in Fig.~\ref{fig:orth_ckl_rolling}. The parameter $\alpha$ takes values between $0$ and $1$, with $\alpha = 0$ representing untwisted structures. 
The value of $\tau$ as suggested by the roll-up construction \citep{CNT_5, Dumitrica_James_OMD} is $5.137$ \mbox{\normalfont\AA} and $8.897$ \mbox{\normalfont\AA} for undeformed armchair and zigzag CKNTs, respectively.

Helical DFT uses a higher order finite difference discretization scheme in \textit{helical coordinates} \citep{My_PhD_Thesis, banerjee2021ab} to solve the following symmetry adapted equations of Kohn-Sham DFT over the fundamental domain:
\begin{align}
\mathfrak{H}^{\text{KS}}\,\psi_j(\bfx;\eta,\nu) = \lambda_j(\eta,\nu)\,\psi_j(\bfx;\eta,\nu)\,.
\label{Eqn:ks_eqn_1}
\end{align}
The Kohn-Sham Hamiltonian operator:
\begin{align}
\mathfrak{H}^{\text{KS}} = -\frac{1}{2}\Delta+V_{\text{xc}}+\Phi+ \mathcal{V}_{\text{nl}},,    
\end{align}
consists of kinetic energy, exchange correlation, electrostatic contribution (i.e., both electron-electron and electron-nucleus interactions) and non-local psudopotential \cite{kleinman1982efficacious} terms. The symmetry-adapted quantum numbers $\eta \in [-\half,\half)$, and $\nu \in\{0,1,2,\ldots,\mathfrak{N}-1\}$ serve to label the eigenstates and electronic occupation numbers of the system (analogous to `k-points' in periodic calculations of solids), along helical and cyclic symmetry directions, respectively. At the end of self-consistent solution of the above equations, the system's ground state electronic free energy per unit fundamental domain may be calculated. We  will denote this quantity as ${\calF}_{\substack{\text{Ground}\\ \text{State}}}(\calP, \calD, \calG)$ to signify its explicit dependence on the positions of the representative atoms within the fundamental domain ($\calP$), the fundamental domain itself ($\calD$), and the symmetry group of the structure under consideration ($\calG$). By introducing variations in $\calG$ and by minimizing ${\calF}_{\substack{\text{Ground}\\ \text{State}}}(\calP, \calD, \calG)$ with respect to the coordinates of the representative atoms, Helical DFT can be employed for \textit{ab initio} exploration of the deformation response of a nanomaterial under torsional or axial loads \citep{yu2022density}.

In order to facilitate the large number of ground state electronic structure calculations of various CKNTs as well as their electromechanical response under applied deformations, Helical DFT simulations were carried out in three stages. First, for a given nanotube and applied strain parameters, ab initio structural relaxation calculations were carried out using a mesh spacing of $h = 0.3$ Bohr, and by sampling $15$ reciprocal space points in the $\eta$ direction. These discretization choices are sufficient to produce chemically accurate forces and ground state energies for the norm conserving pseudopotential \cite{kleinman1982efficacious, troullier1991efficient} used to model the carbon atoms in this work \citep{yu2022density}. Atomic relaxation was carried using the Fast Inertial Relaxation Engine \citep{bitzek2006structural} and calculations were continued until each atomic force component on every atom in the simulation cell reached $0.001$ Ha/Bohr or lower. Next, for each relaxed structure, we redid a self-consistent calculation using the finest discretization parameters that could be reliably employed within computational resource constraints. This corresponds to a mesh spacing of $h = 0.25$ Bohr and $21$ reciprocal space points in the $\eta$ direction. This calculation step enables accurate calculations of energetics and stiffness parameters to be carried out \citep{yu2022density}. Finally, using the self consistent densities and potentials obtained through the above step, we performed a single Hamiltonian diagonalization step using $45$ reciprocal space points in the $\eta$ direction. The eigenstates so obtained were used for band diagram and electronic density of states calculations that are presented in Section \ref{sec:results&discussion}.

For all Helical DFT calculations, we used the Perdew-Wang parametrization \cite{perdew1992accurate} of the LDA\cite{KohnSham_DFT}. We did not observe any major qualitative differences in the electronic properties between the LDA results presented here and those produced using gradient corrected functionals \cite{perdew1996generalized}. We also employed $12^{\text{th}}$ order finite differences \cite{chelikowsky1994finite, Chelikowsky_Saad_1,  Chelikowsky_Saad_2, kikuji2005first, ghosh2017sparc_1, ghosh2017sparc_2, ghosh2019symmetry, banerjee2021ab}, vacuum padding of $10$ Bohrs in the radial direction and $1$ milli-Hartree of smearing using the Fermi-Dirac distribution. The pseudopotential employed was generated using the Troullier-Martins scheme \cite{Troullier_Martins_pseudo, troullier1991efficient}, and is identical to the one used for the flat sheet calculations presented earlier (Section \ref{sec:kagome_graphene}).

\subsection{Plane-wave DFT calculations}
\label{sec:Plane_Wave}
We used plane-wave DFT \citep{Hutter_abinitio_MD, Martin_ES} for carrying out a few additional first principles calculations of CKNTs. Specifically, we investigated the dynamic stability of undistorted CKNTs by performing \textit{ab inito} molecular dynamics (AIMD) simulations. The highly scalable PWDFT code \citep{hu2015dgdft, lin2012adaptive, jia2019parallel,hu2017adaptively} was used for this purpose. We investigated two generic CKNTs --- one each of the zigzag and armchair varieties --- with radii of $0.75$ and $0.85$ nm for our simulations. In order to capture long range deformation modes, we considered atoms beyond the minimal periodic unit cell and chose multiple layers of the tubular structures along the axial direction. This resulted in supercells containing $144$ and $216$ atoms for the armchair and zigzag variety tubes respectively. Periodic boundary conditions were enforced along the axial direction and a large amount of vacuum padding ($\sim 35$ Bohr) was included in the other two directions to prevent interactions between periodic images. Optimized Norm Conserving (ONCV) pseudopotentials \citep{hamann2013optimized, schlipf2015optimization}, and LDA exchange correlation were employed. An energy cutoff of $40$ Ha was employed and only the gamma point of the Brillouin zone was sampled. The structures were first relaxed using the Broyden–Fletcher–Goldfarb–Shanno (BFGS) algorithm \citep{liu1989limited} following which AIMD simulations were performed at temperatures of $315.77$ K, $631.55$ K and $947.31$ K using the Nos\'{e}–Hoover thermostat \citep{evans1985nose, martyna1992nose}. Time steps of $1.0$ fs were employed for integration and $5.0 - 7.0$ ps of trajectory data were collected for analysis.

Finally, we used the Quantum Espresso code \citep{Quantum_Espresso_1, giannozzi2017advanced, giannozzi2020quantum} for computing the projected density of states (PDOS) of undistorted armchair and zigzag CKNTs. Pseudopotentials from the Standard Solid State Pseudopotentials (SSSP) library \citep{SSSP, prandini2018precision}, along with an energy cutoff of $40$ Ha, LDA exchange correlation and Gaussian smearing (corresponding to an electronic temperature of $315.77$ K) were employed. Keeping in mind the geometry of the nanotube, the PDOS were calculated in the local atomic coordinate frame, i.e., the projections were taken on atomic orbitals that had been rotated to a basis in which the occupation matrix appears diagonal.
\section{Results and Discussion}
\label{sec:results&discussion}
In this section, we discuss the structural, mechanical and electronic properties of CKNTs as revealed by our simulations.
\subsection{Structural properties: Cohesive energy, sheet bending modulus and dynamic stability}
\label{subsec:structural_properties}
\begin{figure}[htbp]
    \centering
    \includegraphics[clip, width=0.6\linewidth]{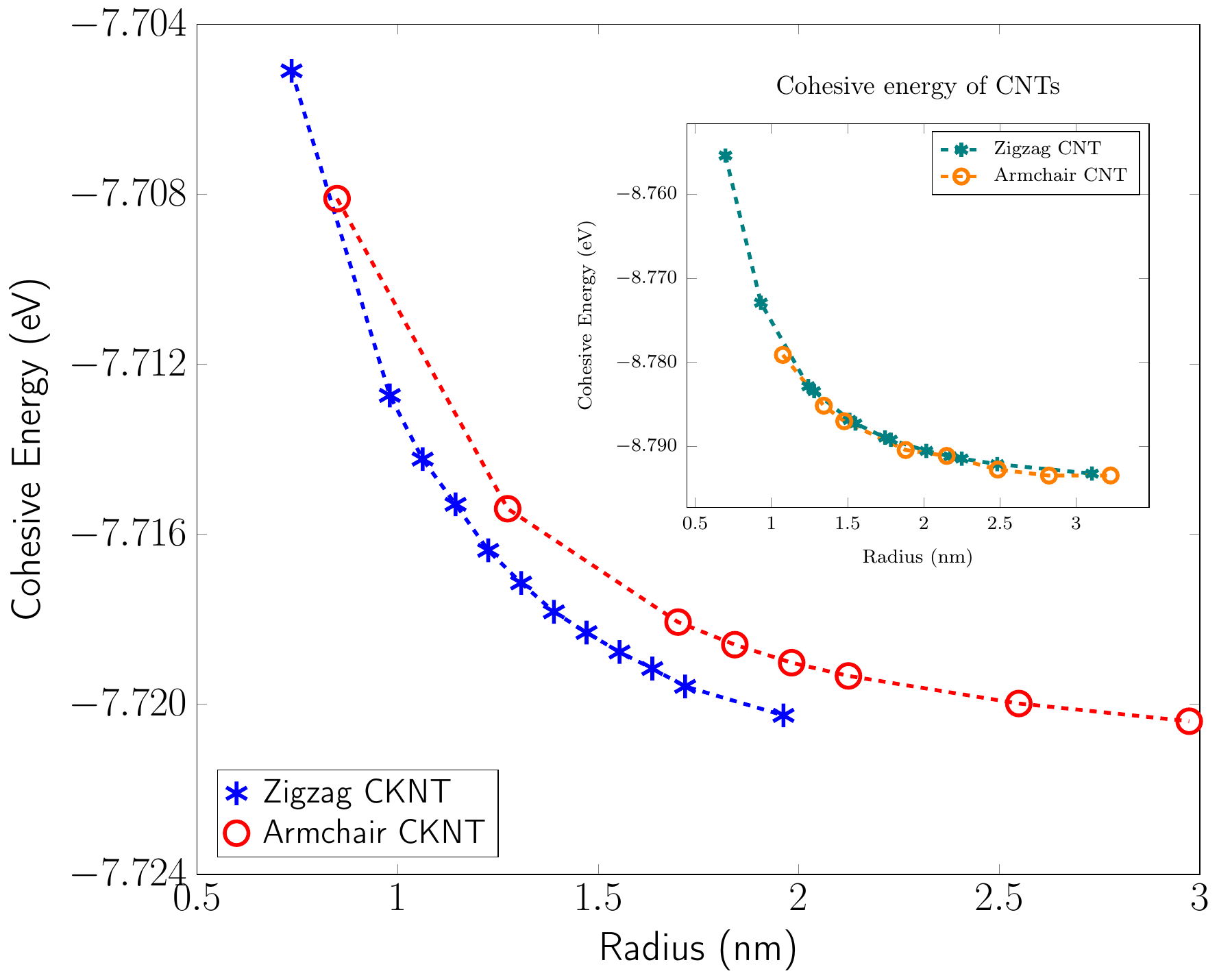}
    \caption{Cohesive energy of zigzag and armchair CKNTs. Inset: Cohesive energy of conventional zigzag and armchair carbon nanotubes (CNTs) presented for comparison.}
    \label{fig:cohesive_energy}
\end{figure}

Fig.~\ref{fig:cohesive_energy} shows the cohesive energy of armchair and zigzag CKNTs as the tube radius varies in the range  $1$ to $3$ nm (approximately). 
Owing to the contribution from the elastic sheet bending energy, the cohesive energy of both types of tubes decrease monotonically as the tube radius increases, i.e., the tubes are energetically more favorable with decreasing sheet curvature. In our calculations, the energy of an atom in Kagome graphene, as calculated in terms of the large radius limit of the energies of CKNTs, agrees with direct calculations of the sheet to better than $1$ milli-eV, thus ensuring overall consistency of the results. For a given radius, the zigzag and armchair CKNTs appear nearly identical energetically, similar to the behavior of conventional CNTs, also shown in Fig.~\ref{fig:cohesive_energy}. Assuming a quadratic dependence of the bending energy on curvature, i.e., Euler-Bernoulli behavior, we evaluated the area-normalized sheet bending modulus of Kagome graphene to be $0.506$ eV and $0.502$ eV in the armchair and zigzag directions, respectively (also see Fig.~\ref{fig:orth_ckl_rolling}). This is about a third of the sheet bending modulus of conventional graphene, estimated to be about $1.5$ eV through similar first principles calculations \citep{ghosh2019symmetry}. 

From Fig.~\ref{fig:cohesive_energy}, it is also evident that for a similar value of the radius, CNTs are energetically more favorable compared to CKNTs (i.e., CNTs have larger cohesive energies). We remark however that this observation in of itself does not preclude the synthesis of CKNTs. Indeed, fullerenes can be readily produced, although they have long been known to have cohesive energies that are somewhat lower than other common allotropes of carbon \citep{Dresselhaus_CNT_textbook, dresselhaus1996science, dresselhaus1995physics, liu2009ab}. More recently $\gamma$-graphyne, which has a significantly lower cohesive energy compared to graphene \citep{shin2014cohesion} has also been chemically synthesized \citep{hu2022synthesis, desyatkin2022scalable,li2018synthesis}. Notably, there has also been success in synthesis of other unusual quasi-one-dimensional allotropes of carbon starting from conventional carbon nanotubes \citep{Zhang2017_y_carbon}, which may be adopted for producing CKNTs. 

\begin{figure}[htbp]
  \centering
      \subfloat[][Armchair CKNT \\(Unit cell with 144 atoms)]{\includegraphics[scale =0.4]{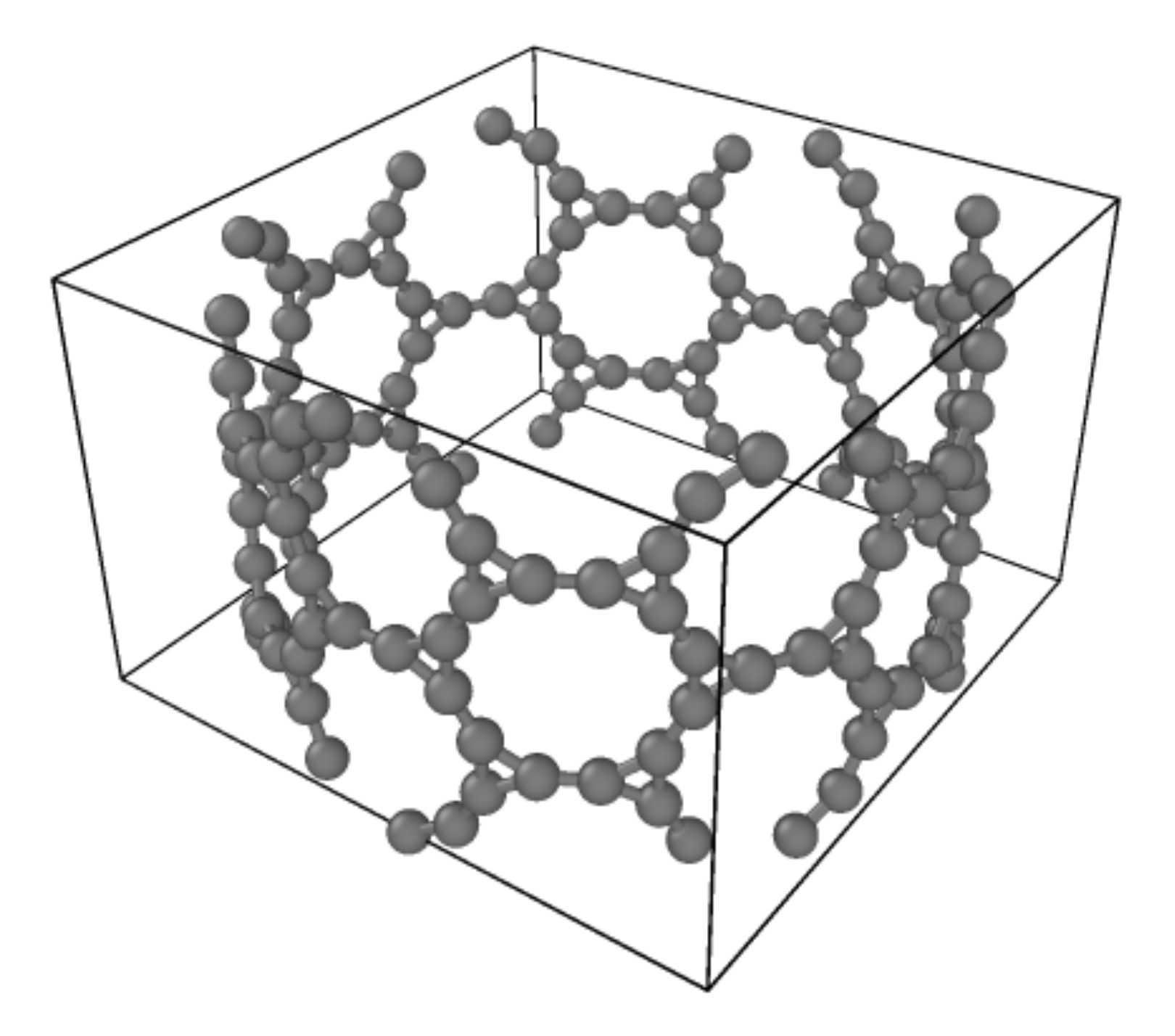}\label{fig:ckl1b_md_315} }
    \hspace{1.5cm}
     \subfloat[][Zigzag CKNT \\(Unit cell with 216 atoms)]{\includegraphics[scale =0.4]{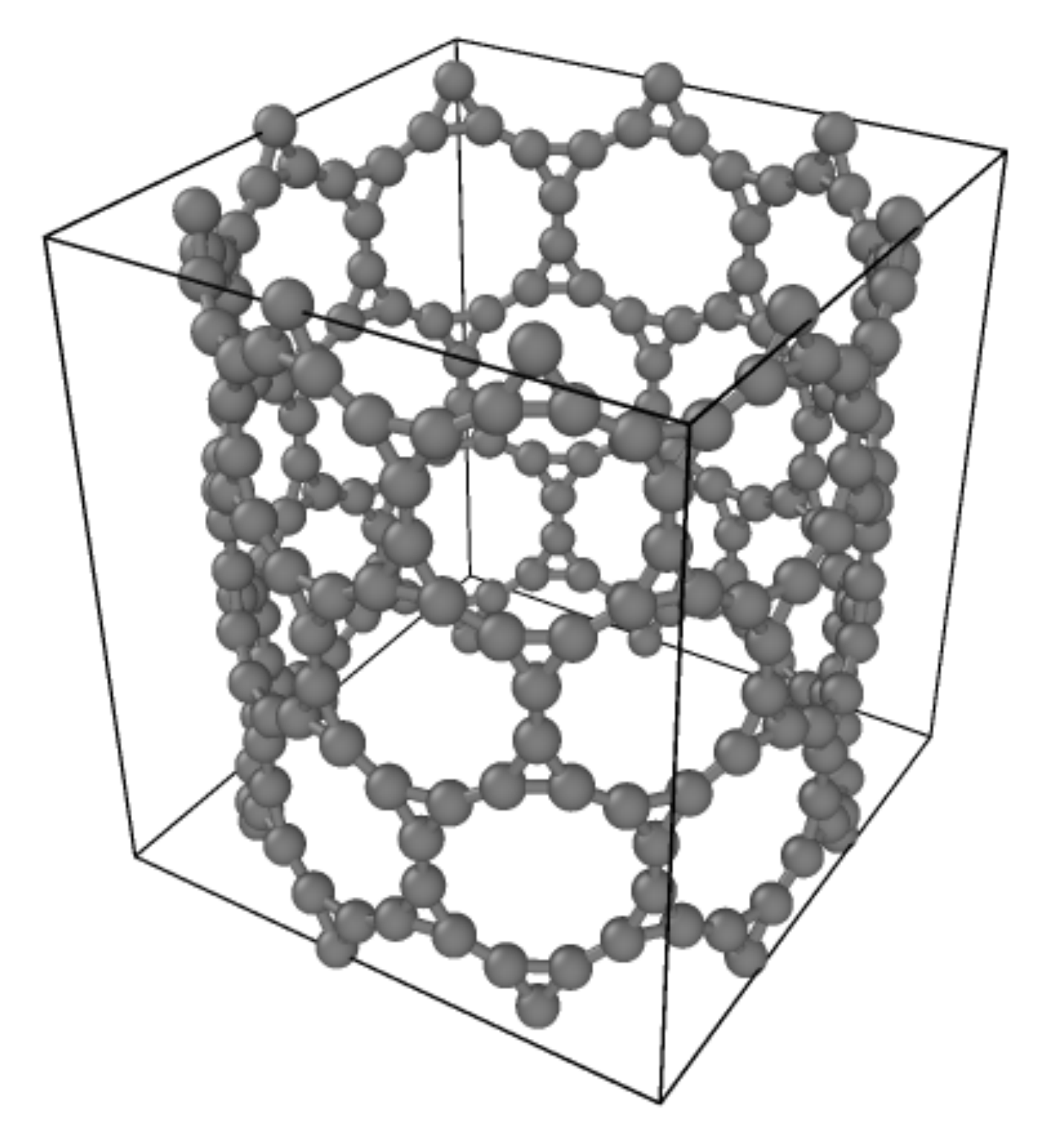}\label{fig:ckl1a_md_315}}
  \caption{Snapshot of AIMD simulations at $315.77$ K for both types of CKNTs.}
  \label{fig:AIMD}
\end{figure}

\begin{figure}[htbp]
  \centering
      \subfloat[][Transitory ellipsoidal cross section of Zigzag CKNT at elevated temperature]{\includegraphics[trim={150 0 150 50},clip,  scale = 0.28]{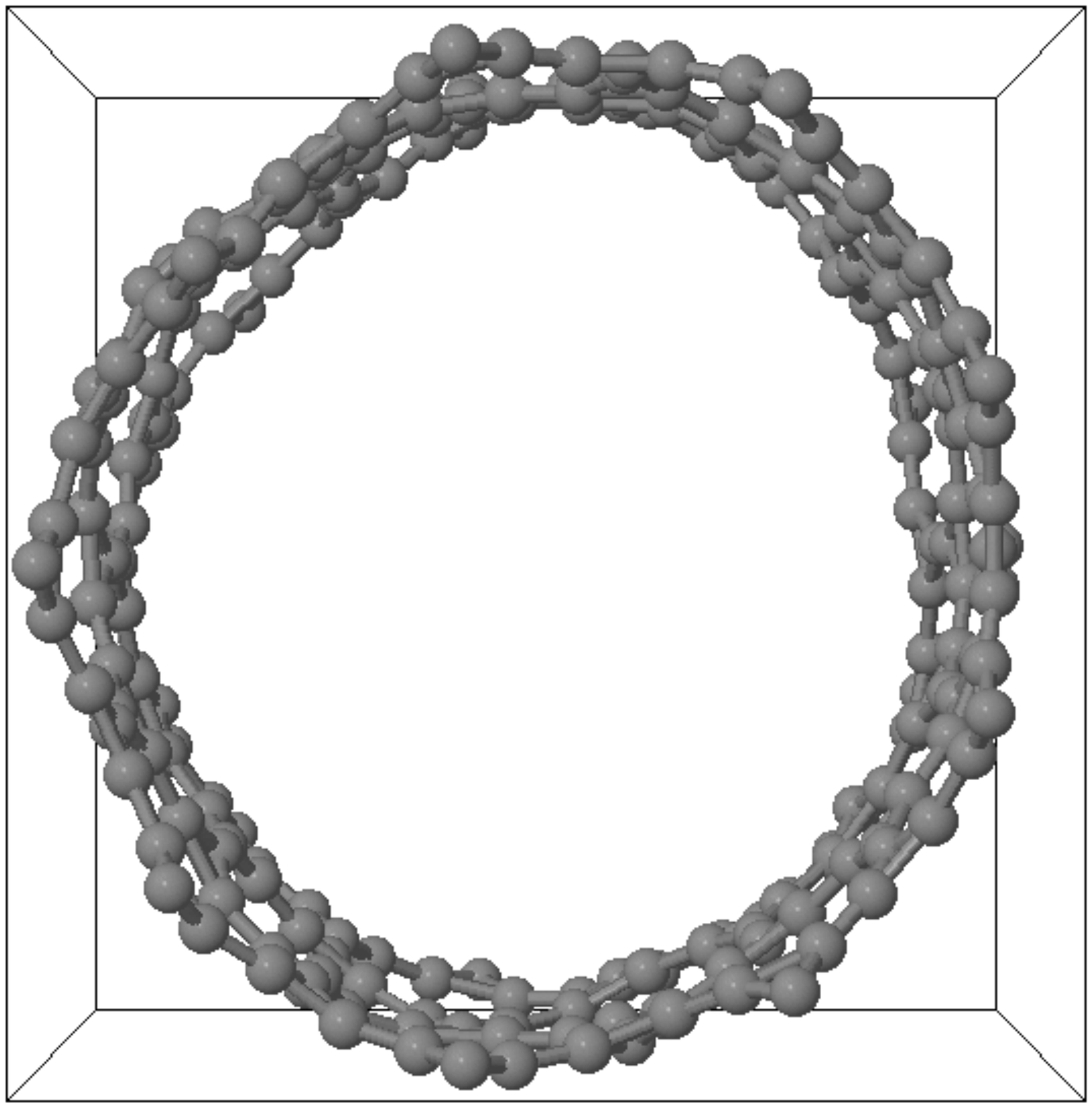}\label{fig:ckl1b_md_647_top} }
   \hspace{2cm}
     \subfloat[][Side view of the same structure showing overall structural integrity and persistence of dodecagonal rings]{\includegraphics[trim={100 0 150 50},clip, scale = 0.29]{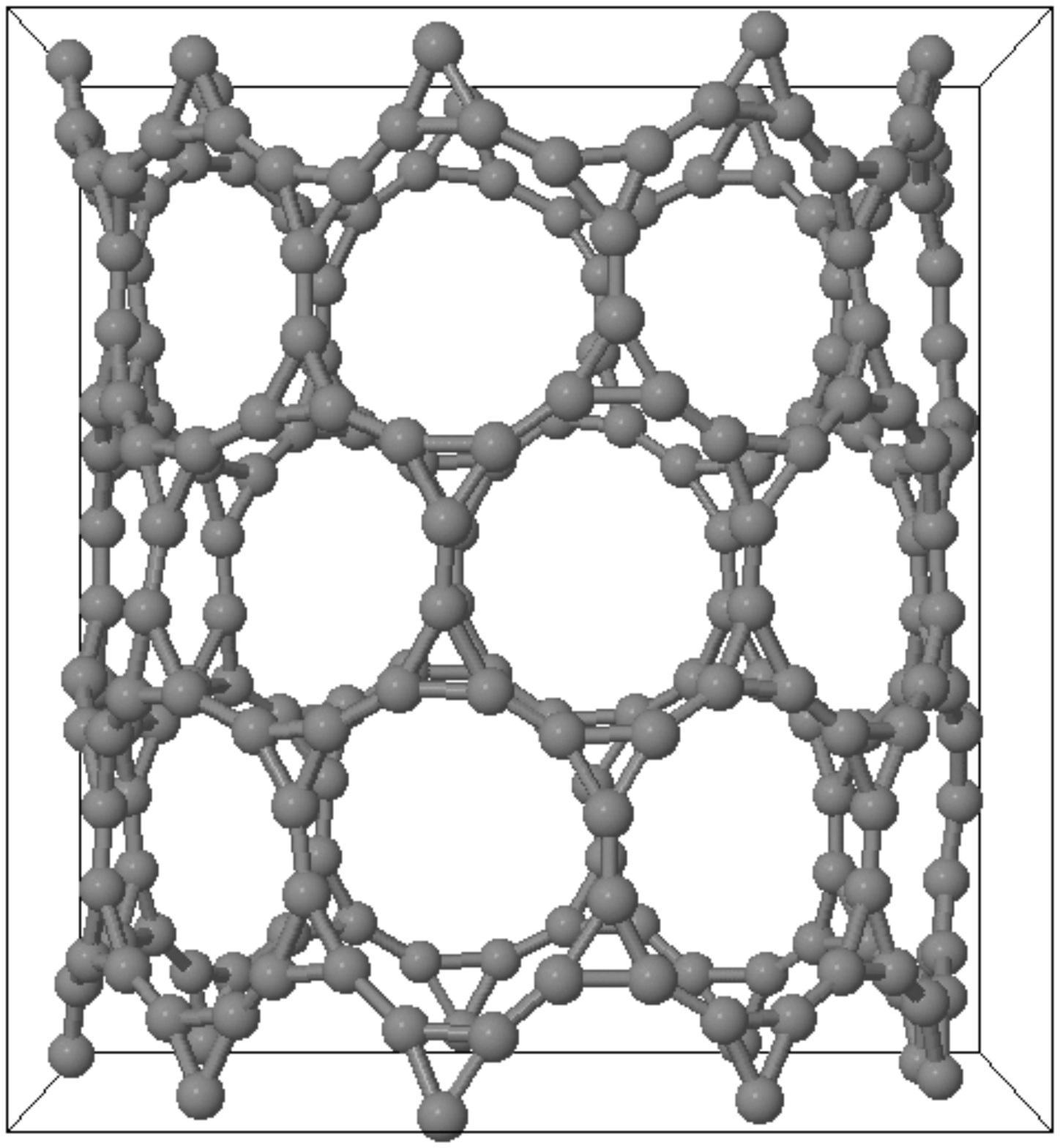}\label{fig:ckl1b_md_647_side}}
  \caption{Snapshot of AIMD simulations of a Zigzag CKNT at an elevated temperature of $631.554$ K. The cross-section shows a propensity for developing transitory distortions (left image). However, the overall structural integrity and the 12-fold rings continue to be maintained (right image). 
  }
  \label{fig:AIMD_Elevated}
\end{figure}

\begin{figure}[htbp]
  \centering
    \subfloat[Energy variation over ab-initio molecular dynamics  trajectories over $7$ ps of simulation time at $315.77$ K (blue), $631.54$ K (red) and $947.31$ K (yellow) of an armchair CKNT. Black broken line represent the mean energy.]{\includegraphics[trim={0cm 0cm 0cm 0cm}, clip, width=0.4\linewidth]{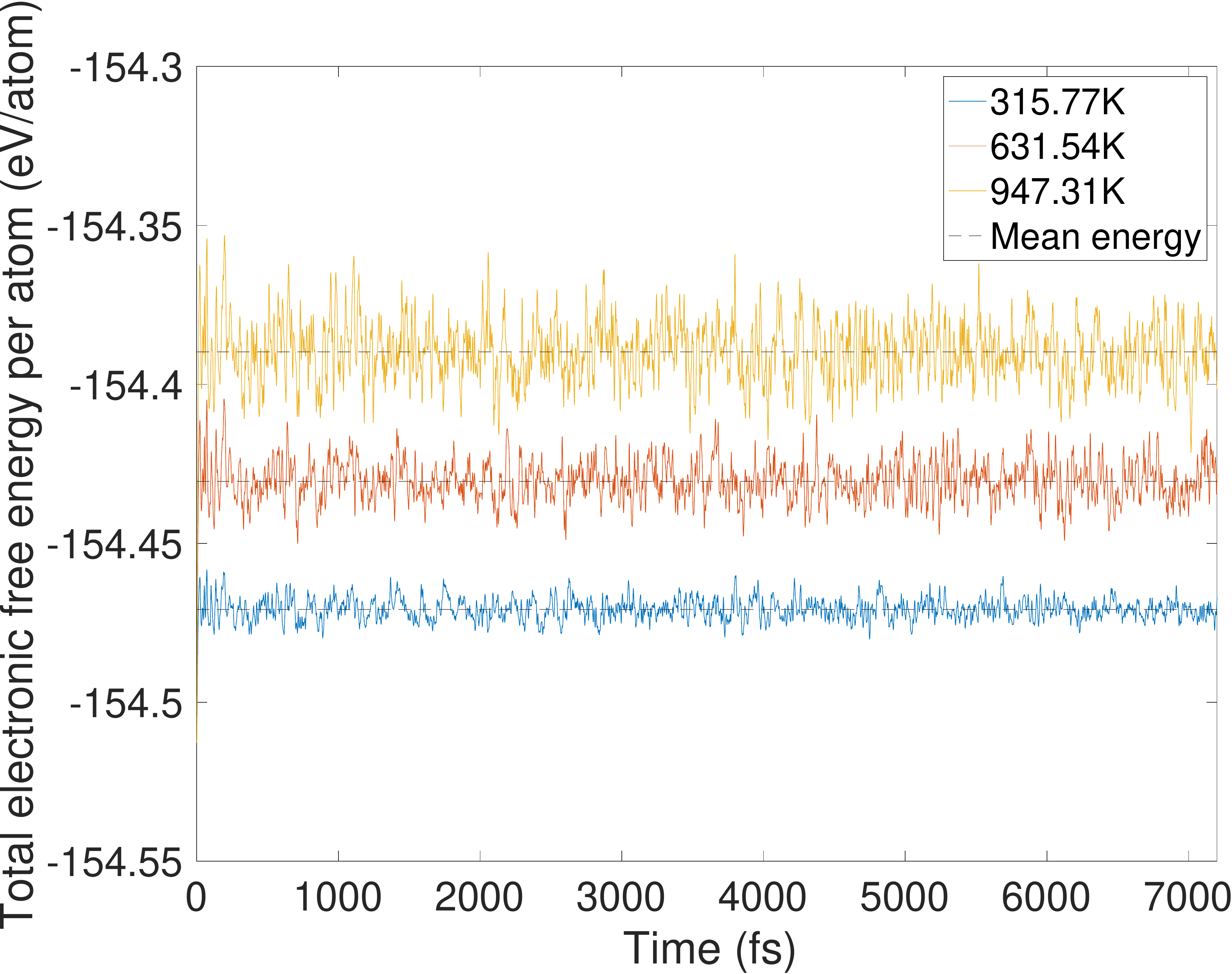}\label{fig:CKL_armchair_aimd}}\hspace{0.3cm}
    \subfloat[Energy variation over ab-initio molecular dynamics trajectories over $5$ ps of simulation time, at $315.77$ K (blue), $631.54$ K (red) and $947.31$ K (yellow), of a zigzag CKNT. Black broken line represent the mean energy.]{\includegraphics[trim={0cm 0cm 0cm 0cm}, clip,width=0.4\linewidth]{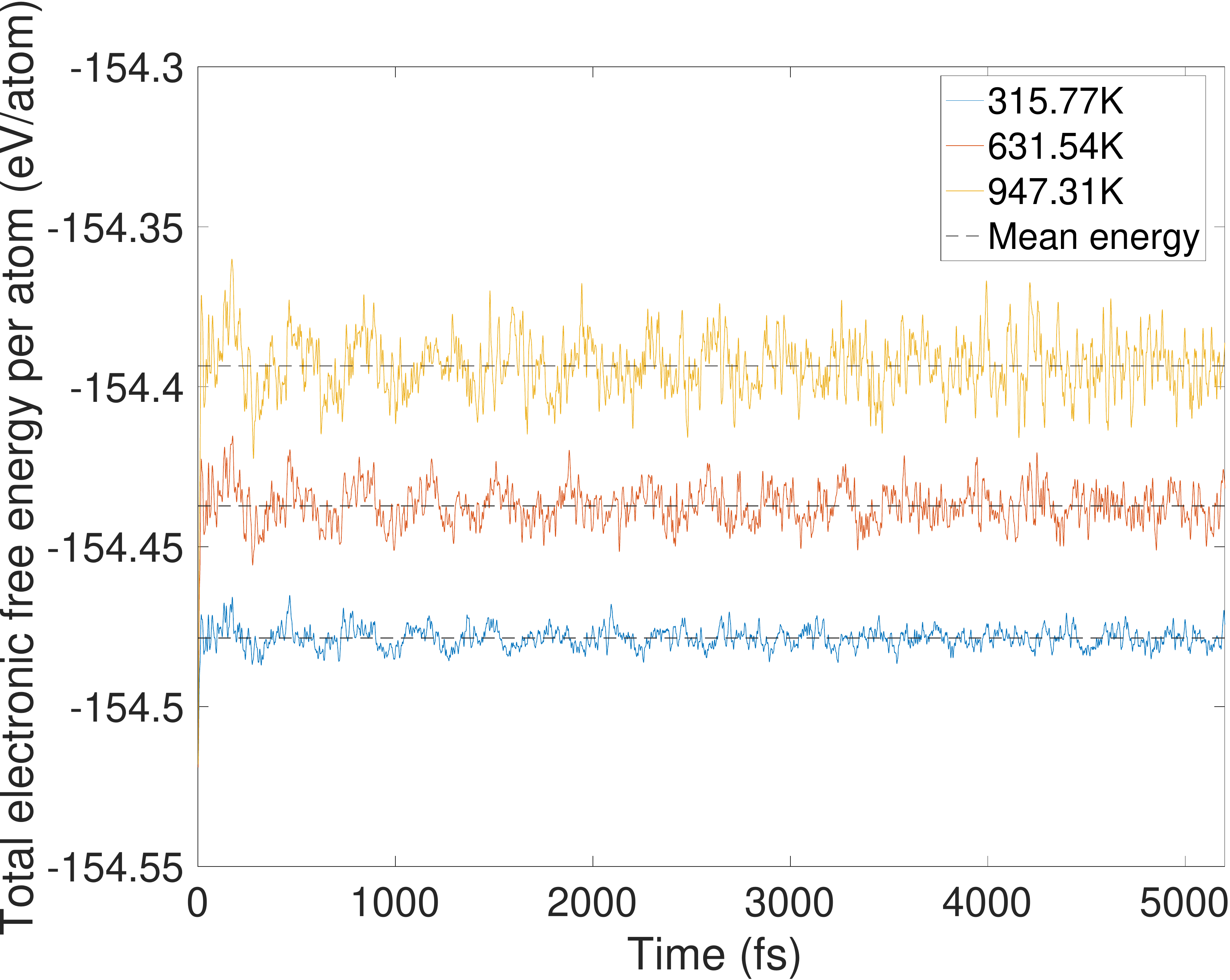}\label{fig:CKL_zigzag_aimd}}
  \caption{System energy variation over ab-initio molecular dynamics (AIMD) trajectories at three different temperatures for (a) an Armchair CKNT and (b) a Zigzag CKNT. The AIMD simulations reveal that the nanotubes maintains their overall structural integrity far above room temperature.}
\label{fig:AIMD_sim}
\end{figure}

The phonon stability of Kagome graphene sheets has been investigated earlier \citep{FerromagnetismAndWignerCrystallization}. Based on band-folding considerations \citep{zhang2007phonon, jishi1993phonon, saito1998raman}, such calculations are also likely to be indicative of the stability of CKNTs at zero temperature. {To investigate the finite temperature structural stability of CKNTs, we carried out   
AIMD simulations at three different temperatures --- $315.77$ K, $631.54$ K and $947.31$ K. The system energy is observed to be stable throughout each of these simulations (Fig.\ref{fig:AIMD_sim}).} We conclude that CKNTs are able to maintain their overall structural integrity at room temperature, and beyond, thus making them physically realistic nanostructures. Notably, during the course of the simulations, the structures appear to undergo dynamic distortions similar to conventional CNTs \citep{kurti1998first, Dumitrica_James_OMD, yao2001mechanical}, and show a propensity for developing transitory ellipsoidal cross sections, especially at elevated temperature (Fig.~\ref{fig:AIMD_Elevated} (a)). Nevertheless, the dodecagonal rings which make up CKNTs and which are crucially related to their fascinating electronic properties (discussed in Section \ref{sec:electronic_prop}), continue to be maintained (Fig.~\ref{fig:AIMD_Elevated} (b)). Given the relatively low sheet bending stiffness of Kagome graphene (as compared to conventional graphene, e.g.), it is quite possible that large diameter CKNTs, like their CNT counterparts \citep{tang2005collapse, xiao2007collapse, elliott2004collapse} have a tendency to collapse.  From this perspective, the distorted cross sections described above are possible indicators of this kind of structural transition, and warrant further investigation in the future. Snapshots of the AIMD simulations is provided in Figs.~\ref{fig:AIMD} and \ref{fig:AIMD_Elevated}, and the entire simulation trajectories are available as Supplementary Materials. 

\subsection{Mechanical properties: Torsional and extensional stiffness}
\label{subsec:mechanical_properties}
We focus on mechanical properties of CKNTs, namely their torsional and extensional responses in the linear elastic regime. As described earlier, Helical DFT allow such calculations to be carried out by introducing changes in the nanotube symmetry group parameters.  We consider $(12,12)$ armchair (radius ~$1.7$ nm) and $(12,0)$ zigzag (radius ~$0.98$ nm) CKNTs as representative examples. For both these tubes, we start from the undistorted, relaxed configurations. 

\begin{figure}[htbp]
    \centering
    \includegraphics[clip, width=0.6\linewidth]{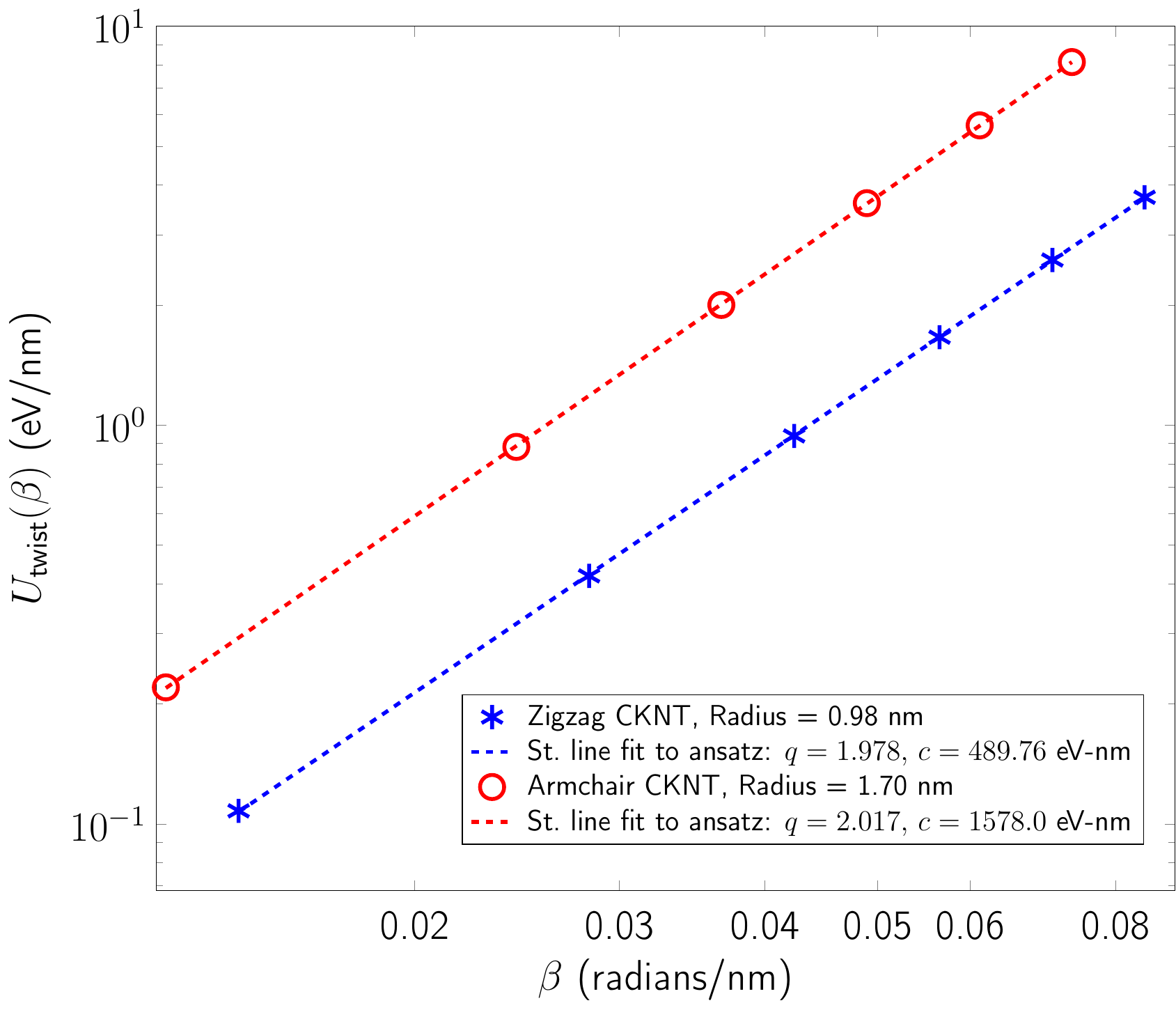}
    \caption{Twist energy per unit length as a function of angle of twist per unit length for two representative nanotubes (both axes logarithmic). Dotted lines indicate straight line fits of the data to an ansatz of the form $U_{\text{twist}}(\beta) = c\times\beta^q$. The exponent $q$ is nearly $2$ in both cases, suggesting linear elastic behavior.}
 \label{fig:linear_twist}
\end{figure}

For simulations involving torsion, we increment the parameter $\alpha$ in regular intervals, imposing up to about $\beta = 4.5^{\circ}$ of twist per nanometer, the limit of linear response for conventional CNTs \citep{Dumitrica_James_OMD}. For each twisted configuration, we relax the atomic forces and compute the twisting energy per unit length of the nanotubes as the difference in the ground state free energy (per fundamental domain) of the twisted and untwisted structures, i.e.:
\begin{align}
U_{\text{twist}}(\beta) = \frac{\mathfrak{N}}{\tau} \bigg({\calF}_{\substack{\text{Ground}\\ \text{State}}}(\calP^{**} , \calD, \calG|_{\beta}) - {\calF}_{\substack{\text{Ground}\\ \text{State}}}(\calP^{*}  , \calD, \calG|_{\beta=0})\bigg)\,.
\end{align}
In the equation above, $\calG|_{\beta}$ and $\calG|_{\beta = 0}$ denote the symmetry groups associated with the twisted and untwisted structures, respectively and (as before) $\mathfrak{N}$ denotes the cyclic group order. Additionally, $\calP^{**}$ and $\calP^{*}$ denote the collections of relaxed positions of the atoms in the fundamental domain in each case. Thereafter, the torsional stiffness is computed as:
\begin{align}
k_{\text{twist}} = \hpd{U_{\text{twist}}(\beta)}{\beta}{2}\bigg\rvert_{\beta = 0}\,.
\end{align}
We observed that for the range of torsional deformations considered here, the behavior of $U_{\text{twist}}(\beta)$ is almost perfectly quadratic with respect to $\beta$, consistent with linear elastic response (see Fig.~\ref{fig:linear_twist}). Moreover, the value of $k_{\text{twist}}$ is estimated to be $3156.1$ eV-nm and $979.51$ eV-nm for the armchair and zigzag tubes, respectively. In the linear elastic regime, the torsional behavior of nanotubes is well approximated by continuum models which suggest that $k_{\text{twist}}$ should vary with the tube radius in a cubic manner \citep{yu2022density, Dumitrica_James_OMD, timoshenko1968elements}. By use of this scaling law, we were able to estimate that conventional CNTs with radii comparable to the CKNTs considered above are expected to be significantly more rigid with respect to twisting (with $k_{\text{twist}}$ values equal $2.8977\times10^4$ eV-nm and $5525.0$ eV-nm for $1.7$ nm and $0.98$ nm radius conventional CNTs, respectively).
 
 Next, for simulations involving axial stretch and compression, we proceed in a manner similar to the torsion simulations. For a given value of axial strain $\epsilon$, we modify the pitch of the helical symmetry group as $\tau = \tau_0(1+\epsilon)$, with $\tau_0$ denoting the equilibrium, undistorted values. Subsequent to this kinematic prescription, we relax the atomic forces, and compute the extensional energy per unit length of the nanotubes as the difference in the ground state free energy per fundamental domain, between stretched and unstretched structures, i.e.:
\beqs
U_{\text{stretch}}(\epsilon) = \frac{\mathfrak{N}}{\tau_0} \bigg({\calF}_{\substack{\text{Ground}\\ \text{State}}}(\calP^{**} , \calD,  \calG|_{\tau=\tau_0(1+\epsilon)}) - \nonumber\\ 
\; \; \; \;\; {\calF}_{\substack{\text{Ground}\\ \text{State}}}(\calP^{*}  , \calD, \calG|_{\tau=\tau_0})\bigg)\,.
\eeqs
In the equation above, $\calG|_{\tau=\tau_0(1+\epsilon)}$ and $\calG|_{\tau=\tau_0}$ denote the symmetry groups associated with the stretched and unstretched structures, respectively. Additionally, $\calP^{**}$ and $\calP^{*}$ denote the collections of relaxed positions of the atoms in the fundamental domain in each case. The stretching stiffness of the nanotubes may be then calculated as:
\begin{align}
k_{\text{stretch}} = \hpd{U_{\text{stretch}}(\epsilon)}{\epsilon}{2}\bigg\rvert_{\epsilon = 0}\,.
\end{align}

\begin{figure}[htbp]
    \centering
    \includegraphics[clip, width=0.6\linewidth]{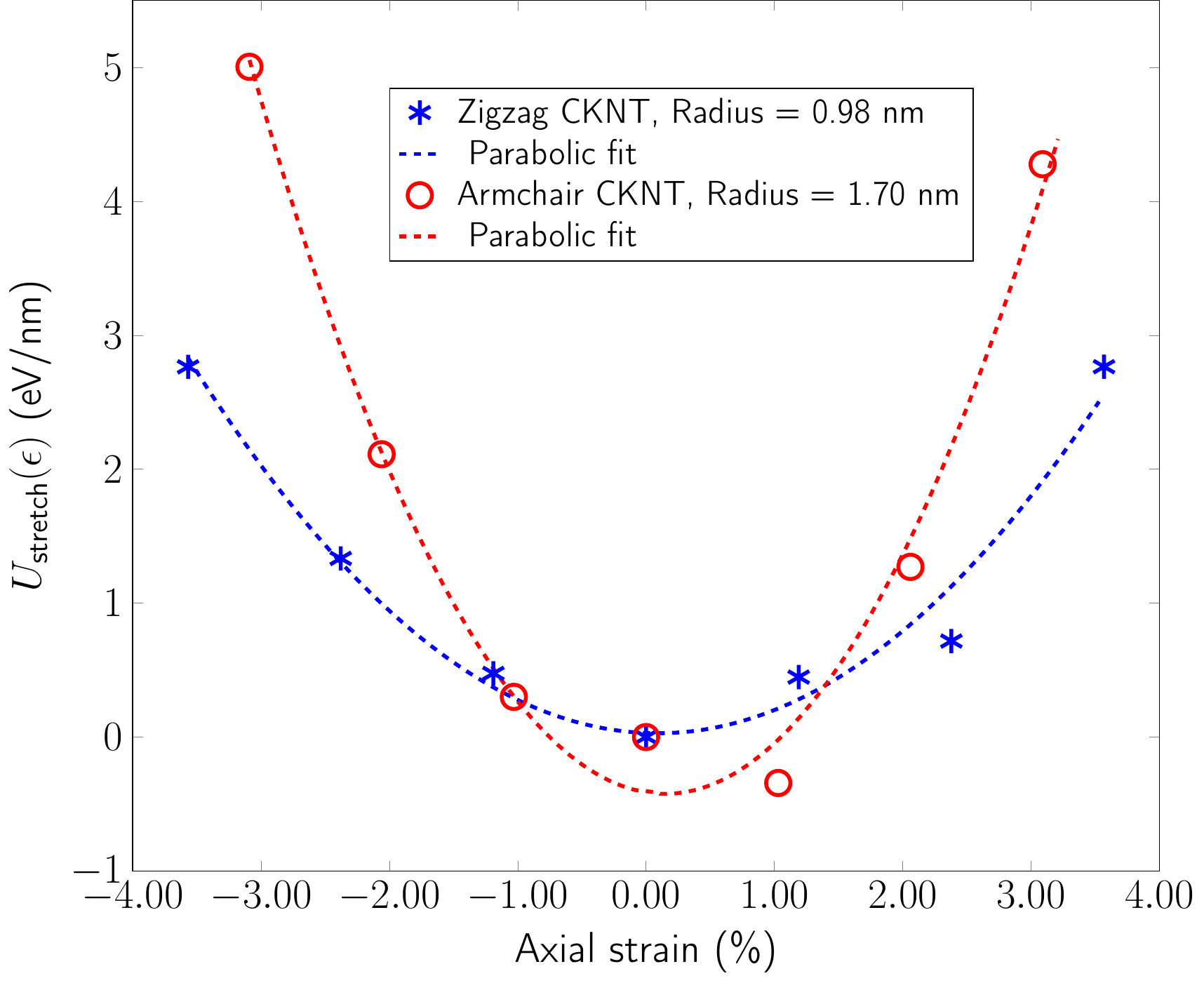}
    \caption{Extensional energy per unit length as a function of axial strain for two representative CKNTs. Dotted curves indicate parabolic fits of the data to an ansatz of the form $U_{\text{stretch}}(\epsilon) = c\times\epsilon^2$.}
\label{fig:extension_stiffness_fit}
\end{figure}

In our simulations, we restricted $\epsilon$ to be between $+3.6\%$ and $-3.6\%$. In this range, $U_{\text{stretched}}(\epsilon)$ is found to depend in a quadratic manner on $\epsilon$, consistent with linear elastic behavior (see Fig.~\ref{fig:extension_stiffness_fit}). We also observed a small Poisson effect, which we have ignored in subsequent analyses. For armchair $(12,12)$ and zigzag $(12,0)$ CKNTs, we estimated $k_{\text{stretch}}$ to be $5220.9$ eV/nm and $2093.4$ eV/nm  respectively. Based on scaling laws arising from continuum theory \citep{timoshenko1968elements}, we also estimated that conventional armchair CNTs with the same radii as the CKNTs considered above would be noticeably stiffer to axial deformations ($k_{\text{stretch}}$ values equal to $1.2213\times10^4$ eV/nm and $7048.2$ ev/nm for $1.7$ nm and $0.98$ nm radius conventional CNTs, respectively).

Overall, these results suggest that CKNTs are significantly more pliable with respect to torsional and axial deformations, as compared to their conventional CNT counterparts. Coupled with the lower bending stiffness of Kagome graphene as compared to conventional graphene, they are indicative of the fact that Kagome graphene has a lower value of in-plane (thickness normalized) Young's modulus and shear modulus.

\subsection{Electronic properties}
\label{sec:electronic_prop}
We discuss the electronic properties of CKNTs as revealed by first principles simulations. We start from a discussion of the properties of undistorted tubes, following which we discuss the electronic response of the tubes when subjected to torsional and axial strains. In \ref{sec:app_TBM} we describe a symmetry adapted tight-binding (TB) model that is able to explain these electronic properties. 
\subsubsection{Electronic properties of undeformed CKNTs}
\label{subsec:electronic_prop_undistorted}
\begin{figure}[htbp]
\centering
\subfloat[Complete band diagram for an undistorted zigzag CKNT of radius 0.98 nm with a zoomed-in view of flat bands and quadratic band crossing point.]{\scalebox{0.4}
{
\includegraphics[ clip, width=\textwidth]{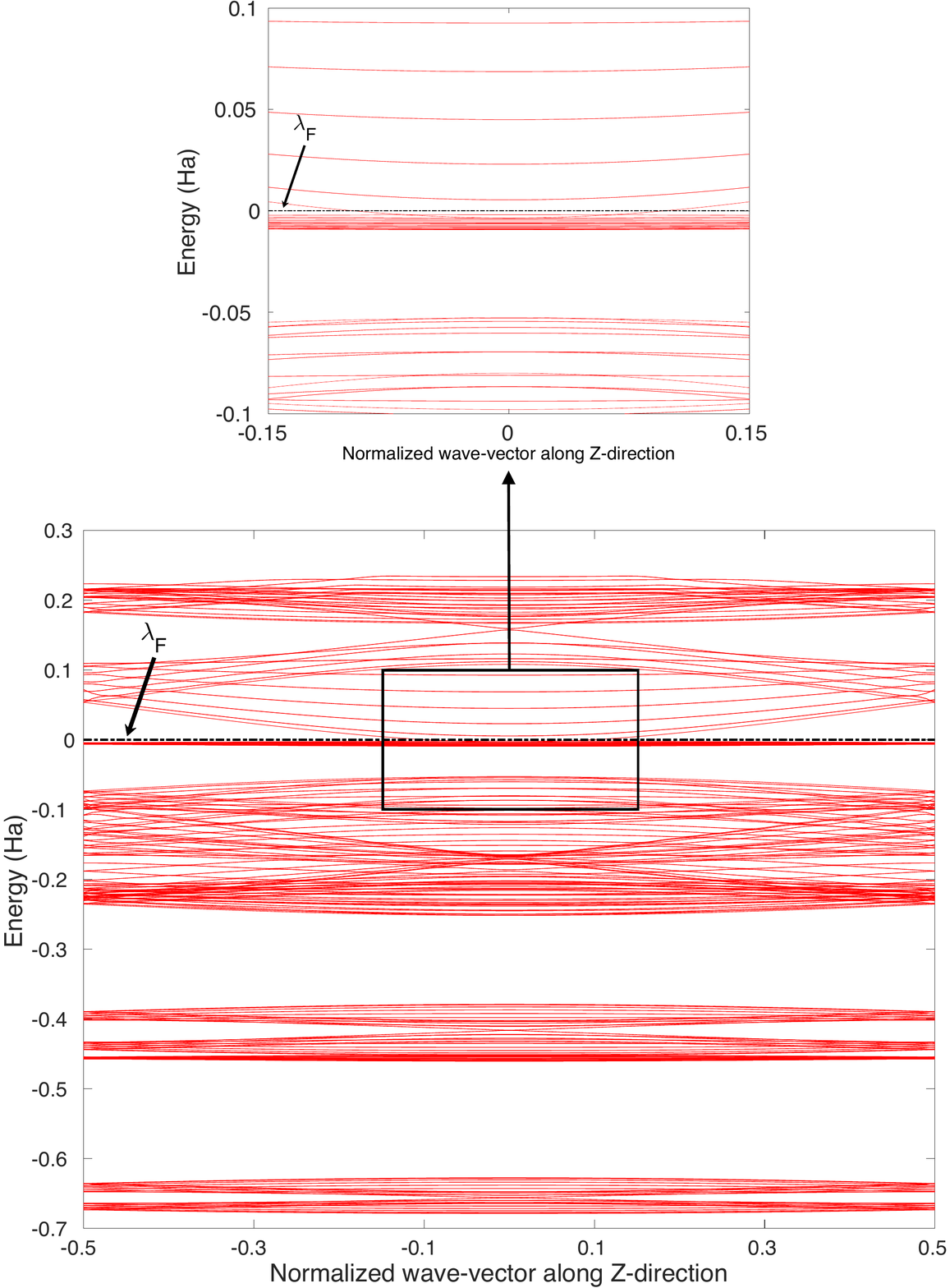}
}\label{fig:banddiag_zigzagfull_DFT}}\quad
\subfloat[Electronic density of states of the same material.]{\scalebox{0.4}
{
\includegraphics[ clip,trim=0.5cm 8.5cm 0.5cm 7.5cm,width=\textwidth]{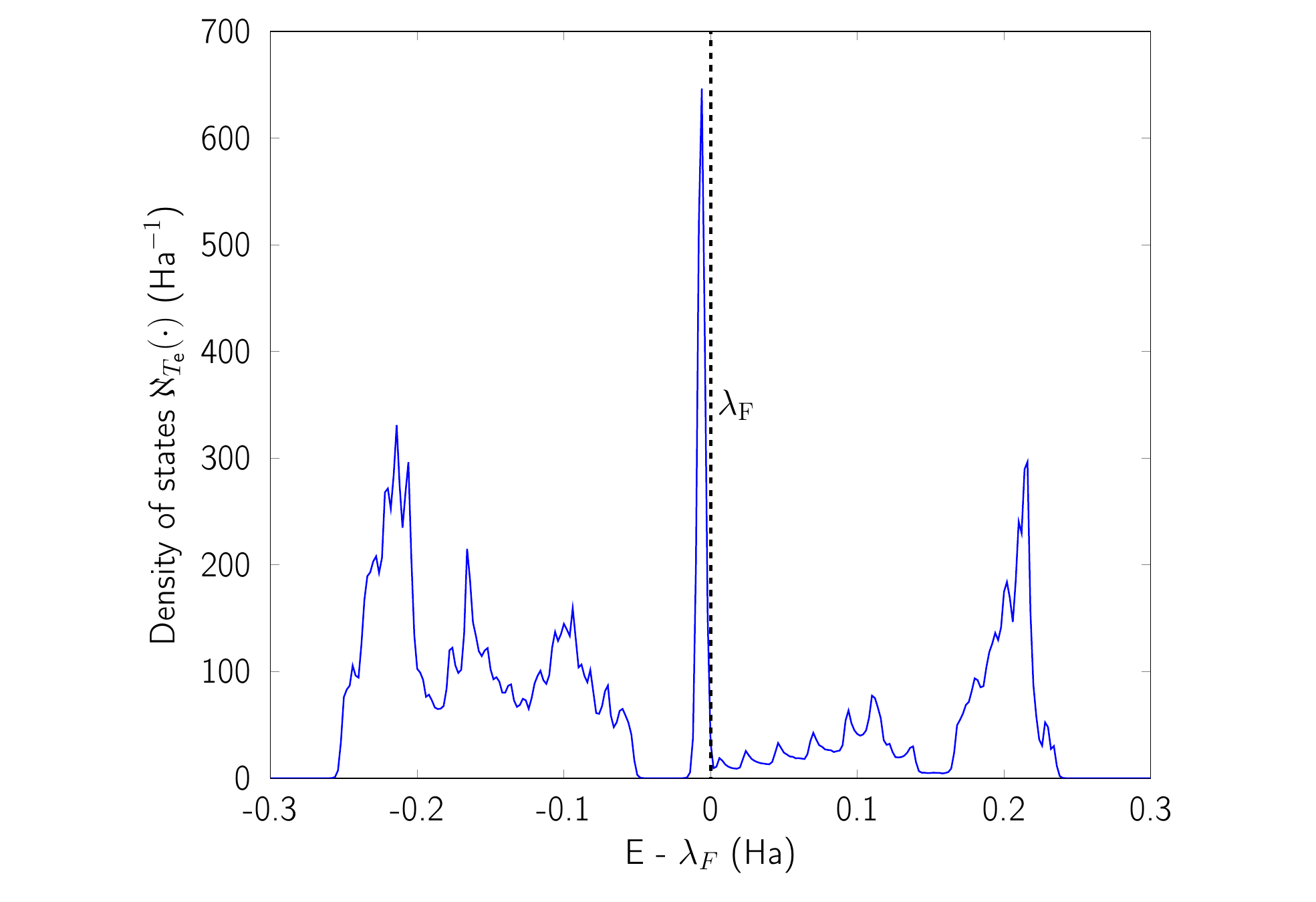}
}
\label{fig:DOS_untwist_zigzag}}
\caption{(a) Complete band diagram and (b) Electronic density of states near the Fermi level of an undistorted zigzag CKNT (radius $0.98$ nm). $\lambda_{\text{F}}$ denotes the Fermi level.}
\label{fig:full_band_zigzag_untwisted}
\end{figure}

\begin{figure}[htbp]
\centering
\subfloat[Symmetry adapted band diagram in $\eta$, along $\nu = 4$. 
]{\scalebox{0.9}
{\includegraphics[ clip, width=0.45\textwidth]{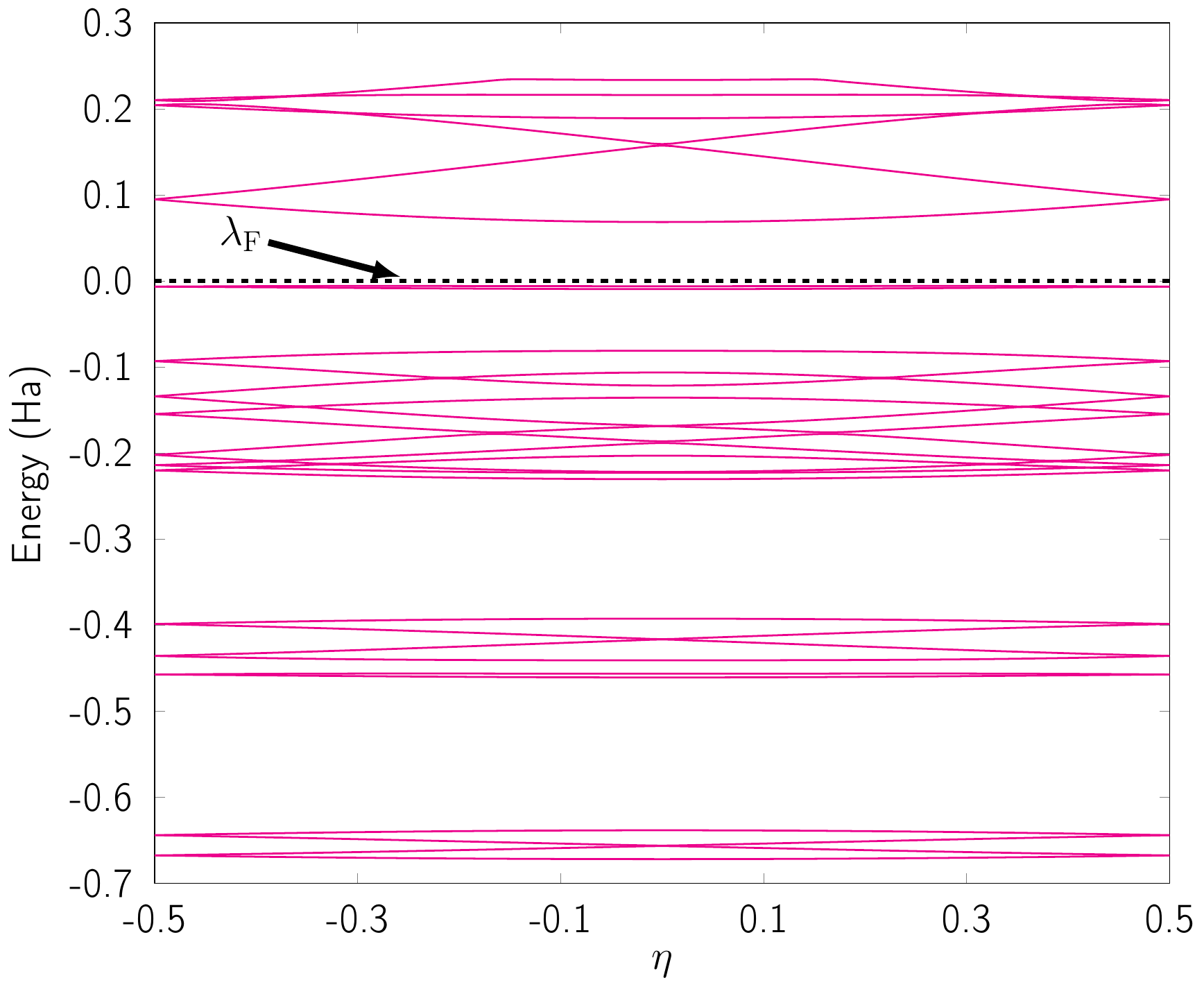}} 
}\hspace{1cm}
\subfloat[Symmetry adapted band diagram in $\nu$, along $\eta = 0$.  Quadratic band crossing at $\nu = 0$ can be observed. 
]{\scalebox{0.9}
{\includegraphics[ clip, width=0.45\textwidth]{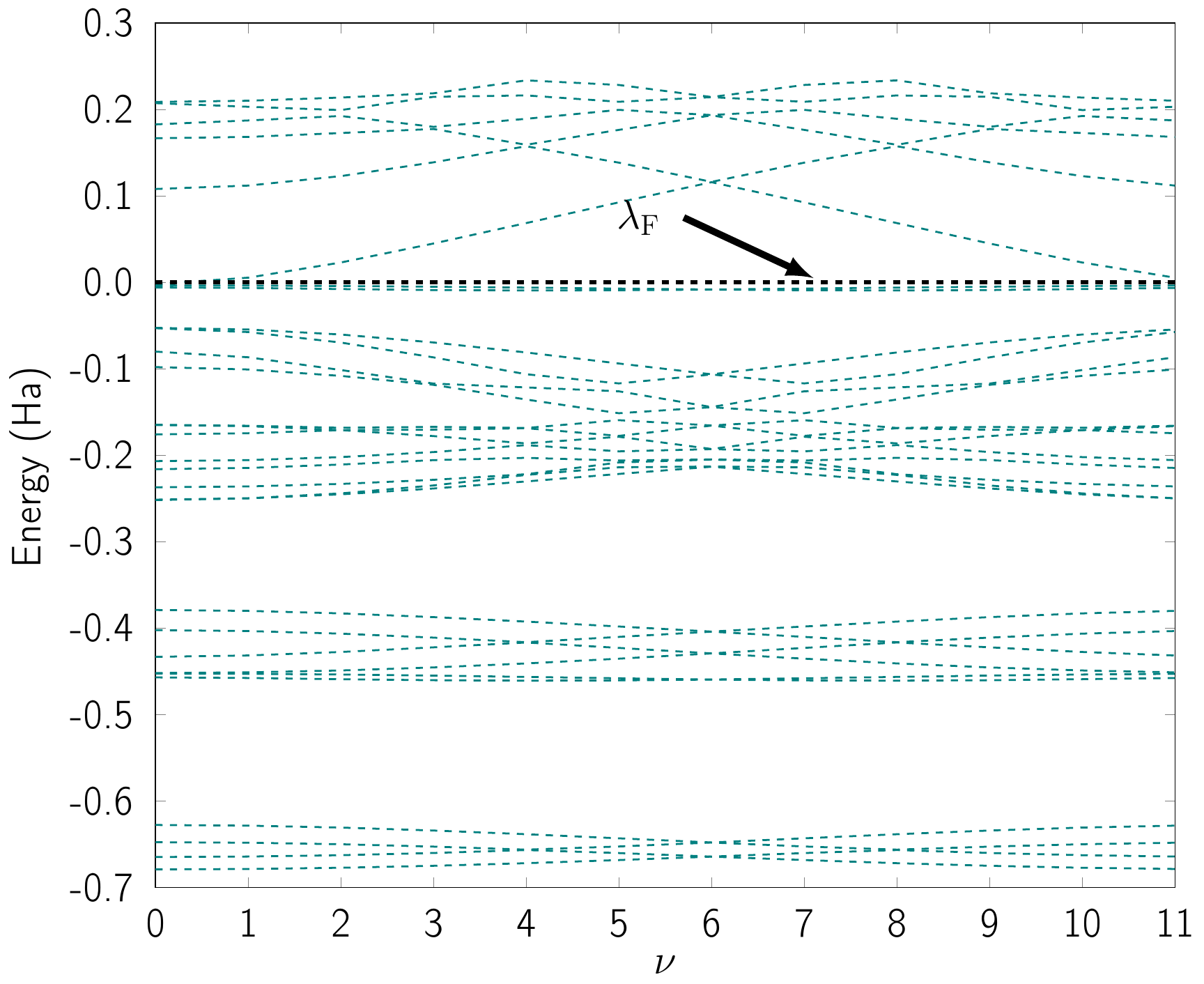}} 
}
\caption{Symmetry adapted band diagrams of an undistorted zigzag CKNT (radius $0.98$ nm) obtained using Helical DFT \citep{banerjee2021ab, yu2022density}. $\lambda_{\text{F}}$ denotes the Fermi level.}
\label{fig:symmetry_band_zigzag_untwisted}
\end{figure}

\begin{figure}[htbp]
\centering
\subfloat[Complete band diagram for an undistorted armchair CKNT of radius 1.70 nm, with zoomed-in view of flat bands and quadratic band crossing point.]{\scalebox{0.4}
{
\includegraphics[ clip, width=\textwidth]{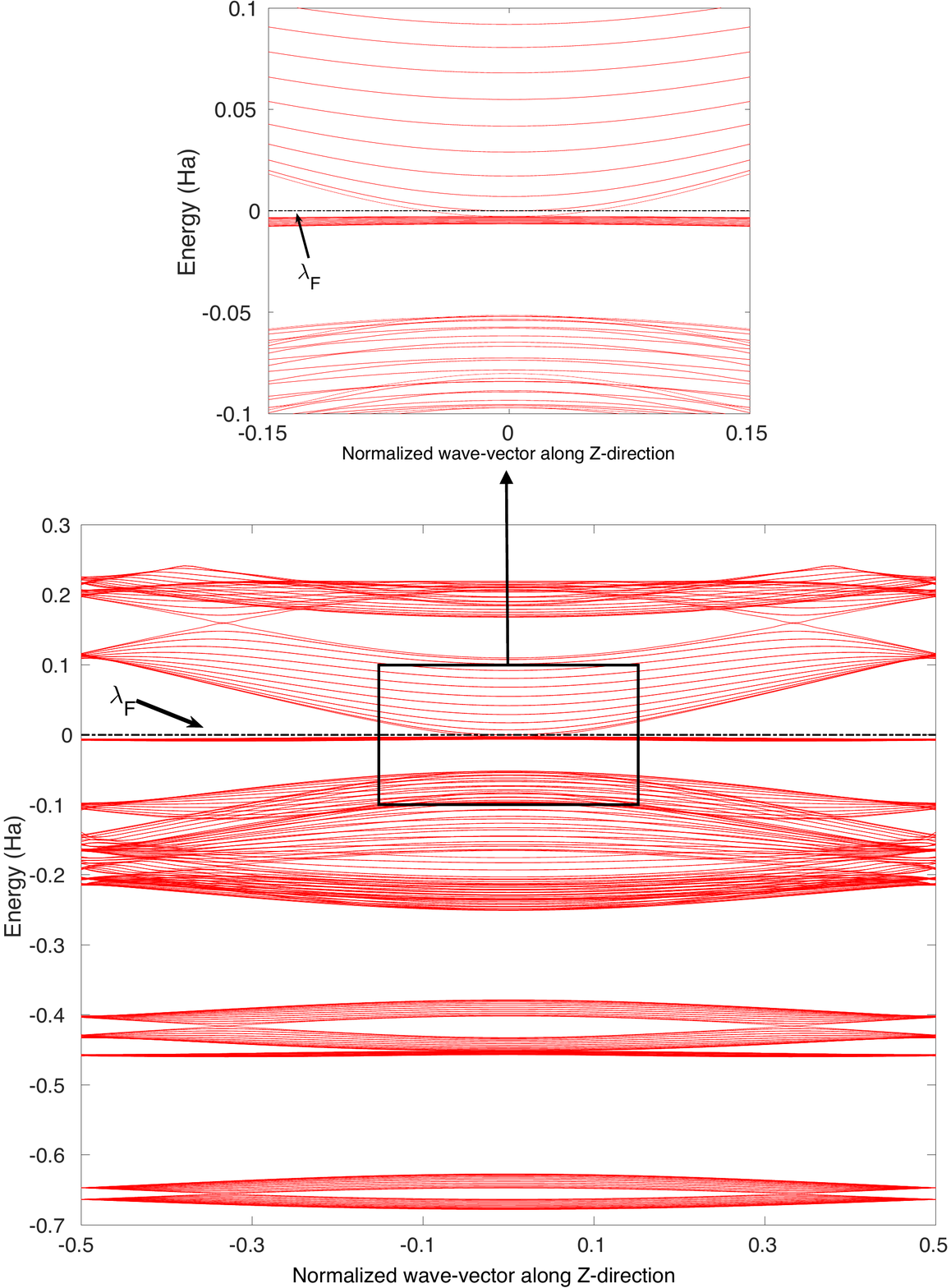}
}\label{fig:banddiag_armchairfull_DFT}}\quad
\subfloat[Electronic density of states of the same material.]
{\scalebox{0.4}
{
\includegraphics[ clip,trim=0.5cm 8.5cm 0.5cm 7.5cm,width=\textwidth]{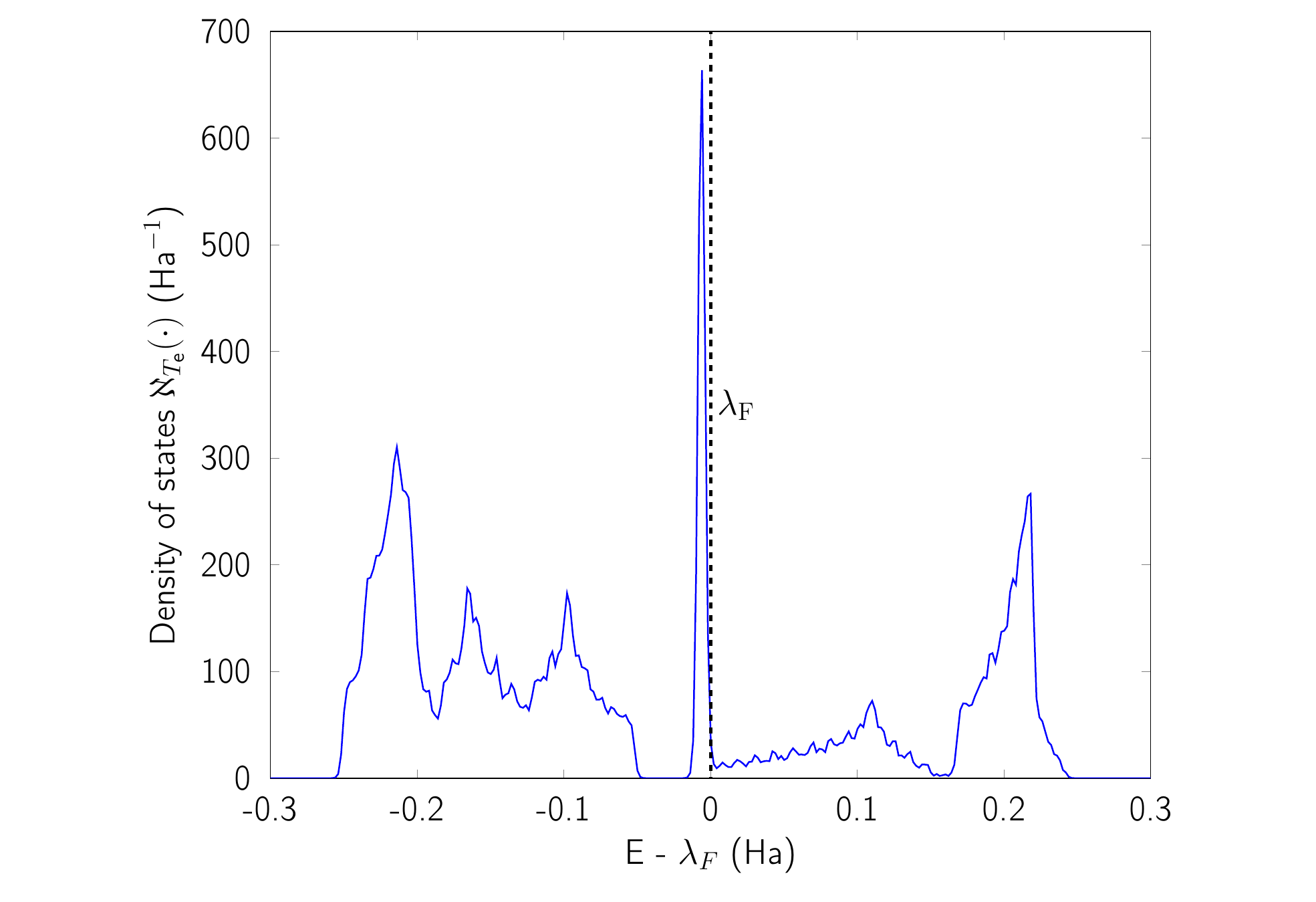}
}
\label{fig:DOS_untwisted_armchair_CKNT}}
\caption{(a) Complete band diagram and (b) Electronic density of states near the Fermi level for the undistorted armchair CKNT (radius $1.70$ nm). $\lambda_{\text{F}}$ denotes the Fermi level.}
\label{fig:full_band_armchair_untwisted}
\end{figure}

\begin{figure}[htbp]
\centering
\subfloat[Symmetry adapted band diagram in $\eta$, along $\nu = 0$. Quadratic band crossing at $\eta = 0$ can be observed.
]{\scalebox{0.9}
{\includegraphics[ clip, width=0.45\textwidth]{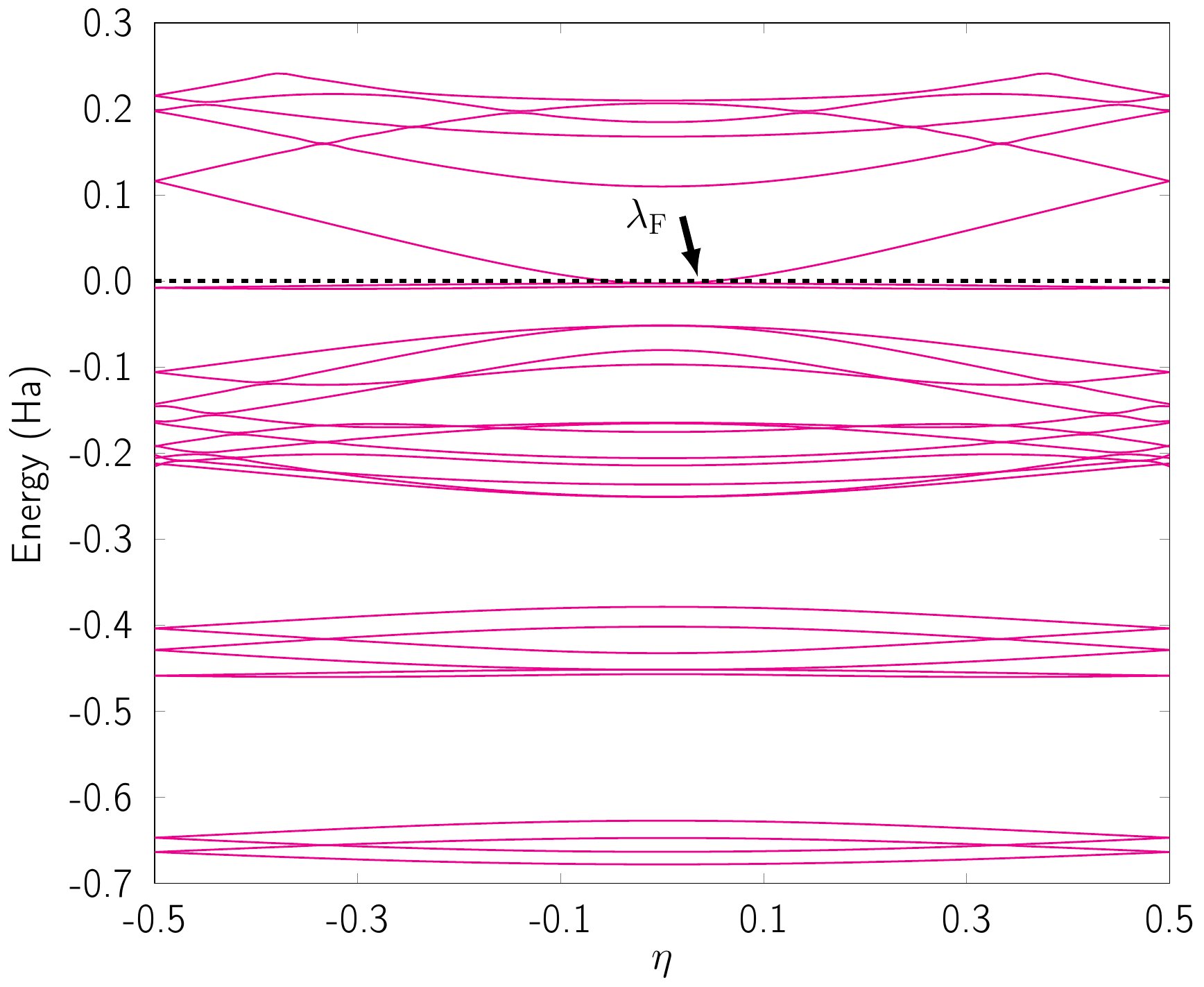}} 
}\hspace{1cm}
\subfloat[Symmetry adapted band diagram in $\nu$, along $\eta = 1/3$.  
]{\scalebox{0.9}
{\includegraphics[ clip, width=0.45\textwidth]{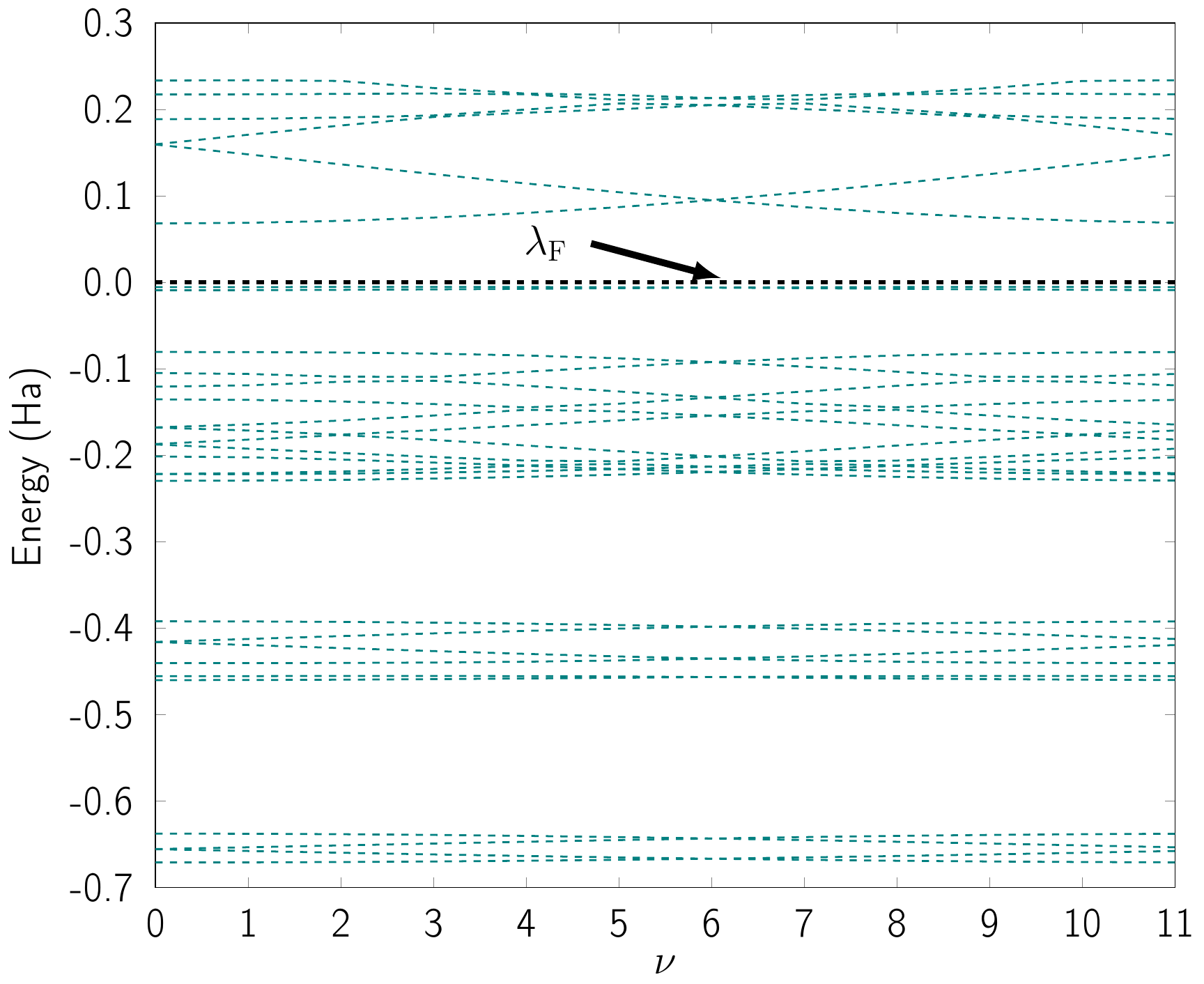}}
}
\caption{Symmetry adapted band diagrams of an undistorted armchair CKNT (radius $1.70$ nm) obtained using Helical DFT \citep{banerjee2021ab, yu2022density}. $\lambda_{\text{F}}$ denotes the Fermi level.}
\label{fig:symmetry_band_armchair_untwisted}
\end{figure}

Conventional CNTs can be metallic or semiconducting depending on whether they are armchair (all tubes metallic) or zigzag (tubes with cyclic group order $\mathfrak{N}$ divisible by 3 are metallic), and the electronic band diagrams of these materials prominently feature Dirac points near the Fermi level \citep{ghosh2019symmetry, yu2022density,  CNT_5, Dresselhaus_CNT_textbook, ding2002analytical}. In contrast, our simulations reveal \textit{all CKNTs} to be metallic, with their electronic band diagrams prominently featuring dispersionless electronic states, or \textit{flat bands}, close to the Fermi level. Figs.~\ref{fig:banddiag_zigzagfull_DFT} and \ref{fig:banddiag_armchairfull_DFT} show complete band diagrams of undistorted CKNTs (i.e., all electronic states corresponding to allowable values of reciprocal space parameters $\eta,\nu$ are plotted), while Fig.~\ref{fig:symmetry_band_zigzag_untwisted} and \ref{fig:symmetry_band_armchair_untwisted} show symmetry adapted versions of these plots (i.e., band diagrams with chosen  reciprocal space parameters along cyclic or helical directions). Notably, a CKNT of cyclic group order $\mathfrak{N}$ is found to feature $2\mathfrak{N}$ nearly degenerate flat bands near the Fermi level.  An associated singular peak in the electronic density of states (Figs.~\ref{fig:DOS_untwist_zigzag}, \ref{fig:DOS_untwisted_armchair_CKNT}) is also observed\footnote{Our simulations suggest that the singular peak in the DOS of CKNTs occurs very close to the Fermi level --- about $0.006$ Ha away for the examples discussed here. In experimental situations, the peak in the DOS can be brought to the Fermi level exactly, by application of an external electric field or by doping \citep{FerromagnetismAndWignerCrystallization}, thus making the associated electronic states more readily accessible.}, and both zigzag and armchair tubes are found to feature quadratic band crossing (QBC) points  at $\eta = 0$ (corresponding to the gamma point of the flat sheet). These features make CKNTs striking examples of realistic quasi-one-dimensional materials that are likely to exhibit strongly correlated electronic states. The detailed investigation of exotic materials phenomenon in CKNTs resulting from such strong electronic correlations --- including e.g., Wigner crystallization, flat-band ferromagnetism and the emergence of superconducting, nematic or topological phases \citep{flat_band_heikkil__2011, FerromagnetismAndWignerCrystallization, dora2014occurrence} --- is the scope of future work. Considering that such phenomena have been studied primarily in bulk and two dimensional materials, the role that the quasi-one-dimensional morphology of CKNTs might play in them makes these investigations particularly interesting. 

\begin{figure}[htbp]
  \centering
    \subfloat[][VBM wavefunction]{\includegraphics[width=.4\textwidth]{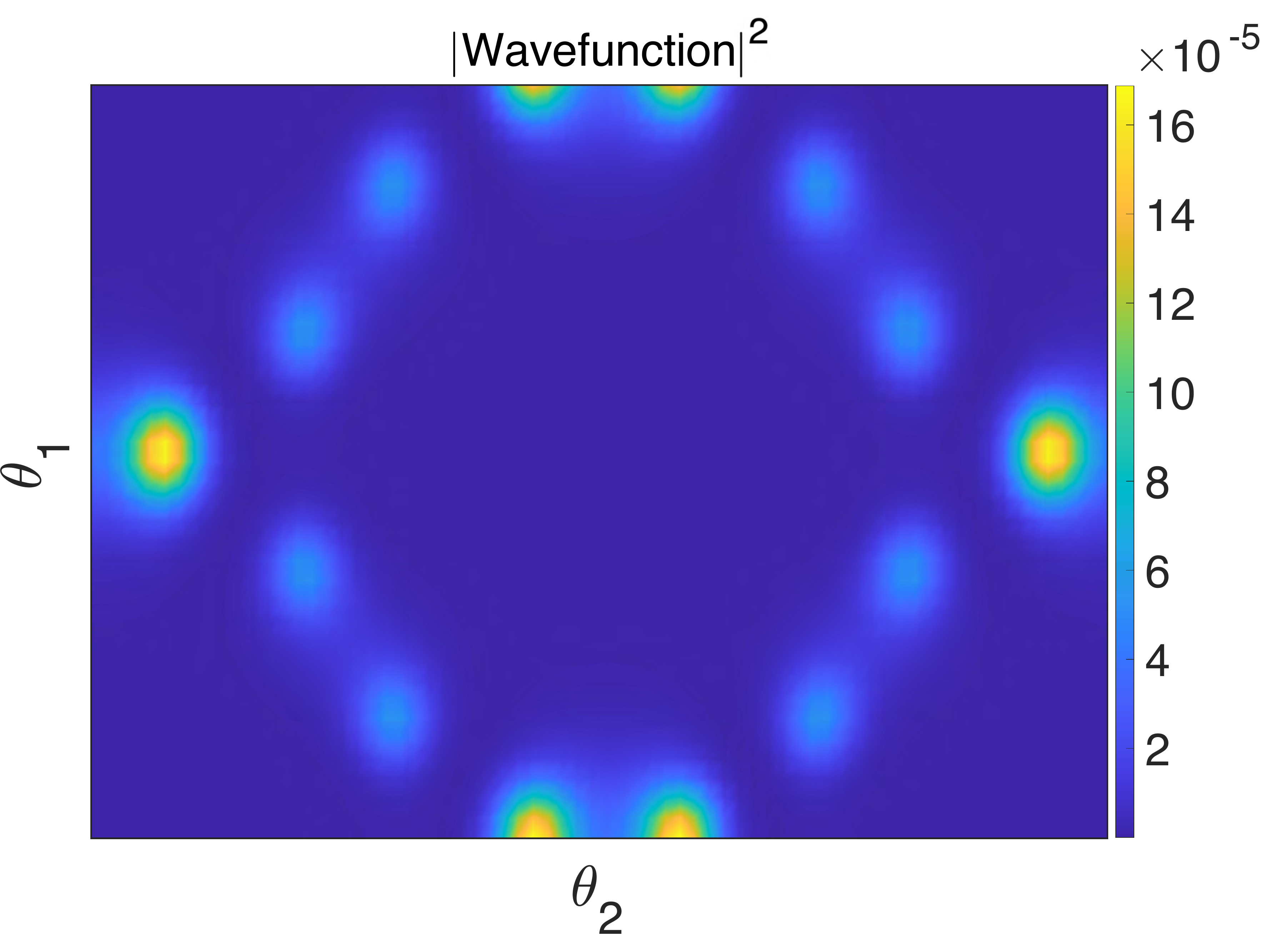}\label{fig:CKL_armchair_untwisted_wavefunc}} \hspace{0.5cm}
    \subfloat[][Electron density]{\includegraphics[width=.4\textwidth]{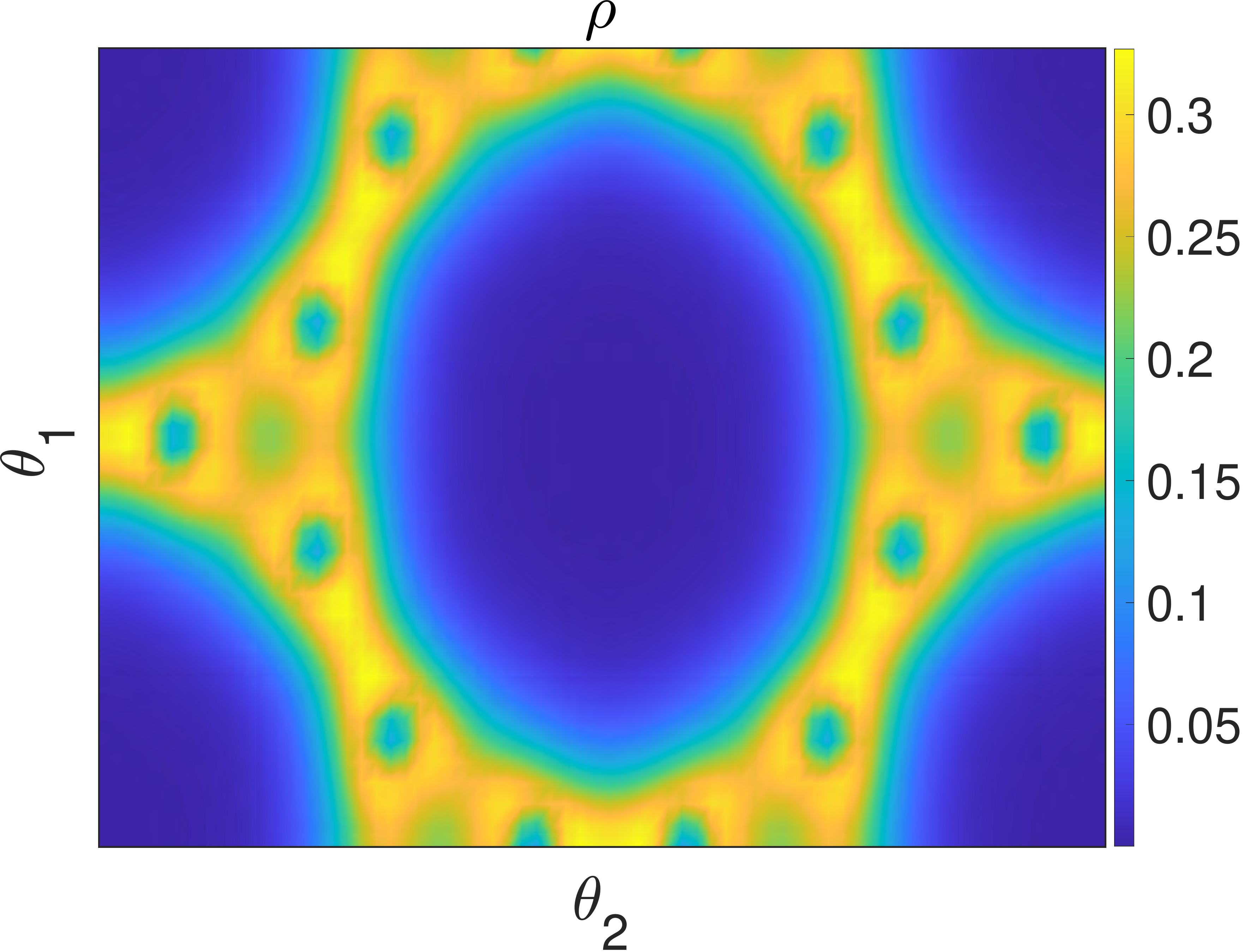}\label{fig:CKL_armchair_untwisted_rho}} 
  \caption{Valence Band Maximum (VBM) wavefunction and the electron density of an undeformed armchair $(12,12)$ CKNT. A slice of the electronic fields at the average radial coordinate of the atoms in the computational domain (represented using helical coordinates \citep{banerjee2021ab}) is shown.}
\label{fig:armchair_rho_psi}
\end{figure}

The dispersionless states in CKNTs are caused by destructive interference,  resulting from geometric and orbital frustration, as has also been shown to occur in other Kagome lattice systems \citep{kagome_flatband_guo,kagome_flatband_tang,kagome_flatband_yudin}. The electron effective mass is arbitrarily large at the flat band and the diminished electronic kinetic energy allows the Columbic interactions to dominate, resulting in strong electronic correlation. In turn, this causes electron localization\footnote{The localization mechanism described above is different from Anderson localization, where electronic waves become diffusionless due to disorder/impurities in the system \citep{cutler1969observation, lagendijk2009fifty}.} and the emergence of a sharp peak (i.e., van Hove singularity \citep{van1953occurrence}) in the electronic density of states (DOS) near the Fermi level (as shown in Figs.~\ref{fig:DOS_untwist_zigzag}  and \ref{fig:DOS_untwisted_armchair_CKNT}). The localized states corresponding to an undistorted armchair CKNT are shown in Fig.~\ref{fig:CKL_armchair_untwisted_wavefunc}. A plot of the electron density distribution for that system is also shown (Fig.~\ref{fig:CKL_armchair_untwisted_rho}).

\begin{figure}[htbp]
\centering
\subfloat[Projected density of states for an\\ armchair CKNT $(\mathfrak{N}=6)$]
{\includegraphics[ clip,trim=0.5cm 8.5cm 0.5cm 7.5cm,width=.45\textwidth]{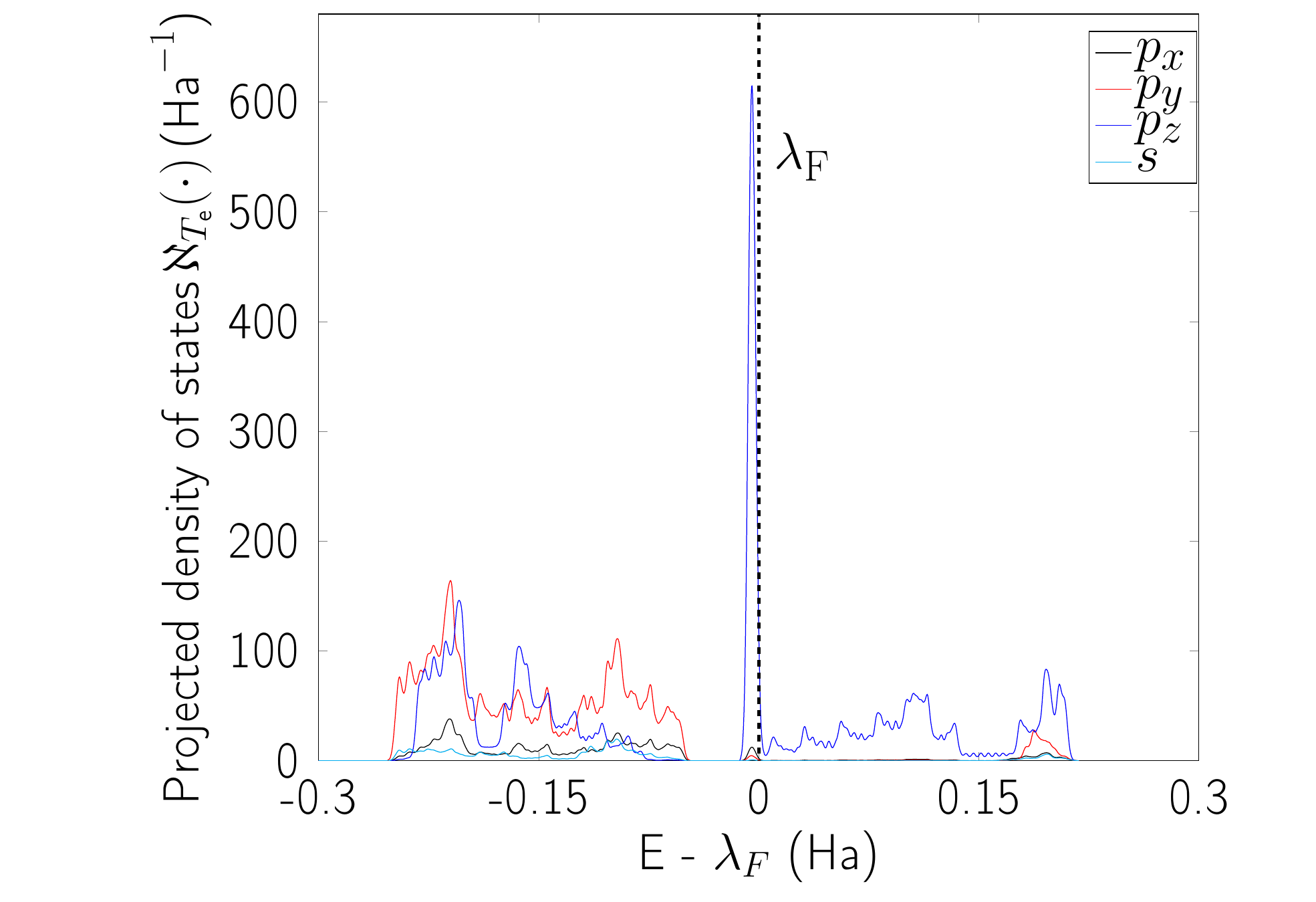}}
\subfloat[Projected density of states for a\\ zigzag CKNT $(\mathfrak{N}=9)$]
{\includegraphics[ clip,trim=0.5cm 8.5cm 0.5cm 7.5cm,width=.45\textwidth]{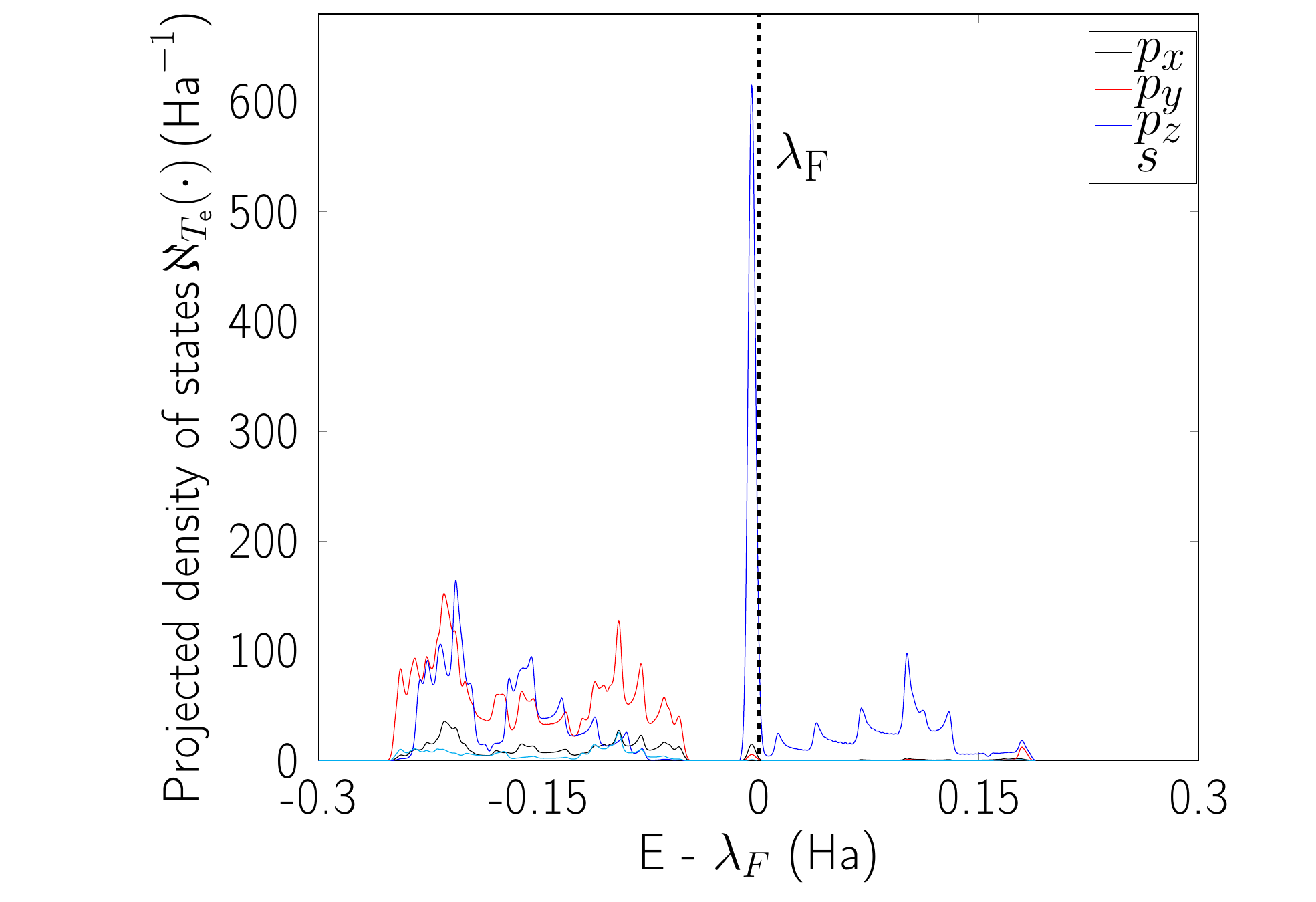}}
\caption{Projected density of states (PDOS) for undistorted armchair and zigzag CKNTs. The largest contribution to the sharp peak near the Fermi level is seen to arise from $\text{p}_{\text{z}}$ orbitals. 
}
\label{fig:pdos_QE}
\end{figure}

At this point, it is worth mentioning some similarities of the electronic properties of CKNTs with their conventional counterparts. Like conventional CNTs, the fascinating electronic properties of CKNTs are largely connected to $\pi$ electrons formed from radially oriented  $\text{p}_{\text{z}}$ orbitals, while the $p_\text{x}$ and $p_\text{y}$ orbitals form in-plane $\sigma$ bonds and are largely electronically inactive \cite{FerromagnetismAndWignerCrystallization}. This is supported by projected density of states (PDOS) calculations for CKNTs  (Fig.~\ref{fig:pdos_QE}), which show that the singular peak in the (total) electronic density of states near the Fermi level is largely attributable to the contributions of the individual (radially oriented) $\text{p}_{\text{z}}$ ($l = 1, m_l = 0$) orbitals, while $p_\text{x}$ and $p_\text{y}$ orbital contributions lie well below the Fermi level.  Moreover, the band diagrams of CKNTs have some similarities in appearance with those of conventional CNTS, e.g., the presence of Dirac points at $\eta = 0$ for zigzag tubes and at $\eta = \pm \frac{1}{3}$ for armchair ones (Figs.~\ref{fig:banddiag_zigzagfull_DFT}, \ref{fig:banddiag_armchairfull_DFT}). However, unlike conventional CNTs, these Dirac points do not appear at the Fermi level in undistorted  CKNTs, but are prominently featured as a part of the excited states of the system.  

\subsubsection{Electromechanical response of CKNTs: Effect of torsional and axial strains on electronic properties}
\label{subsec:electromechanical_prop}
\begin{figure}[htbp]
  \centering
   \subfloat[][Armchair $(12,12)$ CKNT with $4.5^\circ/\text{nm}$ twist]{\includegraphics[width=.50\textwidth]{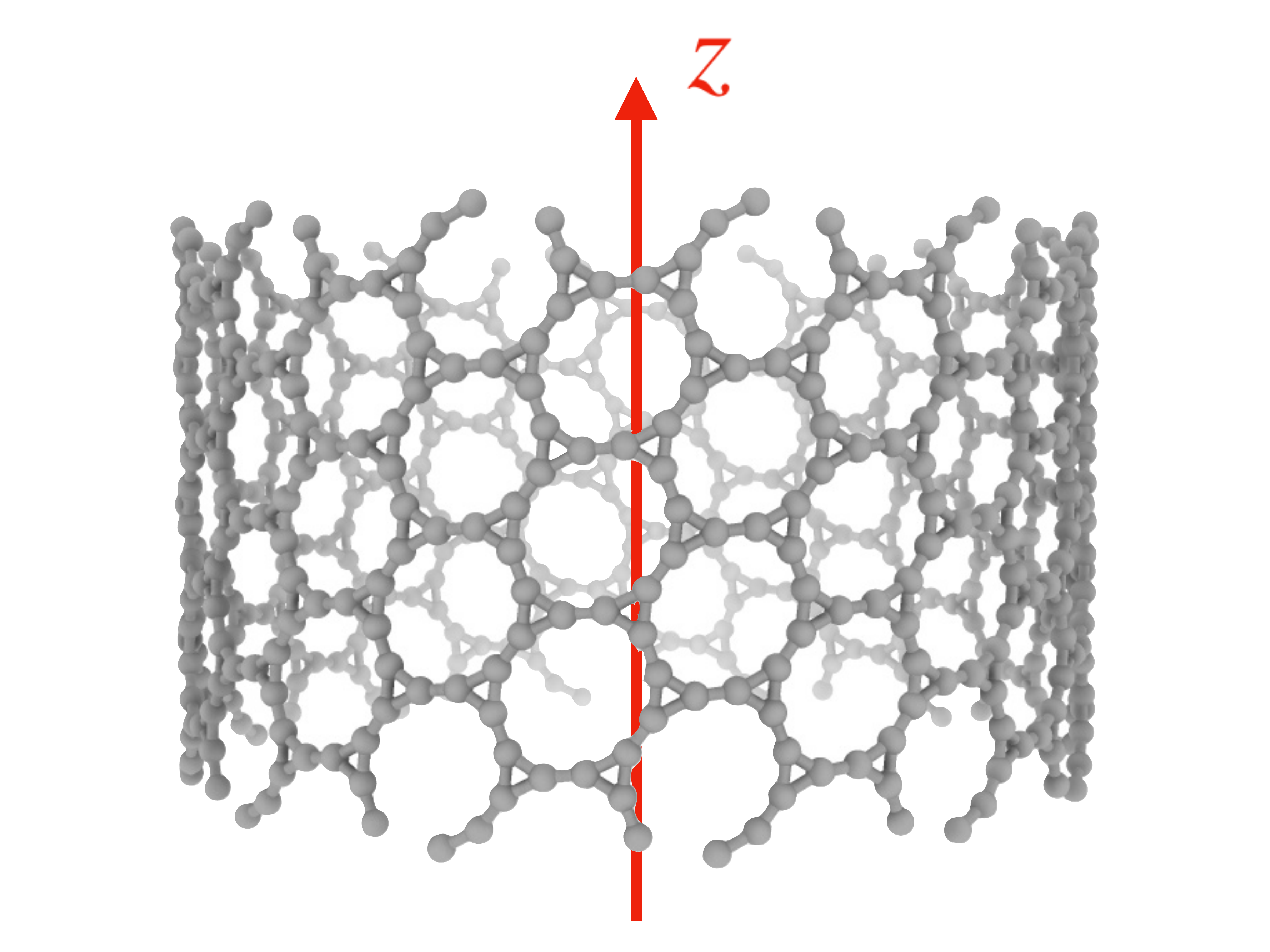}\label{fig:CKL_armchair_twisted}}\\
    \subfloat[][VBM wavefunction]{\includegraphics[width=.49\textwidth]{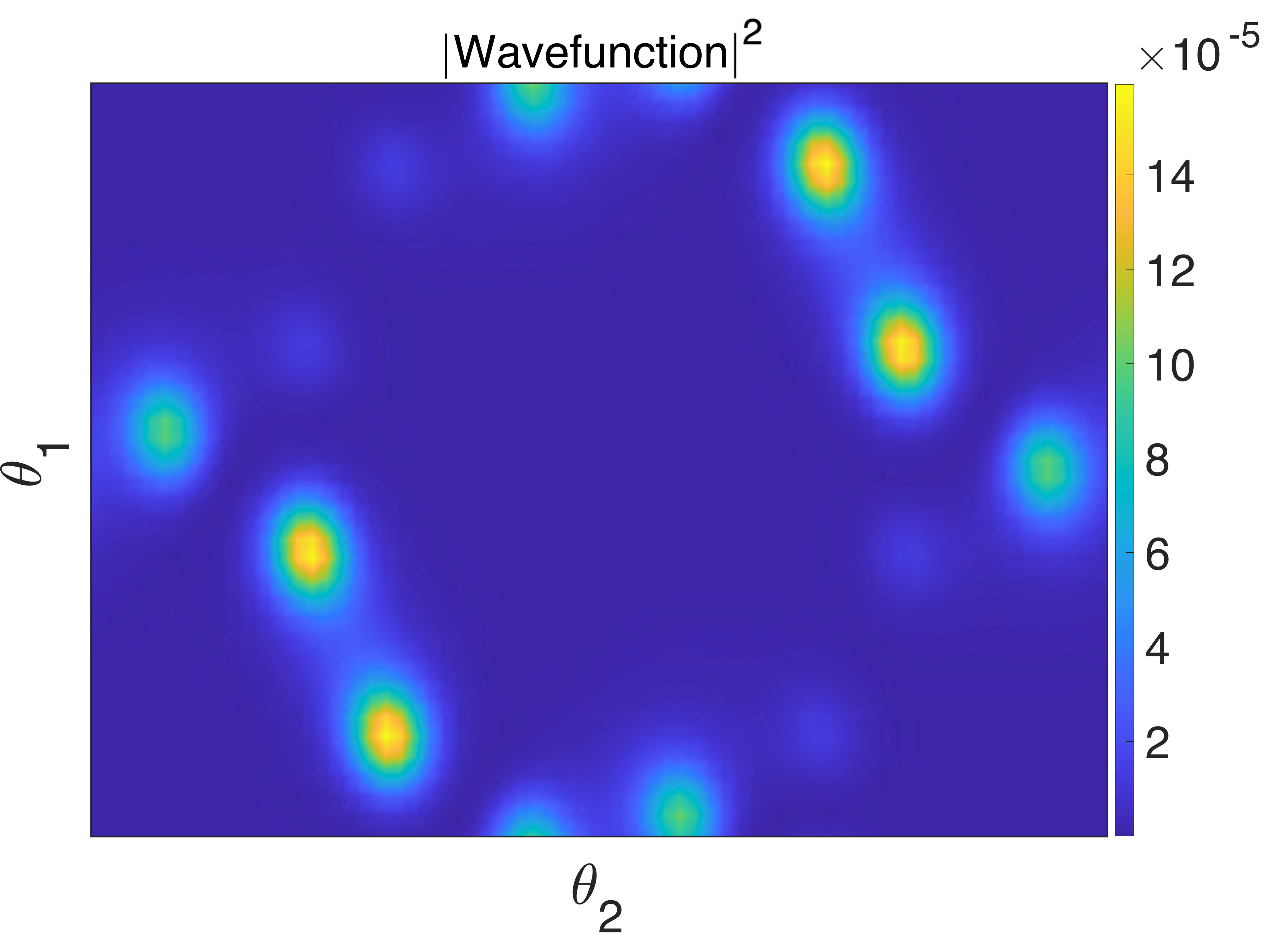}\label{fig:CKL_armchair_twisted_wavefunc}}\quad
    \subfloat[][Electron density]{\includegraphics[width=.46\textwidth]{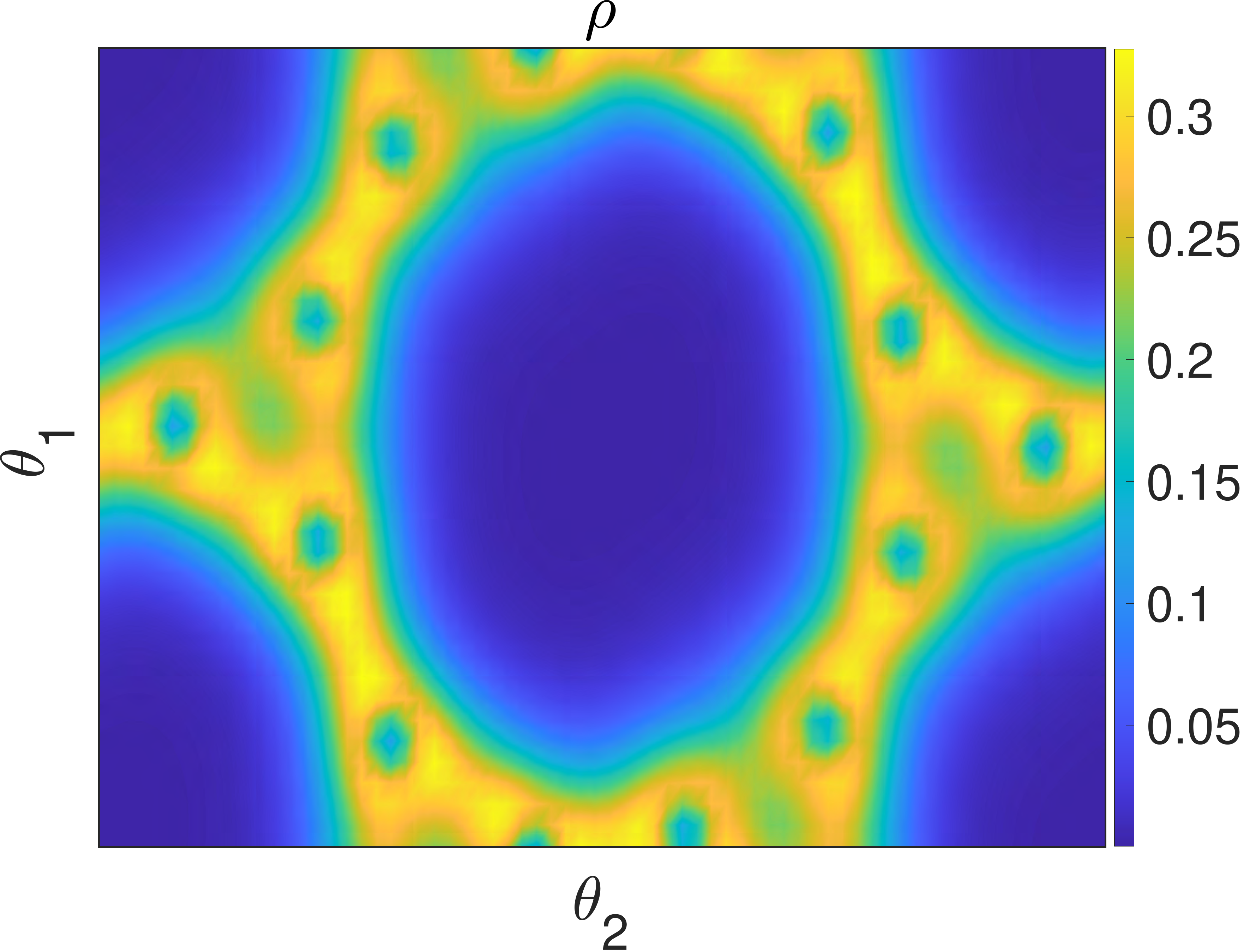}\label{fig:CKL_armchair_twisted_rho}}\\
  \caption{Atomic configuration, Valence Band Maximum (VBM) wavefunction and the electron density of an armchair $(12,12)$ CKNT with $\beta = 4.5^\circ/\text{nm}$ applied twist. A slice of the electronic fields at the average radial coordinate of the atoms in the computational domain (represented using helical coordinates \citep{banerjee2021ab}) is shown.}
\label{fig:twisted_armchair_rho_psi}
\end{figure}

\begin{figure}[htbp]
\centering
\subfloat[Complete band diagram for a twisted armchair CKNT of radius $1.70$ nm and a rate of twist of  $4.20^{\circ}$/nm. Complete and partial flat bands, along with emergent Dirac points are featured.]{\scalebox{0.4}
{
\includegraphics[clip, width=\textwidth]{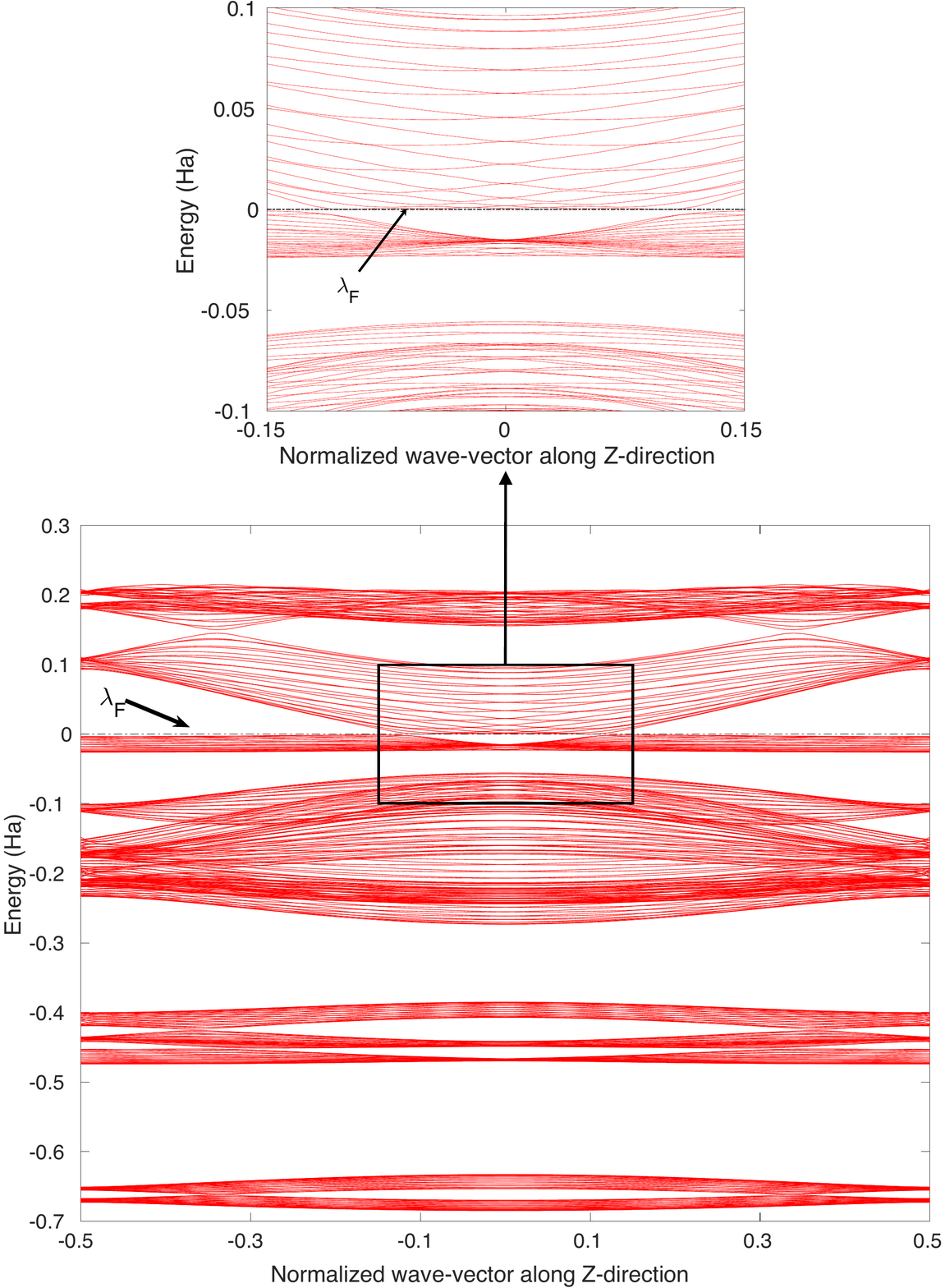}
}\label{fig:twisted_armchair_banddiag_DFT}}\quad
\subfloat[Variation of the electronic density of states of the same material with twisting.]
{\scalebox{0.4}
{
\includegraphics[ clip,trim=0.5cm 7.5cm 0.5cm 7.5cm,width=\textwidth]{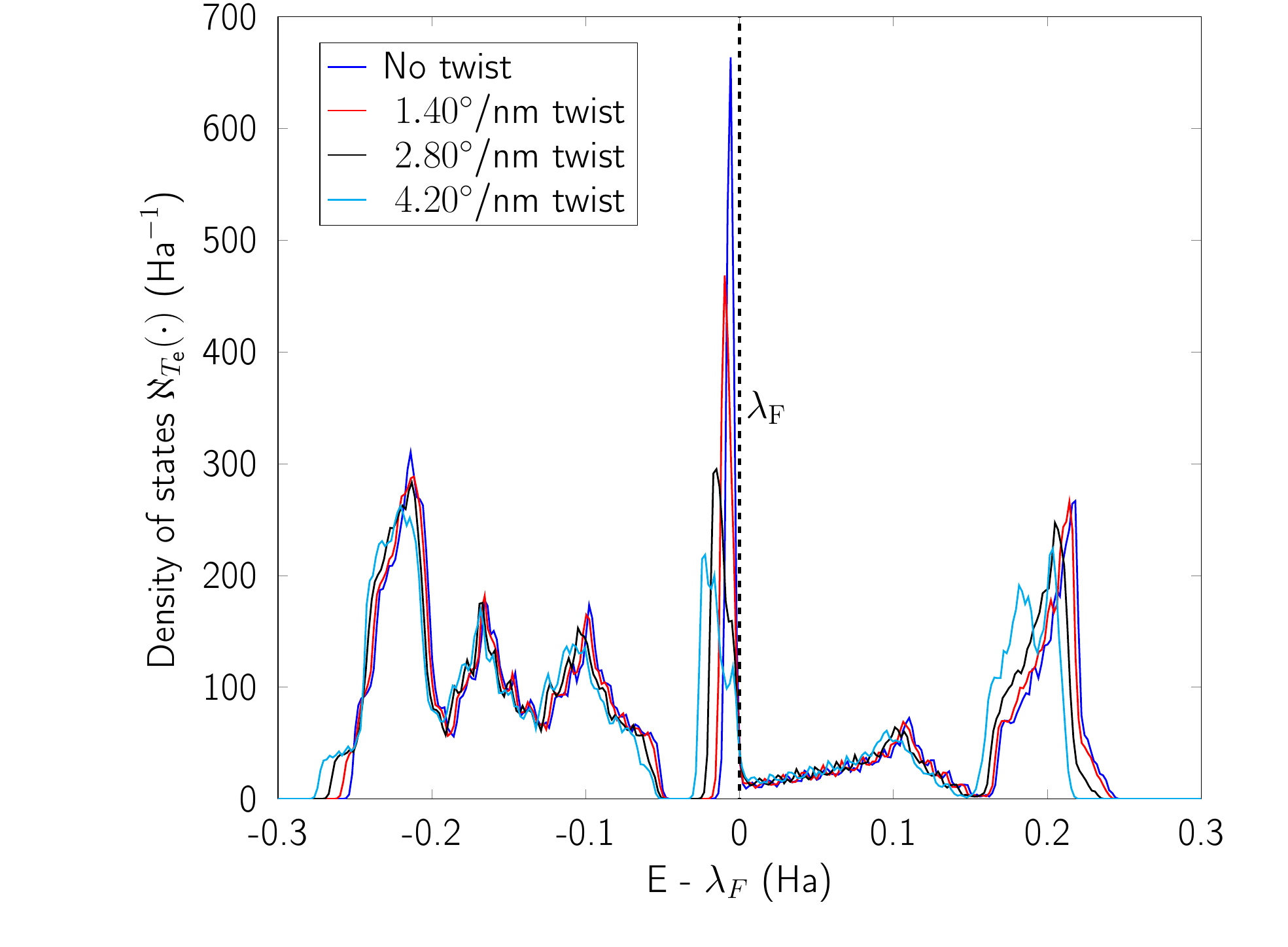}
}
\label{fig:density_of_states_twist}}
\caption{(a) Complete band diagram and (b) Electronic density of states near the Fermi level for the twisted armchair CKNT (radius $1.70$ nm). $\lambda_{\text{F}}$ denotes the Fermi level.}
\label{fig:armchair_twisted_band}
\end{figure}

We discuss the changes to electronic properties of CKNTs to applied deformations. First, we discuss the effect of torsional strains. We use the $(12,12)$ armchair CKNT as a prototypical example; zigzag CKNTs are found to have similar behavior. As shown in Fig.~\ref{fig:CKL_armchair_twisted}, application of twist to the CKNT destroys the in-plane $C_{6}$ symmetry of the underlying Kagome  graphene lattice structure, while maintaining its $C_{2}$ symmetry. Consequently \cite{sun2009topological, milicevic2019type, rhim2019classification}, the quadratic band crossing (QBC) point at $\eta = 0$ splits into a pair of Dirac points (Fig.~\ref{fig:twisted_armchair_banddiag_DFT})\footnote{The ``tilted'' nature of the linearly dispersive electronic bands near these Dirac points appears to suggest connections with (quasi-one-dimensional) Weyl semimetals \citep{milicevic2019type, gomes2021tilted}.}. Furthermore, the degeneracy in the $2\mathfrak{N}$ flat bands at the Fermi level appear to be lifted, and a number of dispersionless states appear to give way to bands that are only \textit{partially flat}. The change of completely flat bands to ones which have some dispersion near $\eta = 0$ is also evidenced by the electronic density of states plot in Fig.~\ref{fig:density_of_states_twist}, which shows that the sharp peak near the Fermi level decays as the rate of applied twist increases. Despite these twist induced changes, a number of dispersionless states survive (Fig.~\ref{fig:CKL_armchair_twisted}) and spatially localized wavefunctions associated with such states continue to be hosted by the nanotube   (Fig.\ref{fig:CKL_armchair_twisted_wavefunc}). Notably, the application of twist results in energy dispersion relations that feature rather dramatic changes in the electronic effective mass as the Brillouin zone is traversed --- from infinitely massive carriers near $\eta = 0$, to massless ones as the Dirac points are reached, and then re-appearance of infinitely massive ones as the edge of the Brillouin zone is approached (i.e., closer to $\eta = \pm \half$). Overall, these observations suggest that torsional strains provide a way of controlling correlated electronic states in CKNTs. Moreover, twisted CKNTs, being chiral, are likely to show asymmetric transport properties \citep{aiello2020chirality, naaman2015spintronics}. Therefore, they provide an interesting, realistic material platform where the combined manifestations of anomalous transport phenomena (the Chiral Induced Spin Selectivity effect \citep{naaman2012chiral}) and flat band physics may be realized and studied.

\begin{figure}[htbp]
\centering
\subfloat[Conventional band diagram for a stretched zigzag CKNT of radius $0.98$ nm and an axial strain of $-3.57$\%.]{\scalebox{0.4}
{
\includegraphics[clip, width=\textwidth]{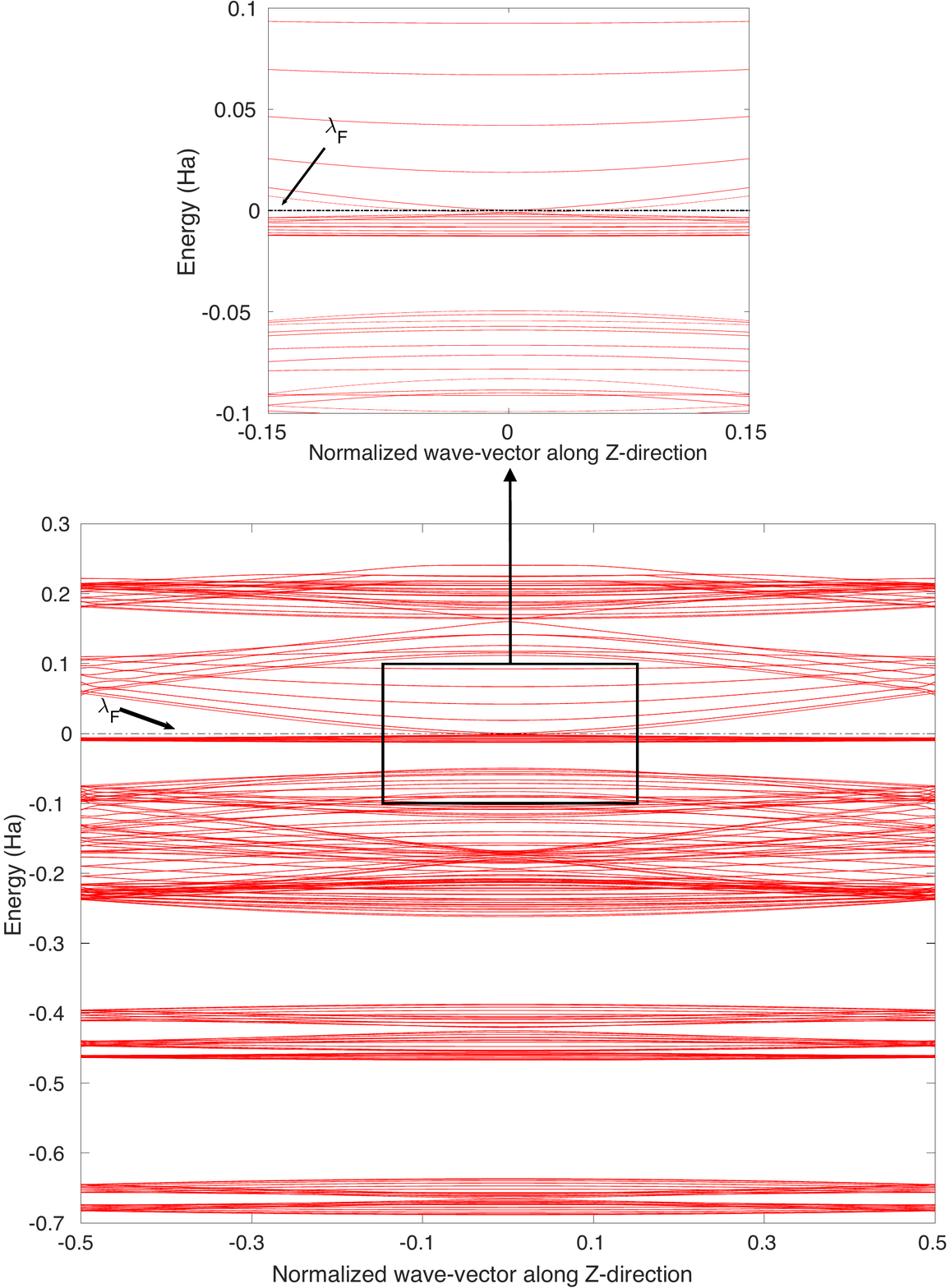}
}\label{fig:zigzag_stretch_full_banddaig}}\quad
\subfloat[Variation in electronic density of states of the same material with applied axial strains.]
{\scalebox{0.4}
{
\includegraphics[ clip,trim=0.5cm 7.5cm 0.5cm 7.5cm,width=\textwidth]{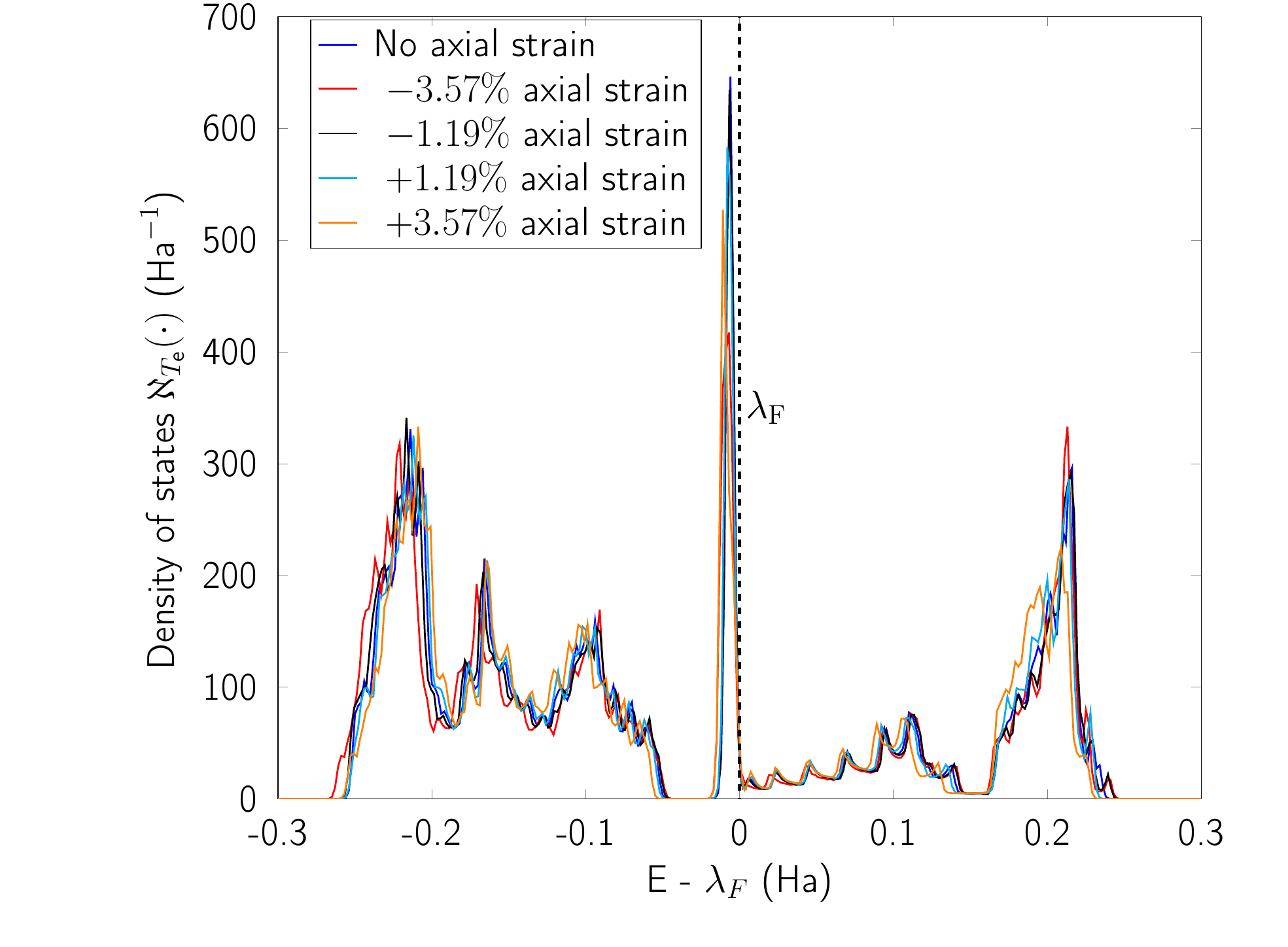}
}
\label{fig:DOS_stetch_zigzag_CKNT}}
\caption{(a) Complete band diagram and (b) Electronic density of states near the Fermi level for the stretched Zigzag CKNT (radius $0.98$ nm). $\lambda_{\text{F}}$ denotes the Fermi level.}
\label{fig:zigzag_stretched_band}
\end{figure}

Next, we discuss the case of axial strains. In general, such deformations also tend to introduce some degree of dispersion into the flat bands of CKNTs near $\eta = 0$, while lifting their degeneracies near the Fermi level. However, their influence appears to be less dramatic than the case of torsional deformations described above. Nevertheless, the axial compression case deserves particular mention. Considering the zigzag $(12,0)$ CKNT for example, we observe (see Fig.~\ref{fig:zigzag_stretch_full_banddaig}) that the dispersion introduced in the flat bands near $\eta = 0$ results in curvature of these states in a manner that is opposite (i.e., convex vs. concave) of the situation encountered while twist is applied (i.e., Fig.~\ref{fig:twisted_armchair_banddiag_DFT}). Thus, a scenario akin to the touching of a pair of parabolic bands \citep{murray2014renormalization, rhim2019classification} emerges. Upon subjecting the tube to larger values of compression, we observed that the parabolic bands near $\eta = 0$ give way to linear dispersion, i.e., the emergence of Dirac points. Commensurate  with these changes, the sharp peak in electronic the density of states (Fig.~\ref{fig:DOS_stetch_zigzag_CKNT}) also diminishes with increasing magnitude of axial strain, although the decrease appears to be less dramatic than the situation encountered with torsion (i.e., Fig.~\ref{fig:density_of_states_twist}).

Overall, the above observations are consistent with literature that suggests that quadratic band crossing points are unstable with respect to strains \citep{uebelacker2011instabilities, chong2008effective}. It is also worthwhile at this point to contrast the electromechanical response of CKNTs to conventional CNTS. 
Zigzag CNTs with cyclic group order divisible by 3 and armchair CKNTs are both metallic \citep{ghosh2019symmetry}, and they are known to be more sensitive to axial and torsional strains respectively. The effect of such deformations, at least for small strains, is to open up a gap at the Dirac points of these materials, resulting in metal-semiconductor transitions \citep{yang1999band, ding2002analytical, Dumitrica_Tight_Binding1, yang2000electronic, yu2022density, heyd1997uniaxial}. As described above, CKNTs appear to show more dramatic electronic transitions when subjected to such strains. At the same time, the simulations above suggest that at least some dispersionless states in CKNTs are robust and continue to be available when the tube is subjected to small torsional and axial strains.

\subsubsection{Tight Binding Model}
\label{subsec:TB_Intro}
Based on the findings of the PDOS calculations (Fig.~\ref{fig:pdos_QE}), we have developed a symmetry adapted tight binding (TB) model of the electronic structure of CKNTs. Our model involves $\pi$-electrons and incorporates up to next-nearest-neighbor interactions. The model is able to capture salient features of the electronic properties of undistorted CKNTs as revealed via first principles calculations. Additionally, by taking into account the variation in the hopping parameters with relative changes in bond lengths (eqs.~\ref{eq:hopping_parameter_NN}, \ref{eq:hopping_parameter_NNN}) the TB model is also able to account for the effects of strain on the electronic properties. We present details of the model in \ref{sec:app_TBM}. The TB model also produces results consistent with a $\pi$-orbital based empirical pseudopotential calculations \citep{mayer2004band, agarwal2022solution}, but has the added advantage of being analytical in nature. Further analysis of the mathematical properties of the TB model is the scope of ongoing work \citep{My_Shivam_Analytical_paper}.

\subsection{Possible routes to the synthesis of CKNTs}
\label{subsec:Synthesis}
Our calculations suggest that CKNTs are kinetically stable, which is often taken to be a promising sign of the synthesizability of carbon allotropes \cite{FerromagnetismAndWignerCrystallization, chen2014carbon}. In the past, a wide variety of metastable carbon allotropes have been fabricated  \cite{tyagi2021handbook}, often with significantly lower cohesive energies than other common stable counterparts. Examples of successful synthesis of such unusual carbon allotropes include $\gamma$-graphyne \citep{hu2022synthesis, desyatkin2022scalable,li2018synthesis, shin2014cohesion} , T-carbon \citep{Zhang2017_y_carbon, qin2019exploring} and nitrogen-doped Kagome graphene \citep{li2022selective, pawlak2021surface}. In particular, various methods of synthesis of conventional carbon nanotubes, including laser ablation and chemical vapor deposition have been explored \citep{guo1995catalytic, kumar2010chemical}. Some of these techniques have also been successfully used in manufacturing other 1D carbon allotropes, e.g. T-carbon nanotubes, which can be created from conventional carbon nanotubes through a picosecond pulsed-laser irradiation induced first order phase transition. Such techniques provide additional routes for synthesizing CKNTs. Irrespective of the method, we anticipate that the analysis presented in this paper is likely to be instrumental in realizing CKNTs experimentally.

\begin{figure}[htbp]
  \centering
    \subfloat[Two possible ways to fabricate 2D flat sheet of  Kagome graphene. Left panel: From tetrabromobicyclopropene to   organometallic intermediate with \ce{Ag} adatoms to the formation of final 2D sheet of carbon kagome lattice on \ce{Au}(111) etchant material on \ce{SiO2} substrate. Right Panel: From cyclopropane or bicycloprpane to 1D chain of cyclopropane to 2D sheet of carbon kagome lattice on \ce{Au}(111) etchant material on \ce{SiO2} substrate.]{\includegraphics[trim={0cm 0cm 0cm 0cm}, clip, width=0.6\linewidth]{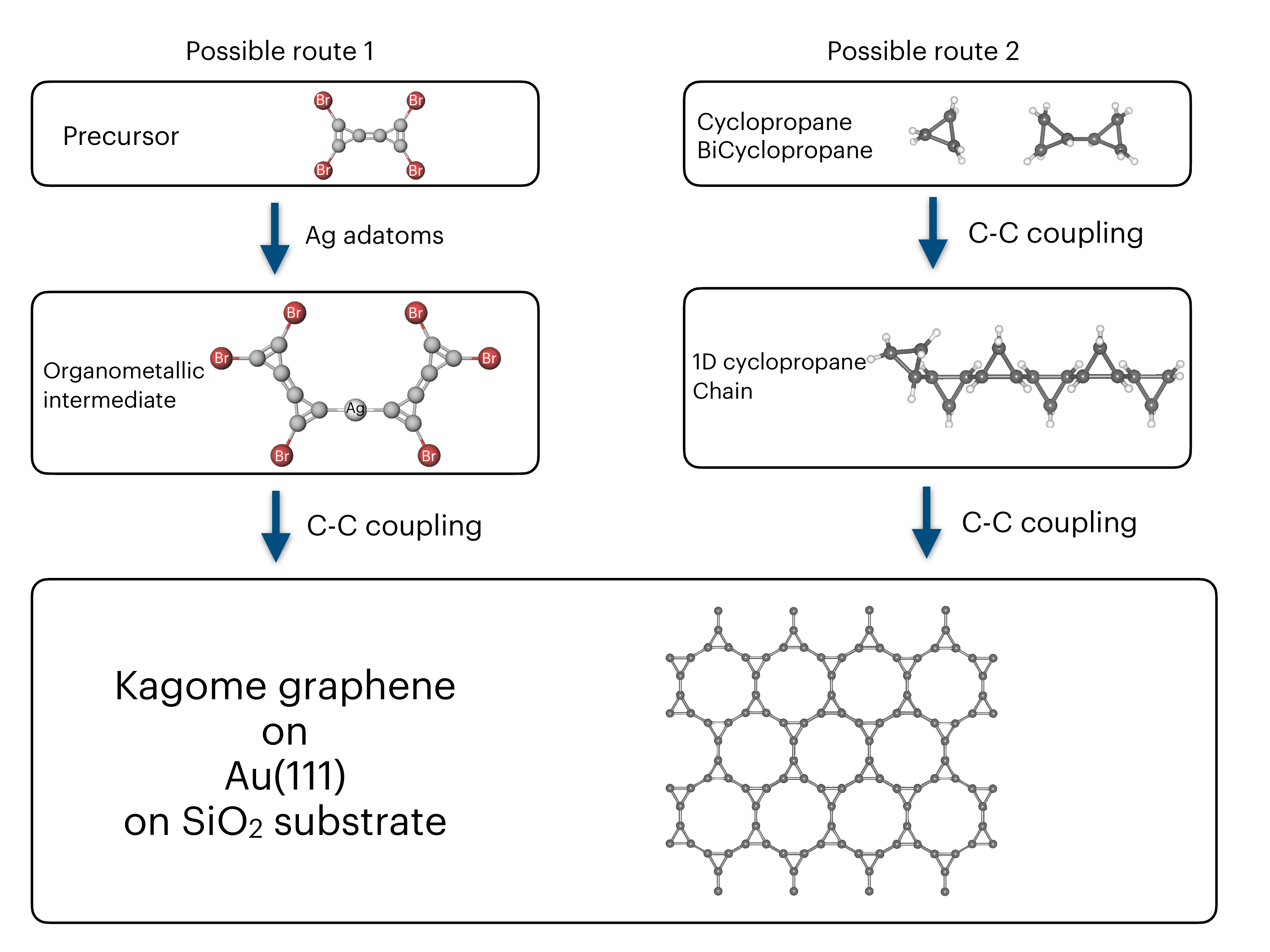}\label{fig:sch1}} 
    \hspace{1cm}
    \subfloat[Formation of Carbon kagome nanotube from Kagome graphene. After the target etching of Kagome graphene layer on \ce{Au} etchant sensitive material on \ce{SiO2} substrate surface the layer is curled up on to the sample surface where it can form bonds with itself.]{\includegraphics[trim={0cm 0cm 0cm 0cm}, clip,width=0.6\linewidth]{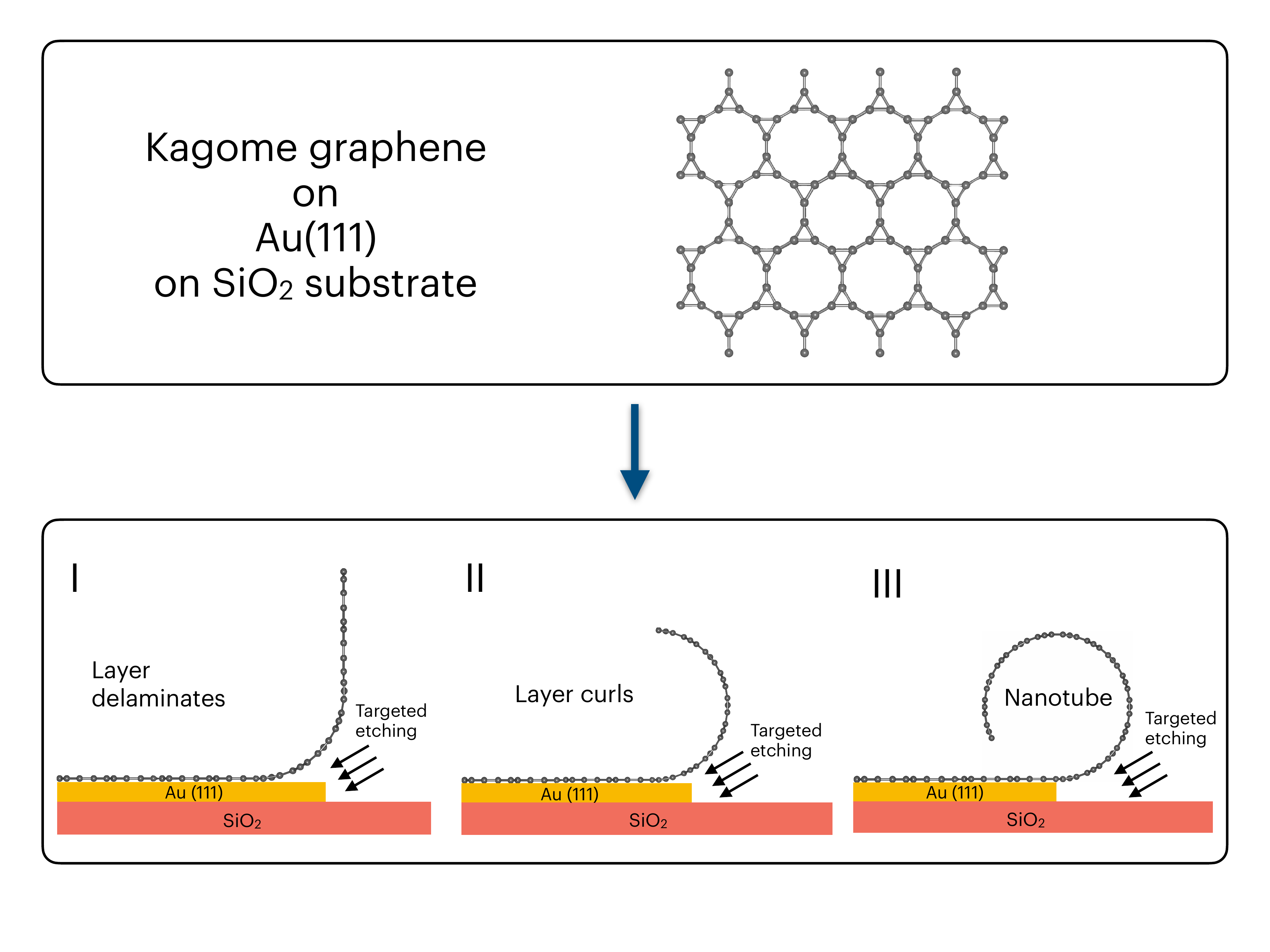}\label{fig:sch2}}
  \caption{Possible routes of synthesis of CKNTs. (a) Two possible routes to synthesis Kagome graphene. (b) Rolling up of a layer of Kagome graphene by target itching to form CKNTs.}
\label{fig:schematics}
\end{figure}

Since a variety of routes have been exploited for synthesizing different allotropes of carbon, multiple avenues also possibly exists for realizing CKNTs. We suggest two possibilities here, both based on the organic synthesis of Kagome graphene and subsequent roll-up of this material to form CKNTs. One possibility is to use silver adatoms to  transform tetrabromobocyclopropene to intermediate organometallic complex and to then form Kagome graphene on the surface of \ce{SiO2} substrate with etchant sensitive gold layer in between \cite{pawlak2021surface}, as represented in left panel of Fig. \ref{fig:sch1}. The second possibility is through the use of cyclopropane or bicyclopropane \cite{FerromagnetismAndWignerCrystallization, chen2014carbon}, wherein tailoring of ligand chemistry can be used to form self-assembled kagome graphene on the surface of gold $(111)$, deposited on a \ce{SiO2} substrate (shown in right panel of Fig. \ref{fig:sch1}).  This latter method is similar to recently demonstrated self-assembly procedures in metastable carbon nanowiggles \cite{PhysRevLett.107.135501, han2015self}. After the synthesis of Kagome graphene, targeted etching of the gold layer \cite{schmidt2001thin} can result in the curling of the 2D material into CKNTs, as desired. The lower bending stiffness of kagome graphene in contrast to conventional graphene (Section \ref{subsec:structural_properties})  will likely assist in this step (Fig.\ref{fig:sch2}).

\section{Conclusion}
\label{sec:conclusions}
In summary, we have introduced carbon Kagome nanotubes (CKNTs) --- a new allotrope of carbon formed through a roll-up construction of Kagome graphene. These nanotubes are unique in the sense that contemporarily, they are perhaps the only known example of realistic, elemental, 1D nanostructures that can host dispersionless electronic states, or \textit{flat bands}, in the entirety of their (one-dimensional) Brillouin zone. They also feature Dirac points in their band diagram, as well as a singular peak in their electronic density of states. Thus, they provide an attractive  platform for studying and exploiting strongly correlated electronic phenomena in nanomaterials with wire-like geometries.

To characterize CKNTs, we have carried out an extensive series of first principles simulations, including specialized calculations based on symmetry adaptation. We have investigated both zigzag and armchair varieties of CKNTs. Cohesive energy computations  and ab initio molecular dynamics simulations suggest CKNTs should exist as stable structures at room temperature and beyond. Our simulations reveal that it is easier to roll Kagome graphene into CKNTs than it is to roll conventional graphene into CNTs. Moreover,  CKNTs are found to be significantly more pliable with respect to torsional and extensional deformation as compared to CNTs of similar radii. Both zigzag and armchair varieties of CKNTs are found to be metallic, and are seen to feature multiple degenerate flat bands and a corresponding singular peak in the electronic density of states, near the Fermi level. We studied the response of electronic properties of CKNTs to applied torsional and axial strains and showed that such deformations are very effective in inducing large changes in the electronic states of this material. We also developed a tight binding model which is able to capture many of the electronic properties of CKNTs observed in first principles calculations, and also reproduces correctly, the effect of strains in the material.

Given the relative stability of CKNTs and the fact that there has been recent successes in experimentally synthesizing various novel carbon allotropes, it appears likely that CKNTs can be produced and studied under laboratory conditions in the near future. Such pursuits, as well as more detailed characterizations of electronic phenomena in CKNTs at a level beyond Density Functional Theory, and the exploration of anomalous transport phenomena in distorted CKNTs, are all attractive directions of future research.
\begin{center}
---
\end{center}
\appendix
\section{Tight binding model for CKNTs}
\label{sec:app_TBM}
We have developed a $p_{\pi}$-orbital based tight-binding (TB) model for understanding the electronic structure of CKNTs. As described below, to correctly reproduce the findings of the ab initio calculations, especially the effects of strain, we found it necessary to include up to next-nearest-neighbor (NNN) interactions and to make the hopping amplitudes dependent on relative changes in bond lengths.  The TB Hamiltonian in second quantization framework \citep{fetter2012quantum, girvin2019modern}  is written as: 
\beqs
\bfH = \sum_{i,\gamma}\varepsilon_{i\gamma}\mathsf{a}^\dagger_{i\gamma}\mathsf{a}_{i\gamma} + \sum_{\gamma}\sum_{\langle i,j\rangle}t_{1(i\gamma,j\gamma)}\mathsf{a}^\dagger_{i\gamma}\mathsf{a}_{j\gamma} \nonumber \\
+\sum_{\gamma}\sum_{\langle\langle i,j \rangle\rangle}t_{2(i\gamma,j\gamma)}\mathsf{a}^\dagger_{i\gamma}\mathsf{a}_{j\gamma} + \text{h.c.}\,.
\eeqs
Here, $\varepsilon_{i\gamma}$ denotes the onsite energy of site $i$ and orbital $\gamma$, $t_{1(i\gamma,j\gamma)}$ and $t_{2(i\gamma,j\gamma)}$ are the hopping amplitudes between orbitals $\gamma$ of the nearest-neighbors (NNs) $\langle i,j\rangle$ and the next-nearest-neighbors (NNNs) $\langle\langle i,j \rangle\rangle$, respectively, and h.c. is the hermitian conjugate. The annhilation and creation operators are denoted by $\mathsf{a}_{i\gamma}$, $\mathsf{a}^\dagger_{i\gamma}$, respectively.  The onsite energy, $\varepsilon_{i\gamma}$, is considered zero for convenience. To calculate the TB band structure of CKNTs, we adopt the Dresselhaus method \cite{Dresselhaus_CNT_textbook}. This procedure involves development of a TB formulation for the flat sheet (i.e., Kagome graphene), and subsequent mapping of the atoms of the two-dimensional lattice to a cylinder, so as to invoke boundary conditions appropriate to the nanotube. Details of these steps are discussed below.

\begin{figure}[htbp]
\centering
\subfloat[Zigzag $(12,0)$ CKNT.]{\scalebox{0.9}
{\includegraphics[ clip, width=0.5\textwidth]{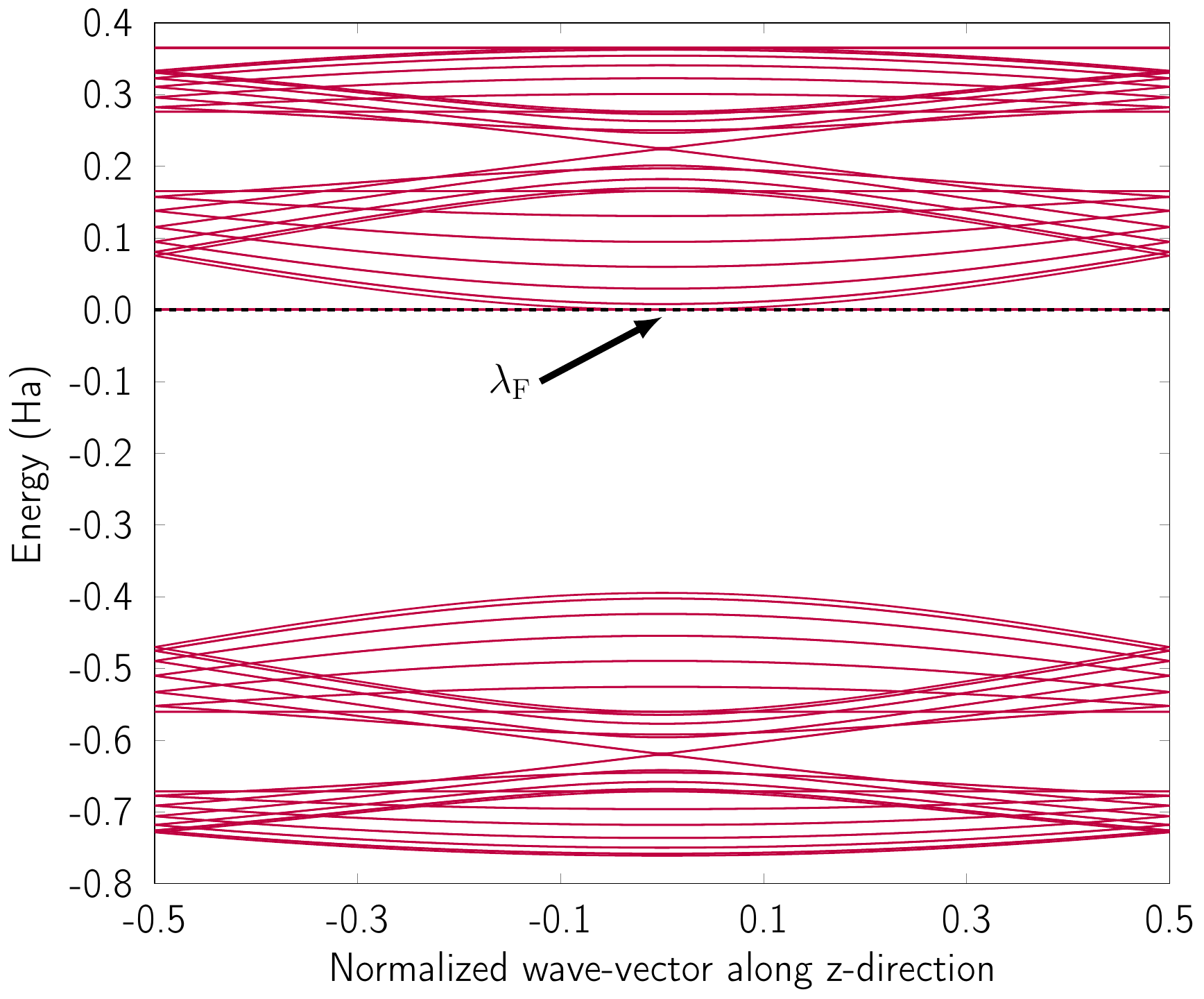}} 
\label{fig:TB_zigzag_band_CK}}$\;$
\subfloat[Armchair $(12,12)$ CKNT.]{\scalebox{0.9}
{\includegraphics[ clip, width=0.5\textwidth]{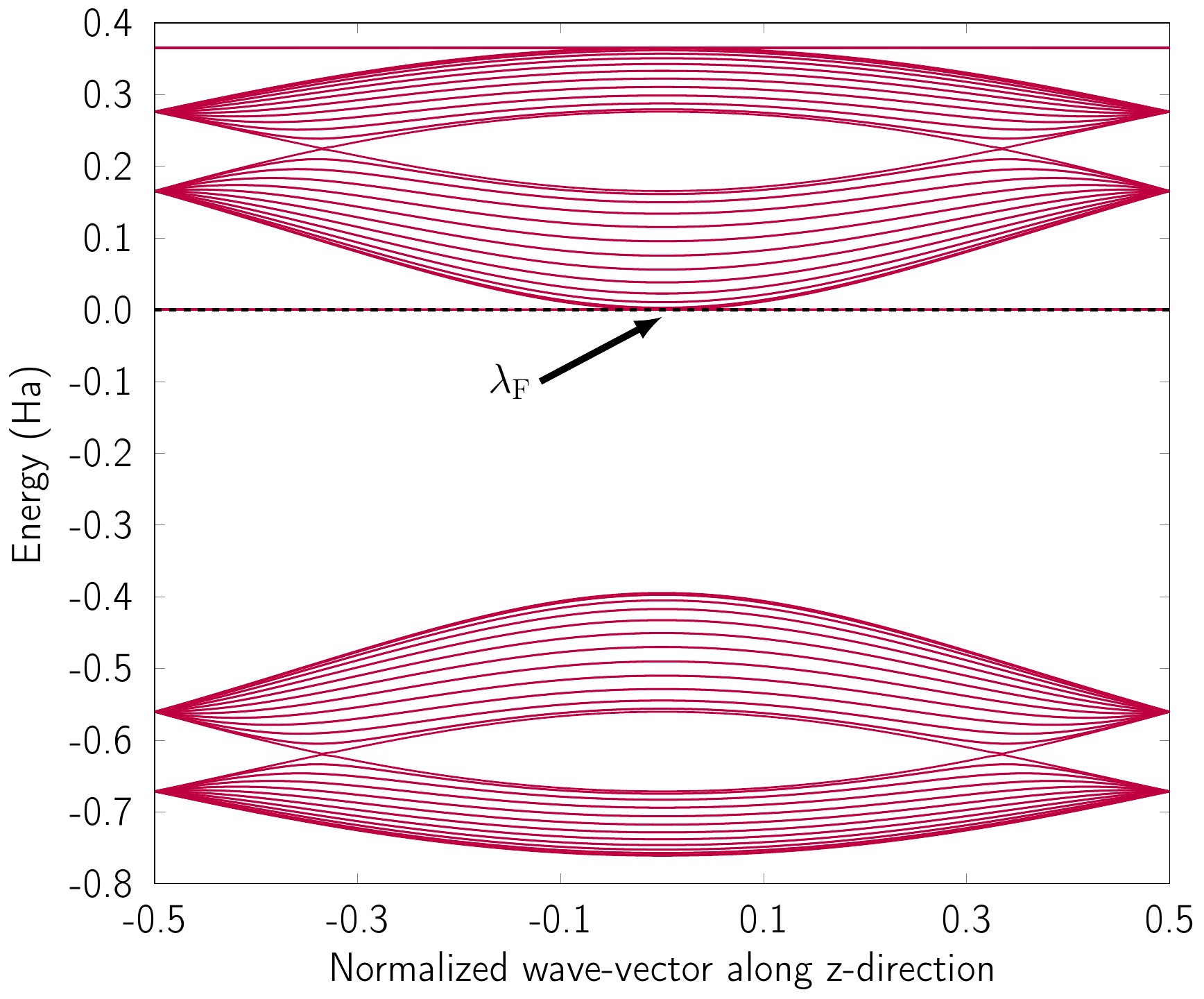}} 
\label{fig:TB_armchair_band_CK}}
\caption{Band diagram of undistorted CKNTs obtained from the tight binding model: (a) Armchair and (b) Zigzag cases shown. $\lambda_{\text{F}}$ denotes the Fermi level.}
\label{fig:CKL_TB_band_diag}
\end{figure}

\begin{figure}[htbp]
\centering
\subfloat[Zigzag $(12,0)$ CKNT]{\scalebox{0.9}
{\includegraphics[ clip, width=0.5\textwidth]{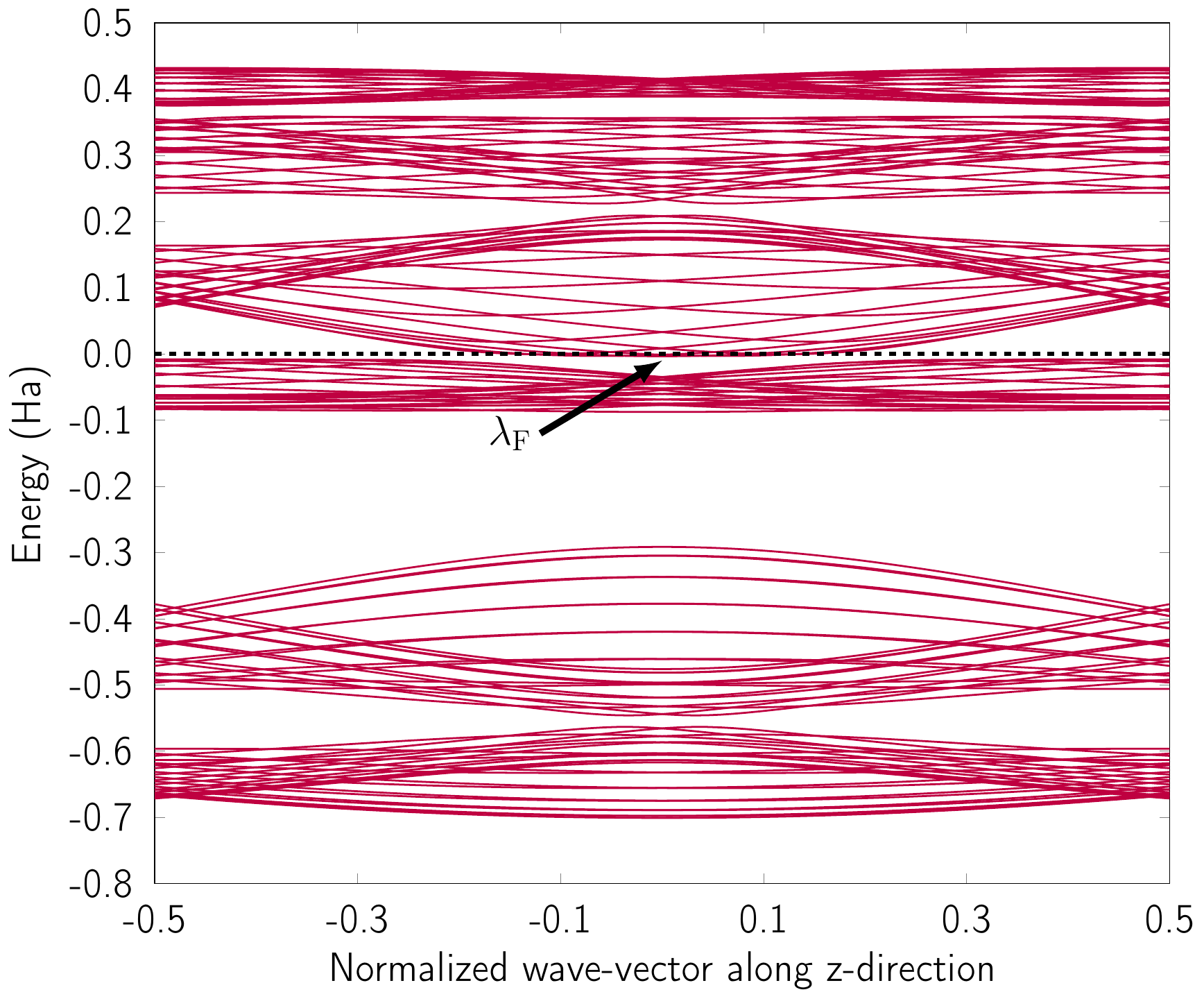}} 
\label{fig:TB_band_diag_zigzag_twisted_CKNT}}$\;$
\subfloat[Armchair $(12,12)$ CKNT]{\scalebox{0.9}
{\includegraphics[ clip, width=0.5\textwidth]{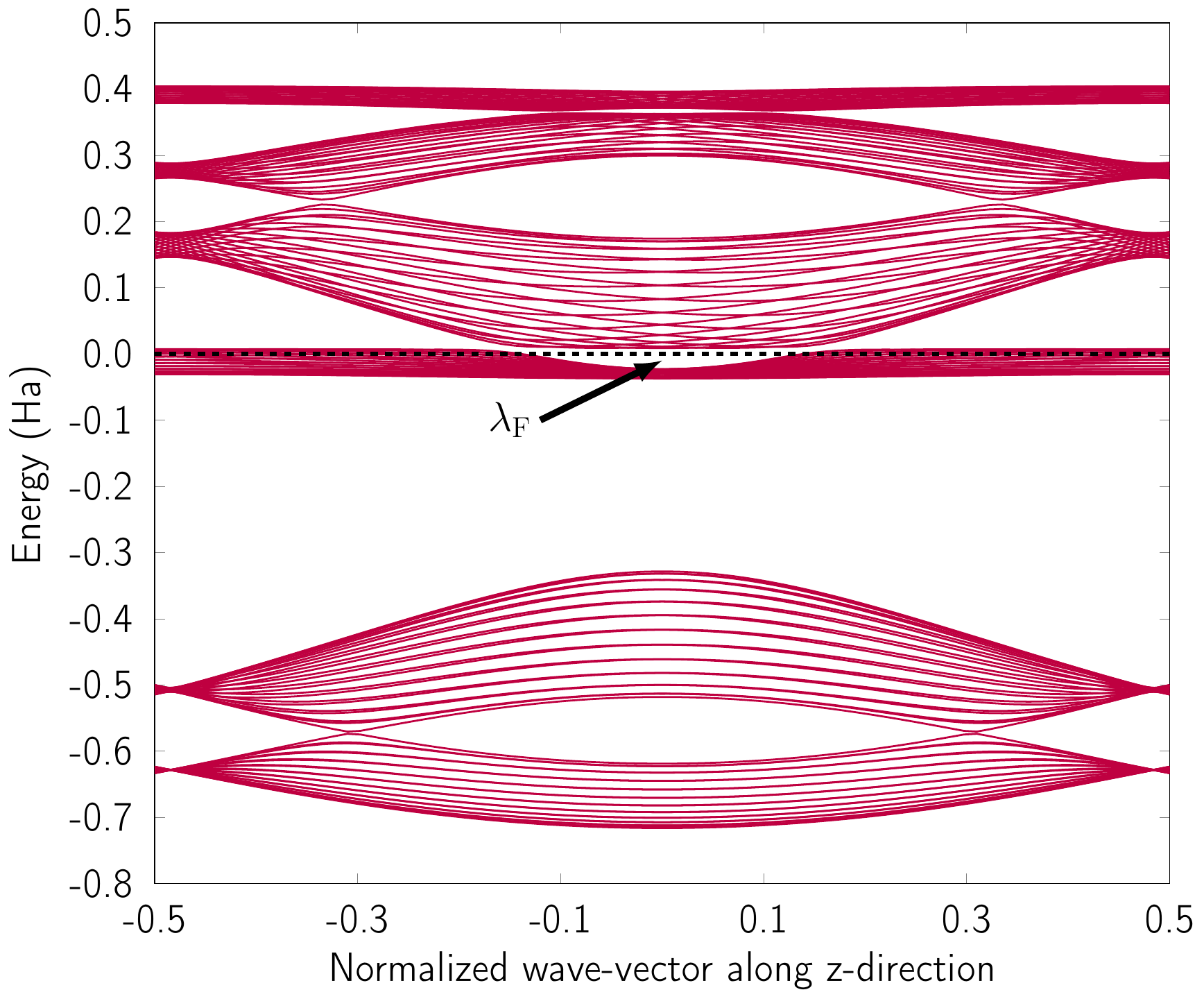}} 
\label{fig:TB_banddiag_armchair_CK_Twisted}}
\caption{Band diagram of twisted CKNTs obtained from the tight binding method, with about $4.8^\circ/\text{nm}$ of applied twist. $\lambda_{\text{F}}$ denotes the Fermi level.}
\label{fig:CKL_TB_band_diag_twisted}
\end{figure}

Results from our TB calculations are presented in Figs.~\ref{fig:CKL_TB_band_diag} and \ref{fig:CKL_TB_band_diag_twisted}. As can be seen, there is excellent qualitative agreement between these results and the first principles data presented earlier. 

\subsection{Tight Binding model for Kagome graphene}
\label{subsec:app_TB_kagome_graphene}
\begin{figure}[htbp]
    \centering
    \includegraphics[trim={1cm 1cm 3cm 0.5cm}, clip, width=0.6\linewidth]{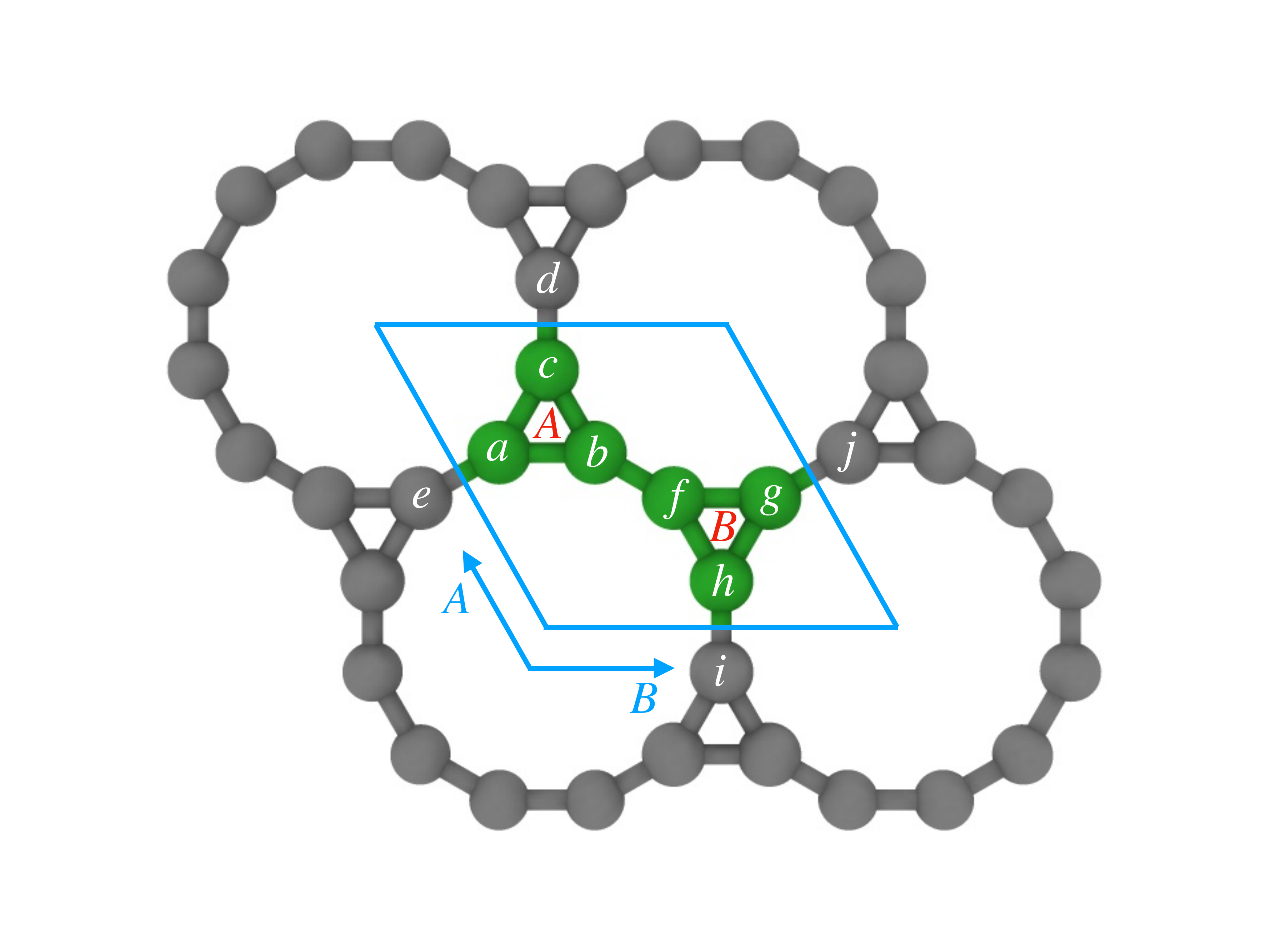}
    \caption{Kagome Graphene structure with two sublattices $A$ and $B$ consisting of total six carbon atoms in the unit cell.}
    \label{fig:ckl_unitcell_TB}
\end{figure}

The hexagonal unit cell of Kagome Graphene consist of two triangular sublattices A and B (Fig.~\ref{fig:ckl_unitcell_TB}), consisting of a total of six Carbon atoms per unit cell. We consider only $p_z$ orbitals in the TB method which yield a $6 \times 6$ TB hamiltonian, written in momentum space for NN interactions as \citep{FerromagnetismAndWignerCrystallization}:
\begin{equation} 
\bfH^{\text{NN}}_{\text{Kagome Graphene}} = 
\begin{pmatrix}
\bfH_{AA} && \bfH_{AB} \\ 
\bfH_{BA} && \bfH_{BB}
\end{pmatrix},
\end{equation}
Axial or torsional strains in CKNTs can be mapped to appropriate deformations in Kagome graphene. For example, torsional strains in CKNTs have the effect of inducing a shear strain in the underlying Kagome graphene lattice. Generally, such distortions transform the sublattices in Kagome graphene from being perfect equilateral triangles to scalene triangles, so that the hopping matrix of sublatice $B$ will no longer be the hermitian conjugate of that of subattice $A$. The NN hopping matrix $\bfH_{AA}$ of the sublattice $A$ can be written as (refer to Fig.~\ref{fig:ckl_unitcell_TB}):
\begin{align}
  \bfH_{AA} =
\begin{pmatrix}
0 && t_{1ab} e^{ix_{2ab}\bfa_1\cdot\bfk} && t_{1ac} e^{ix_{2ac}\bfa_2\cdot\bfk}  \\
t_{1ab} e^{-ix_{2ab}\bfa_1\cdot\bfk} && 0 && t_{1bc} e^{ix_{2bc}\bfa_3\cdot\bfk} \\
t_{1ac} e^{-ix_{2ac}\bfa_2\cdot\bfk} && t_{1bc} e^{-ix_{2bc}\bfa_3 \cdot \bfk} && 0
\end{pmatrix}, 
\end{align}
while the inter-triangular hopping matrix $\bfH_{AB}$ can be written as:
\begin{align}
\begin{split}
\bfH_{AB}  = \text{Diag}[t^{\prime }_{1ae}e^{ix_{1ae}(\bfa_1+\bfa_2)\cdot \bfk},  t^{\prime }_{1cd}e^{-ix_{1cd}(\bfa_2-\bfa_3)\cdot \bfk}, \\ t^{\prime }_{1bf}e^{-ix_{1bf}(\bfa_1+\bfa_3)\cdot \bfk}],
\end{split}
\end{align}
The matrices $\bfH_{BB}$ and $\bfH_{BA}$ can be written in the similar manner, and are not shown here for brevity. Consistent with the geometry of Kagome graphene and the mapping of strains from CKNTs to the Kagome graphene lattice, we choose $\bfa_1 = [1,0,0], \; \bfa_2 = [\cos(\theta_0+2\pi\alpha), \sin(\theta_0 + 2\pi\alpha),0]$ and $\bfa_3 = \bfa_1- \bfa_2$ as the unit vectors in real space, and we further set $\theta_0 = \frac{\pi}{3}$. We denote $\bfk = [k_x,k_y]$ as the unit vector in reciprocal space.

The hopping amplitudes depend on the structure of the lattice and the applied deformations, and can be written as the function of relative change in bond length  \cite{jiang2019topological}:
\beq
\begin{split}
t_{1uv} &= t^o_{1}\left(\exp{\left(\frac{x_2 - x_{2uv}}{x_2}\right)}\right)^n \;\;\; \text{and} \\ t^\prime_{1uv} &= t^{\prime o}_{1}\left(\exp{\left(\frac{x_1 - x_{1uv}}{x_1}\right)}\right)^n\,.
\label{eq:hopping_parameter_NN}
\end{split}
\eeq

Here  $t^{\prime 0}_1 = -5 \;\text{eV}$,  $t^o_{1} = -7 \; \text{eV}$ are the NN hopping paramters \cite{you2019flat}, and  $x_1$ and $x_2$ are inter- and intra-triangular bond lengths in the undeformed structure (Fig.~\ref{fig:hexagonal_ckl}). The bond lengths $x_{2uv}$ and $x_{1uv}$ are in the deformed system, and $u,v$ run over the NN atoms as shown in Fig.~\ref{fig:ckl_unitcell_TB}. The parameter $n$ controls the magnitude of the hopping parameter upon deformation of the lattice and parameterizes the ``flatness'' of the dispersionless states. Here, $\alpha$ is the nanotube's applied twist parameter, as explained earlier in section \ref{sec:helical dft}. Following  literature \cite{jiang2019topological}, we use $n =8$ for getting both relatively flat bands and smooth evolution of band structure upon deformation. 

The NNN interactions in our TB Hamiltonian take the following form:
\beq
\bfH^{\text{NNN}}_{\text{Kagome Graphene}} = 
\begin{pmatrix}
\mathbf{0} && \bfH_{\text{NNN}} \\ 
\bfH_{\text{NNN}}^{\dagger} && \mathbf{0}
\end{pmatrix}\,,
\eeq
here, hopping matrix $\bfH_{\text{NNN}}(u,u) =0$ and  $\bfH_{\text{NNN}}(u,v) = t_{2uv}e^{i\bfk\cdot \bfx_{3uv}}, \; \text{for}\;  u\neq v$. The NNN hopping parameter is given as:
\beq
t_{2uv} = t_2^{o}\left(\exp\left(\frac{x_3 - x_{3uv}}{x_3}\right)\right)^n.
\label{eq:hopping_parameter_NNN}
\eeq

Here $x_3$ is the distance between the NNN in the original structure and $x_{3uv}$ is the same distance in the deformed nanotube; $u,v$ denote the NNN atoms; and $t_2^o = -0.0125\; \text{eV}$.

\subsection{\texorpdfstring{$(n,m)$}{(n,m)} nanotube and boundary conditions}
\label{subsec:app_nanotube_BCs}
To adapt the TB model of Kagome graphene described above to CKNTs, we follow the method developed by Dresselhaus \cite{Dresselhaus_CNT_textbook}. The general procedure starts by considering a semi-infinite two-dimensional sheet, whereby the two opposite edge sides are glued together to form an infinite tube, with the axis of the tube chosen to be the $\textbf{e\textsubscript{Z}}$-direction. The periodicity along the circumferential direction naturally imposes periodic boundary conditions on the circumferential wave vector, denoted $\boldsymbol{k_{\perp}}$ below. In what follows, we briefly describe the procedure for imposing such periodic boundary conditions on a nanotube constructed from an arbitrary two-dimensional lattice. Axial and torsional deformations in the nanotube can be naturally incorporated by considering the effect of applied strains to the two-dimensional lattice.

We first describe the construction of an $(n,m)$ nanotube by rolling up a flat sheet along the direction of a \textit{chiral vector}. The flat sheet is assumed to be generated using a 2D hexagonal lattice, with lattice vectors $\bfa_1$ and $\bfa_2$, each of length $a_0$ (the lattice vectors for Kagome graphene are given in \ref{subsec:app_TB_kagome_graphene} above). The chiral vector for a nanotube with chirality $(n,m)$ is then $\textbf{C}_h = n \textbf{a}_1 + m \textbf{a}_2$, where $ n,m$ are non-negative integers. The \textit{translation vector} which is parallel to the axis of the nanotube is expressed as $\textbf{T} = T_1 \textbf{a}_1 + T_2 \textbf{a}_2 $. Furthermore, $\gcd(T_1,T_2) = 1$, $T_1,T_2$ $\in \Z$, and:
\beq
T_1 = \frac{2m+n}{d_r}\,,\,T_2 = -\frac{2n+m}{d_r}\,,\,d_r = \gcd(2m+n,2n+m)\,.\nonumber
\eeq
The translation vector determines the axial periodicity of the nanotube, such that in real space, the unit cell of the nanotube is described by a rectangle generated by $\bfC_h \times \bfT$. The reciprocal lattice vectors corresponding to $\bfC_h$ and $\bfT$ are $\bfk_\perp$ and $\bfk_\parallel$, respectively. Since, $\bfC_h$ and $\bfT_h$ are orthogonal vectors, this gives $\bfC_h \cdot \bfk_\perp = 2\pi$, $\bfC_h \cdot \bfk_\parallel = 0$, $\bfT \cdot \bfk_\perp = 0$, and $\bfT \cdot \bfk_\parallel = 2\pi$. By substituting  $\textbf{k}_{\perp}$ and $\textbf{k}_{\parallel}$ into the orthogonality constraints, we get:
\begin{align}
 \textbf{k}_{\perp} = \frac{1}{\calN}(-T_2 \textbf{b}_1 + T_1 \textbf{b}_2)\,,\label{eq:k_perp} \\
 \textbf{k}_{\parallel} = \frac{1}{\calN} (m \textbf{b}_1 - n\textbf{b}_2)\,,\label{eq:k_parallel}
\end{align}
where, $\bfb_1 =\frac{2\pi}{a_0}(\bfa_2 \times \bfz)$ and $\bfb_2 = \frac{2\pi}{a_0}(\bfz \times \bfa_1)$  are the unit vectors of the reciprocal lattice of the two-dimensional lattice, $a_0$ is the lattice parameter, and $\bfz$ is perpendicular to both $\bfa_1$ and $\bfa_2$. Furthermore, $\calN$ is the number of primitive unit cells per nanotube unit cell, and can be expressed as:
\beq
\calN = \frac{|\bfC_h \times \bfT|}{|\bfa_1 \times \bfa_2|}\,.
\eeq
Note that, for the zigzag and armchair tubes considered here, $\calN$ is related to the cyclic group order as $\calN = 2\,\mathfrak{N}$.

The energy dispersion relations (i.e., electronic band diagrams) can be obtained \cite{Dresselhaus_CNT_textbook} by first writing the TB Hamiltonian of the flat sheet, and then applying the periodic boundary condition along the direction of chiral vector, $\bfC_h$, i.e., $\psi(\bfx + \bfC_h) = \psi(\bfx)$. This leads to the condition  $\exp(i\bfk \cdot \bfC_h) = 1$ and quantizes the wave vector $\bfk_\perp$, thus giving  rise to $\calN$ discrete wavevectors parallel to $\bfk_\perp$, in the direction of $\bfC_h$. The wavevector $\bfk_\parallel$, which is parallel to the axis of the nanotube, remains continuous, however. Since the two-dimensional sheet is being rolled-up to form a quasi-one-dimensional nanotube, the Brillouin zone (BZ) is one-dimensional, and the condition $\bfT \cdot \bfk_\parallel = 2\pi$ implies that the length of the Brillouin zone is $2\pi/T$.
Thus, the $\calN$ one-dimensional energy dispersion relations for the nanotube can be parametrized as $\epsilon\left(s \textbf{k}_{\perp} + \frac{\textbf{k}_{\parallel}}{|\textbf{k}_{\parallel}|}k \right)$, with $s=0,\dots,\calN-1$ and $\displaystyle -\frac{\pi}{T} \leq k < \frac{\pi}{T}$. 

Note that this construction immediately implies that a zigzag or armchair CKNT of cyclic group order $\mathfrak{N}$ will feature $2\,\mathfrak{N}$ flat bands, since the starting point is the Kagome graphene TB Hamiltonian (\ref{subsec:app_TB_kagome_graphene}) featuring a single flat band across its Brillouin zone. This is consistent with the observations made from first principles calculations (Section \ref{sec:electronic_prop}).
\begin{center}
---
\end{center}
\section*{Acknowledgements}
This work was supported by grant DE-SC0023432 funded by the U.S. Department of Energy, Office of Science. This research used resources of the National Energy Research Scientific Computing Center, a DOE Office of Science User Facility supported by the Office of Science of the U.S. Department of Energy under Contract No. DE-AC02-05CH11231, using NERSC awards BES-ERCAP0025205 and BES-ERCAP0025168. ASB acknowledges startup support from the Samueli School Of Engineering at UCLA, as well as funding from UCLA’s Council on Research (COR) Faculty Research Grant. ASB also acknowledges support through a Faculty Career Development Award from UCLA's Office of Equity, Diversity and Inclusion. ASB would like to thank Felipe Jornada (Stanford Univ.), Giulia Galli (Univ.~of Chicago), Garnet Chan (Caltech), Ellad Tadmor (Univ.~of Minnesota), Lin Lin (Univ.~of California, Berkeley), David J.~Singh (Univ.~of Missouri), Suneel Kodambaka (Virginina Tech.), Yves Rubin (UCLA) and Rahul Roy (UCLA) for insightful discussions and suggestions, and Neha Bairoliya (Univ.~of Southern California) for providing encouragement and support, during the preparation of this manuscript.  SS would like to thank Richard D. James (Univ.~of Minnesota), and acknowledge financial support from MURI project grant (FA9550-18-1-0095). The authors would like to thank UCLA's Institute for Digital Research and Education (IDRE) and the Minnesota Supercomputing Institute (MSI) for making available some of the computing resources used in this work.

\end{document}